\documentclass[fleqn,usenatbib,useAMS]{styles/mnras}
\usepackage{newtxtext,newtxmath}
\usepackage[T1]{fontenc}
\usepackage{amsmath}
\usepackage{xcolor}
\usepackage{xspace}
\usepackage{xurl}
\usepackage{array}
\usepackage{graphicx}
\usepackage{orcidlink}
\usepackage{longtable}
\usepackage{makecell}

\let\oldarcsec\arcsec
\renewcommand{\arcsec}{\oldarcsec\xspace}
\newcommand{\um}{\ensuremath{\mathrm{\mu m}}\xspace}
\newcommand{\kms}{\ensuremath{\mathrm{km~s}^{-1}}\xspace}

\def\referee#1{#1}

\title[ACES Continuum Imaging]{ALMA Central Molecular Zone Exploration Survey (ACES) II: 3mm continuum images}

\newcounter{affcounter}

\newcommand{\defaffiliationlabel}[1]{%
  \refstepcounter{affcounter}%
  \expandafter\xdef\csname #1\endcsname{\theaffcounter}%
}

\newcommand{\affref}[1]{$^{\csname #1\endcsname}$}

\newcommand{\affrefs}[1]{%
  $^{%
    \@for\@ref:=#1\do{%
      \@ref\@ifnextchar\@nil{}{,}%
    }%
  }$%
}

\newcommand{\affrefTwo}[2]{$^{\csname #1\endcsname,\csname #2\endcsname}$}
\newcommand{\affrefThree}[3]{$^{\csname #1\endcsname,\csname #2\endcsname,\csname #3\endcsname}$}
\newcommand{\affrefFour}[4]{$^{\csname #1\endcsname,\csname #2\endcsname,\csname #3\endcsname,\csname #4\endcsname}$}

\newcommand{\printaffiliation}[2]{%
  $^{\csname #1\endcsname}$#2\\%
} %
\defaffiliationlabel{uflorida}
\defaffiliationlabel{ukarcnode}
\defaffiliationlabel{ice_csic}
\defaffiliationlabel{ieec}
\defaffiliationlabel{eso}
\defaffiliationlabel{shao}
\defaffiliationlabel{naoc_key}
\defaffiliationlabel{mpe}
\defaffiliationlabel{cfa}
\defaffiliationlabel{colorado}
\defaffiliationlabel{cab_csic}
\defaffiliationlabel{ucn}
\defaffiliationlabel{cassaca}
\defaffiliationlabel{ljmu}
\defaffiliationlabel{mpia}
\defaffiliationlabel{naoj}
\defaffiliationlabel{nrao}
\defaffiliationlabel{ita_heidelberg}
\defaffiliationlabel{izw_heidelberg}
\defaffiliationlabel{radcliffe}
\defaffiliationlabel{upenn}
\defaffiliationlabel{COOL}
\defaffiliationlabel{iff_csic}
\defaffiliationlabel{umass}
\defaffiliationlabel{kiaa_pku}
\defaffiliationlabel{pku_astro}
\defaffiliationlabel{uconn}
\defaffiliationlabel{iaa_taipei}
\defaffiliationlabel{kansas}
\defaffiliationlabel{chalmers}
\defaffiliationlabel{clap}
\defaffiliationlabel{iop_epfl}
\defaffiliationlabel{oaq}
\defaffiliationlabel{inaf_arcetri}
\defaffiliationlabel{jbca}
\defaffiliationlabel{uva}
\defaffiliationlabel{leiden}
\defaffiliationlabel{iaa_csic}
\defaffiliationlabel{anu}
\defaffiliationlabel{ias}
\defaffiliationlabel{ucl}
\defaffiliationlabel{ari_heidelberg}
\defaffiliationlabel{eso_chile}
\defaffiliationlabel{jao}
\defaffiliationlabel{ipm}
\defaffiliationlabel{nanjing}
\defaffiliationlabel{nanjing_key}
\defaffiliationlabel{villanova}
\defaffiliationlabel{mit}
\defaffiliationlabel{umd}
\defaffiliationlabel{ulaserena}
\defaffiliationlabel{gbo}
\defaffiliationlabel{utokyo}
\defaffiliationlabel{aberystwyth}

\author[A. Ginsburg \& ACES Team]{Adam Ginsburg,\affref{uflorida}\orcidlink{0000-0001-6431-9633}
Daniel~L.~Walker,\affref{ukarcnode}\orcidlink{0000-0001-7330-8856}
Ashley~T.~Barnes,\affref{eso}\orcidlink{0000-0003-0410-4504}
Xing Lu,\affrefTwo{shao}{naoc_key}\orcidlink{0000-0003-2619-9305}
\'Alvaro S\'anchez-Monge,\affrefTwo{ice_csic}{ieec}\orcidlink{0000-0002-3078-9482}
\newauthor
Jaime E. Pineda,\affref{mpe}\orcidlink{0000-0002-3972-1978}
Marc W.~Pound,\affref{umd}
Pei-Ying Hsieh,\affref{naoj}\orcidlink{0000-0001-9155-39}
Katharina Immer,\affref{eso}\orcidlink{0000-0003-4140-5138}
Qizhou Zhang,\affref{cfa}\orcidlink{0000-0003-2384-6589}
Nazar Budaiev,\affref{uflorida}\orcidlink{0000-0002-0533-8575}
Savannah R. Gramze,\affref{uflorida}\orcidlink{0000-0002-1313-429X}
Desmond Jeff,\affrefTwo{uflorida}{nrao}\orcidlink{0000-0003-0416-4830}
Claire Cook,\affref{kansas}
Alyssa Bulatek,\affref{uflorida}\orcidlink{0000-0002-4407-885X}
Elisabeth A.C. Mills,\affref{kansas}\orcidlink{0000-0001-8782-1992}
\textit{and the ACES collaboration:}
\newauthor
John Bally,\affref{colorado}\orcidlink{0000-0001-8135-6612}
Laura Colzi,\affref{cab_csic}\orcidlink{0000-0001-8064-6394}
Pablo Garc\'ia,\affrefTwo{ucn}{cassaca}\orcidlink{0000-0002-8586-6721}
Jonathan D. Henshaw,\affrefTwo{ljmu}{mpia}\orcidlink{0000-0001-9656-7682}
\newauthor
Izaskun Jim\'enez-Serra,\affref{cab_csic}\orcidlink{0000-0003-4493-8714}
Ralf S.\ Klessen,\affrefFour{ita_heidelberg}{izw_heidelberg}{cfa}{radcliffe}\orcidlink{0000-0002-0560-3172}
\newauthor
Simon R. Dicker,\affref{upenn}\orcidlink{0000-0002-1940-4289}
Steven N. Longmore,\affrefTwo{ljmu}{COOL}\orcidlink{0000-0001-6353-0170}
Francisco Nogueras-Lara,\affrefTwo{iaa_csic}{eso}\orcidlink{0000-0002-6379-7593}
V\'ictor M. Rivilla,\affref{cab_csic}\orcidlink{0000-0002-2887-5859}
\newauthor
Miriam G. Santa-Maria,\affrefTwo{uflorida}{iff_csic}\orcidlink{0000-0002-3941-0360}
Q. Daniel Wang,\affref{umass}\orcidlink{0000-0002-9279-4041}
Fengwei Xu,\affrefTwo{kiaa_pku}{pku_astro}\orcidlink{0000-0001-5950-1932}
Cara Battersby,\affref{uconn}\orcidlink{0000-0002-6073-9320}
\newauthor
Paul T. P. Ho, \affref{iaa_taipei}\orcidlink{0000-0002-3412-4306}
J.~M.~Diederik Kruijssen,\affref{COOL}\orcidlink{0000-0002-8804-0212}
Maya Petkova,\affref{chalmers}\orcidlink{0000-0002-6362-8159}
\newauthor
Mattia C. Sormani,\affref{clap}\orcidlink{0000-0001-6113-6241}
Robin G. Tress,\affref{iop_epfl}\orcidlink{0000-0002-9483-7164}
Jennifer Wallace,\affref{uconn}\orcidlink{0009-0002-7459-4174}
J. Armijos-Abenda\~no,\affref{oaq}\orcidlink{0000-0003-3341-6144}
\newauthor
Lucia Armillotta,\affref{inaf_arcetri}
N. Bijas,\affref{jbca}\orcidlink{0000-0002-6398-7530}
Rojita Buddhacharya,\affref{ljmu}\orcidlink{0009-0004-0685-7678}
\newauthor
Laura A. Busch,\affref{mpe}
Natalie O. Butterfield,\affref{nrao}\orcidlink{0000-0002-4013-6469}
M\'elanie Chevance,\affrefTwo{ita_heidelberg}{COOL}\orcidlink{0000-0002-5635-5180}
\newauthor
Samuel Crowe,\affref{uva}\orcidlink{0009-0005-0394-3754}
Ana Karla D\'iaz-Rodr\'iguez,\affref{ukarcnode}\orcidlink{0000-0001-9112-6474}
Katarzyna M. Dutkowska,\affref{leiden}\orcidlink{0000-0003-0980-6871}
Rub\'{e}n Fedriani,\affref{iaa_csic}\orcidlink{0000-0003-4040-4934}
\newauthor
Christoph Federrath,\affref{anu}\orcidlink{0000-0002-0706-2306}
Simon~C.~O.~Glover,\affref{ita_heidelberg}\orcidlink{0000-0001-6708-1317}
Qi-Lao Gu,\affref{shao}
Rebecca J. Houghton,\affref{ljmu}\orcidlink{0000-0002-9723-1088}
Yue Hu,\affref{ias}\orcidlink{0000-0002-8455-0805}
\newauthor
Namitha Issac,\affref{shao}\orcidlink{0000-0002-7881-689X}
Janik Karoly,\affref{ucl}\orcidlink{0000-0001-5996-3600}
Mark R. Krumholz,\affref{anu}
Fu-Heng Liang,\affrefTwo{ari_heidelberg}{eso}\orcidlink{0000-0003-2496-1247}
\newauthor
Sergio Mart\'{i}n,\affrefTwo{eso_chile}{jao}\orcidlink{0000-0001-9281-2919}
Farideh Mazoochi,\affref{ipm}
Xing Pan,\affrefThree{nanjing}{nanjing_key}{cfa}
Dylan Par\'e,\affref{villanova}
Thushara G.S. Pillai,\affref{mit}\orcidlink{0000-0003-2133-4862}
\newauthor
Denise Riquelme-V\'asquez,\affref{ulaserena}\orcidlink{0000-0001-5389-0535}
Anika Schmiedeke,\affref{gbo}\orcidlink{0000-0002-1730-8832}
Yoshiaki Sofue,\affref{utokyo}\orcidlink{0000-0002-4268-6499}
\newauthor
Volker Tolls,\affref{cfa}\orcidlink{0000-0003-1841-2241}
Gwenllian M. Williams,\affref{aberystwyth}\orcidlink{0000-0001-5933-2147}
Suinan Zhang\affref{shao}\orcidlink{0000-0002-8389-6695}
Emily Moravec,\affref{gbo}\orcidlink{0000-0001-9793-5416}
\newauthor
Charles E. Romero,\affref{upenn}\orcidlink{0000-0001-5725-0359}
Brian S.\ Mason,\affref{nrao}\orcidlink{0000-0002-8472-836X}
John Orlowski-Scherer\affref{upenn}
and
H Perry Hatchfield\affref{uconn}\orcidlink{0000-0003-0946-4365}
\\
$^{*}$Author affiliations are listed at the end of the paper
}

\date{Accepted 2026 February 04. Received 2026 February 04; in original form 2025 July 18}

\pubyear{2025}

\begin{document}
\label{firstpage}
\pagerange{\pageref{firstpage}--\pageref{lastpage}}
\maketitle

\begin{abstract}
The ALMA Central Molecular Zone Exploration Survey, ACES, has mapped $\gtrsim1000$ square arcminutes at 3 mm toward the center of our Galaxy.
ACES provides the first large-scale, high-resolution ($\sim2.5\arcsec$) view of the central $\sim200$ parsecs of the Milky Way.
In this work, we describe the continuum data processing and present the continuum data products.
In the combined mosaic of 45 individual ALMA mosaics, the typical \referee{RMS} noise achieved is $\sim0.1$ mJy per $\sim2.5$\arcsec beam, though there is a tail of substantially higher noise toward regions with bright continuum structure, especially around Sgr A* and Sgr B2.
In-band spectral indices are measurable for a small fraction of the brightest and most compact sources, enabling distinction between dust-dominated and free-free- or synchrotron-dominated sources.
To recover emission on large angular scales, we present the GBT MUSTANG-2 Three millimeter Extended Nucleus Survey (TENS), \referee{a new 10\arcsec resolution survey of the CMZ}, which we combine with the ACES image by feathering.
To demonstrate the quality and reliability of the ACES data, we compare to previously-published ALMA data obtained with higher resolution and sensitivity, finding overall good agreement with past results, but some disagreement toward the brightest sources.
\end{abstract}


\section{Introduction} \label{sec:intro}

Galactic centres act as the engines of galaxy evolution, funneling gas inward, converting it to stars and expelling it through energetic feedback \citep[e.g.][]{Veilleux2005}. Because the Milky Way’s Central Molecular Zone (CMZ) is the only nucleus where these mechanisms can be \referee{spatially} resolved from global ($>100$\,pc) to star-forming (0.05\,pc) scales and below, it is the benchmark for understanding nuclear star formation, stellar feedback and the feeding of central supermassive black holes \citep[e.g.,][]{Genzel2010, Longmore2013, EHT2022, Henshaw2023}. Paradoxically, despite its dense reservoir of molecular gas, the CMZ is presently forming stars an order of magnitude more slowly than empirical relations predict \citep[e.g.][]{Barnes2017} and appears to do so episodically, perhaps regulated by turbulence and orbital dynamics \citep{Kruijssen2014, Krumholz2017, Sormani2020}. A decisive explanation requires a contiguous, high‑resolution census of where the dense gas -- and the young stars it spawns -- actually reside within the CMZ.

Conducting such a census is hampered by the fact that the CMZ is hidden behind the Galactic plane, rendering it effectively invisible at short (optical) wavelengths. It has been surveyed extensively at all other wavelengths, including X-ray \citep{Wang2002,Muno2009,Heard2013}, near-infrared \citep[e.g.][]{Nogueras-Lara2018,Wang2010}, far-infrared \citep{Molinari2011}, millimeter \citep{Bally2010,Ginsburg2013,Aguirre2011,Tang2021a,Tang2021b,Battersby2020,Pound2018}, and radio \citep{Law2008a,Law2008b,Heywood2022}. However, at millimeter wavelengths, the spatial resolution, coverage, and sensitivity have been limited before ALMA. 

The ALMA Central Molecular Zone (CMZ) Exploration Survey (ACES) continuum mosaic presented here aims to provide a contiguous, high‑resolution census of dense gas and young high-mass stars throughout the CMZ. ALMA's small native field-of-view means that mosaicking is required to cover areas larger than about an arcminute square.
While mosaicking is a well-established technique, it adds a significant layer of complexity to the image processing, both stretching the approximations used in interferometric imaging \citep{Ekers1979,Cornwell1988,Pety2010} and the limits of computing hardware. While previous works have covered substantial areas, spanning 10's of square arcminutes \citep[e.g.,][]{Ginsburg2022,Leroy2021}, ACES enters new territory by imaging a mosaic-of-mosaics spanning 45 individual fields and $>1000$ square arcminutes. 

In this work, we present the continuum data processing approach for the largest mosaic \referee {the ALMA 12m array} has yet produced\referee{, which resulted in additional technical challenges not faced by smaller mosaics-of-mosaics}.
Covering $>$1000 arcmin$^2$ with \referee{$\sim$2.5}$\arcsec$ (0.05\,pc) resolution and $\sim$0.1\,mJy\,beam$^{-1}$ sensitivity \referee{(detailed explanation of these values is in \S\ref{sec:dataprocessing})}, the ACES 3\,mm map (augmented with MUSTANG‑2 to recover $>$3 arcmin scales\referee{; see \S \ref{sec:gbtmustang2})}) supplies column‑density and spectral‑index information across the CMZ.
By mapping the locations of compact dust cores, HII regions and synchrotron emission, these data are pivotal to the ACES science objectives of building a global understanding of star formation, feedback, and the mass flows and energy cycles across the \referee{CMZ}. 

\referee{
This paper presents the continuum images from ACES.
It is accompanied by an overview paper describing the aims of the project \citep{Longmore2025ACES} and three papers describing the cubes in high-resolution \citep{Walker2025_ACESIII}, medium-resolution \citep{Lu2025_ACESIV}, and low-resolution \citep{Hsieh2025_ACESV} spectral windows.
}

This paper presents a summary of the observations in Section \ref{sec:observations}, a detailed description of the data processing and results in Section \ref{sec:dataprocessing}, and a brief set of analyses to validate the data in Section \ref{sec:analysis}.
We conclude with a brief summary.

\begin{figure*}%
    \centering
    \includegraphics[width=\textwidth]{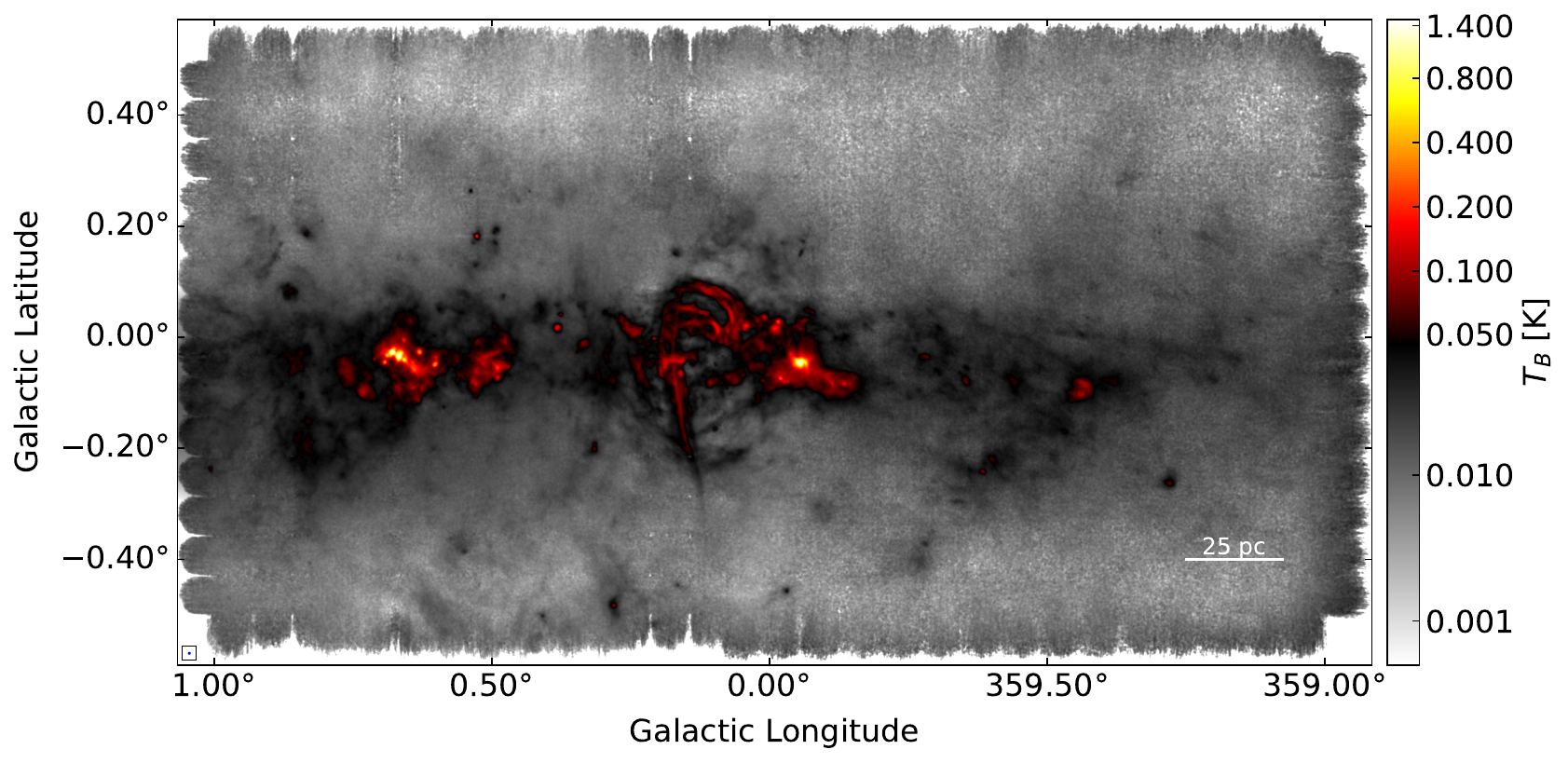}
    \caption{\referee{The GBT MUSTANG-2 TENS image combined with Planck all-sky data that is used to fill in the short spacings missing in the ALMA interferometric data. }}
    \label{fig:gbt}
\end{figure*}

\begin{figure*}%
    \centering
    \includegraphics[width=\textwidth]{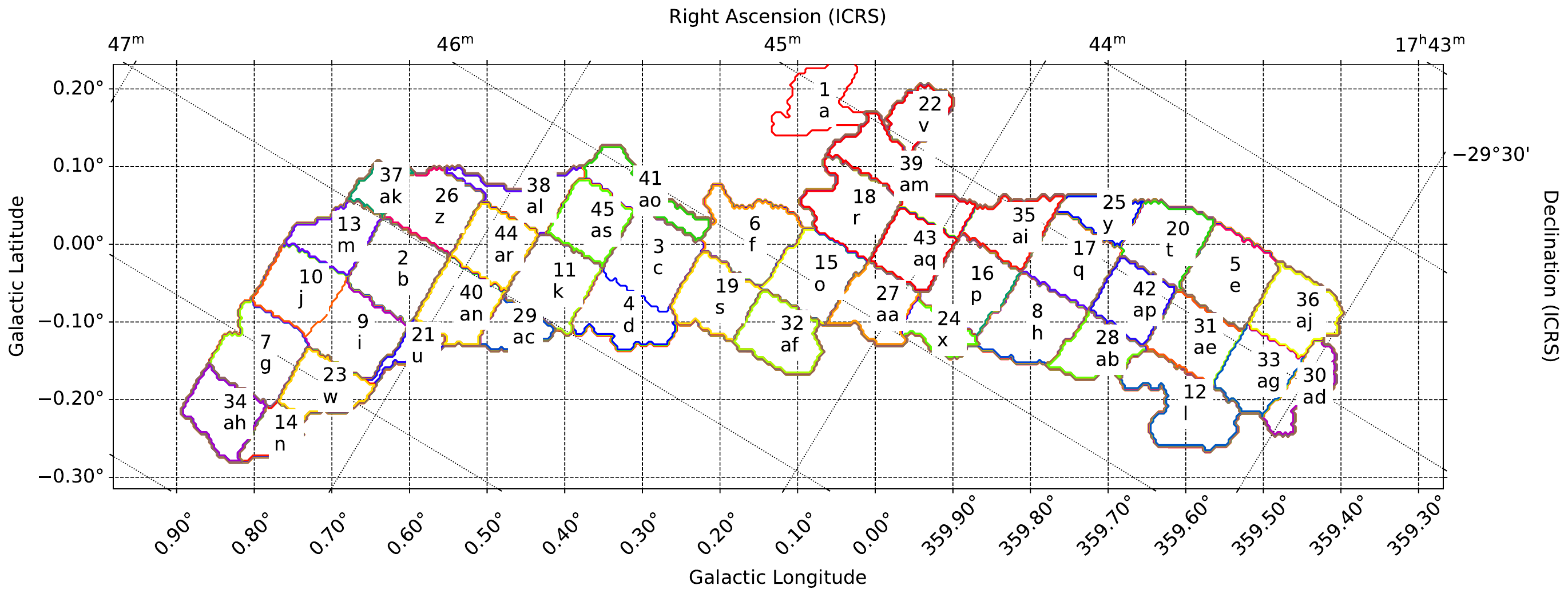}
    \caption{Overview of the ACES fields.  The survey footprint is comprised of 45 mosaics, labeled \texttt{a} through \texttt{as} (1 through 45).
    \referee{The mapping from letters to Galactic coordinates can be found in Table \ref{tab:observation_metadata_12m}.}
    }
    \label{fig:fieldidlabels}
\end{figure*}

\section{Observations}
\label{sec:observations}

Observations in the ACES project were performed in the ALMA Large Program 2021.1.00172.L \citep{Longmore2025ACES}.
ACES used a hybrid spectral window setup, in which two narrow-band windows focused on high-resolution spectral line observations, two medium-band windows compromised between spectral resolution and coverage, and two broad-band windows provide the majority of the continuum sensitivity.
The spectral setup is summarized in Table \ref{tab:spectral_setup}.
The 12m array observations, which are those discussed here (see Section \ref{sec:singledish} for discussion of short spacing recovery), were carried out in
configurations C-3 and C-4 with start dates between January 1 and September 30, 2022\footnote{\url{https://almascience.eso.org/observing/observing-configuration-schedule/prior-cycle-observing-and-configuration-schedule}}.
\referee{A table in the appendix, } Table \ref{tab:observation_metadata_12m},  reports the full details of these
observations, including the field name, time of each execution, the configuration,
the precipitable water vapor (PWV) during the execution, the number of pointings in the scheduling block, the exposure time per mosaic pointing,
and the expected resolution and largest angular scale.

The ACES observations included data from the 7m ACA array, however, we do not present them here.
While the 7m array data fill in the short spacing to some degree, the longest baselines of the 7m array (45m) are smaller than the 100m diameter of the Green Bank Telescope, and there are no ALMA total power (TP) continuum data.
The MUSTANG-2 MGPS90 data (Section \ref{sec:gbtmustang2}) have significantly better angular scale recovery and noise characteristics on the overlap scales; \referee{see Figure \ref{fig:gbt}}.
The 7m data require additional reprocessing and reimaging that we opted not to do for this paper.
The only use case for which the 7m continuum data would be preferred is spectral index measurement on larger angular scales (see Section \ref{sec:spectralindex}).
The other ACES data papers that present images of the line data will include the 7m and TP data \referee{\citep{Walker2025_ACESIII, Lu2025_ACESIV, Hsieh2025_ACESV}}.

The final continuum images from ACES and the feathered ACES plus TENS (see Section \ref{sec:gbtmustang2}) data are shown in \referee{Section \ref{sec:singledish}}.
These images, and the process of producing them, will be discussed in Section \ref{sec:dataprocessing}.
The labeling scheme for the fields is given in Figure \ref{fig:fieldidlabels}.

\begin{figure*}
    \centering
    \includegraphics[width=0.45\textwidth]{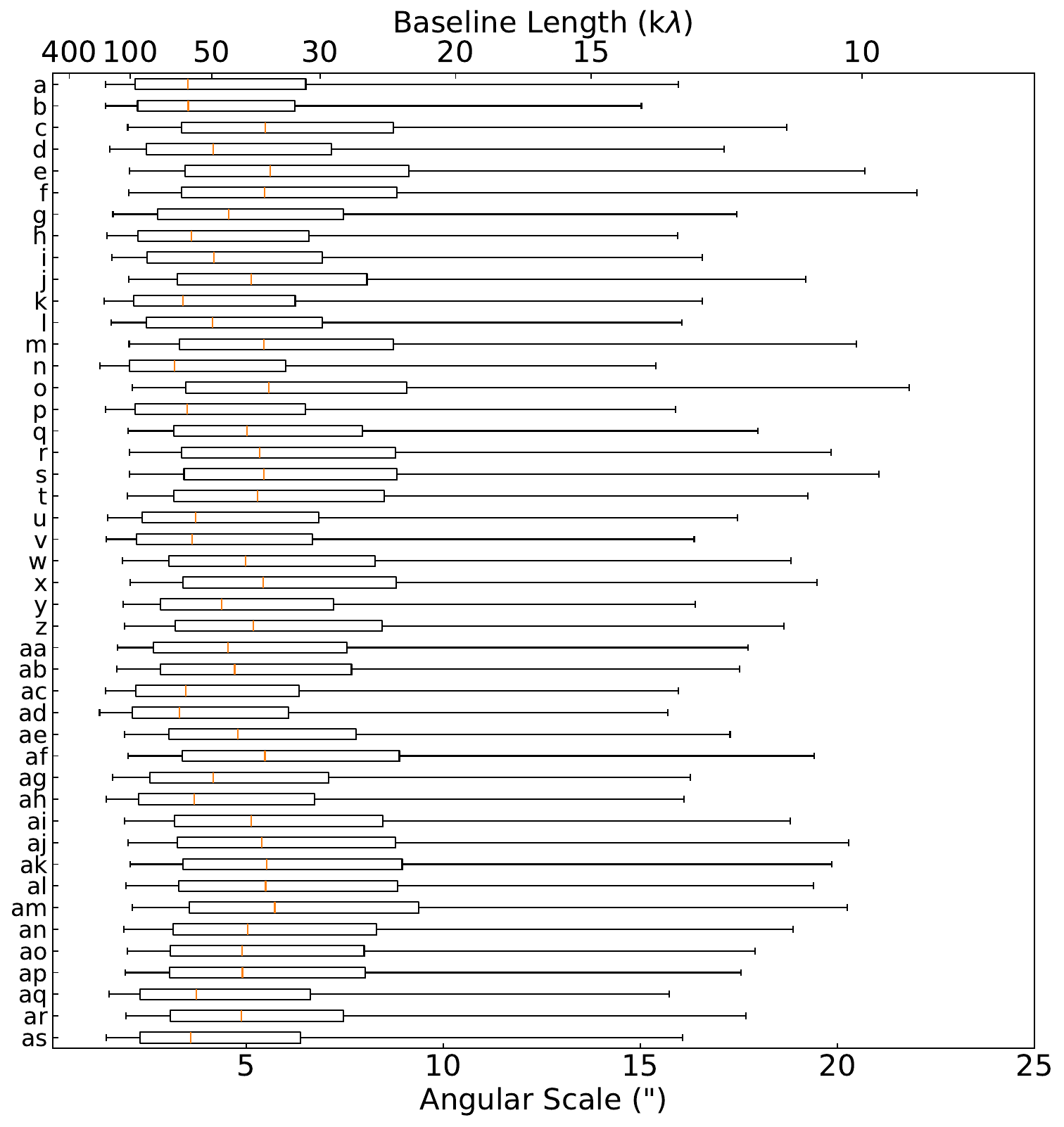}
    \includegraphics[width=0.52\textwidth]{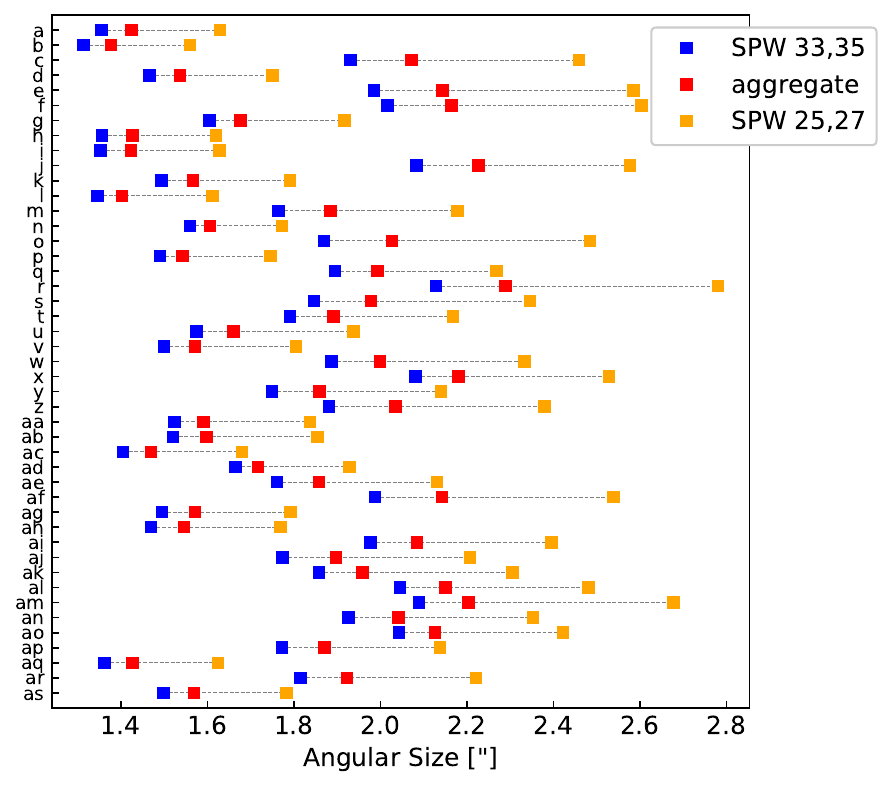}
    \caption{Visualization of the \referee{(u,v)} coverage per field.  (left) The box-and-whisker
    plots in each row show the 5th and 95th percentile as whiskers and the 25th
    and 75th percentile as box ends.  The field identifiers listed on the left 
    side of the plot refer to the labels shown in Figure \ref{fig:fieldidlabels}.
    (right) The beam geometric average sizes for each image product, in arcseconds, for each imaged field.  The points show the high-frequency spectral windows, the aggregate bandwidth image including all spectral windows, and the low-frequency spectral windows in blue, red, and orange, respectively from left to right.
    }
    \label{fig:uvhistograms}
\end{figure*}

The \referee{(u,v)} coverage of ACES was planned to be as uniform as possible. 
However, as a particularly large Large Program, it was necessary to spread ACES observations over several array configurations.
The resulting \referee{(u,v)} coverage is not uniform, as can be seen in Figure \ref{fig:uvhistograms}.
For example, field \texttt{b} has significantly more baselines at larger separation than field \texttt{c}.
The effect of this varying \referee{(u,v)} coverage will be discussed in Section \ref{sec:beams}.

\begin{table*}[htp]
\caption{ACES Spectral Configuration}
\resizebox{\textwidth}{!}{
\begin{tabular}{llllll}
\label{tab:spectral_setup}
SPW & $\nu_{L}$ & $\nu_{U}$ & $\Delta\nu$ & Bandwidth & Lines \\
 & $\mathrm{GHz}$ & $\mathrm{GHz}$ & $\mathrm{MHz}$ & $\mathrm{GHz}$ &  \\
\hline
25 & 85.9656 & 86.4344 & 0.4880 & 0.46875 & HC$^{15}$N 1-0, SO 2(2)-1(1), SiO 2-1 v=1 maser, H$^{13}$CN 1-0\\ 
 &&&&& 86.05496, 86.09395, 86.24337, 86.33992 \\
27 & 86.6656 & 87.1344 & 0.4880 & 0.46875 & H$^{13}$CO+ 1-0, SiO 2-1, HN$^{13}$C 1-0\\ 
 &&&&& 86.7543, 86.84696, 87.09085 \\
29 & 89.1592 & 89.2178 & 0.061030 & 0.05859 & HCO+ 1-0\\ 
 &&&&& 89.18852 \\
31 & 87.8959 & 87.9545 & 0.061030 & 0.05859 & HNCO 4-3\\ 
 &&&&& 87.925238 \\
33 & 97.6625 & 99.5375 & 0.97656 & 1.875 & CS 2-1, CH$_3$CHO 5(1,4)–4(1,3) A–, H40$\alpha$, SO 3(2)-2(1)\\ 
 &&&&& 97.98095, 98.90095, 99.02295, 99.29987 \\
35 & 99.5625 & 101.438 & 0.97656 & 1.875 & HC$_3$N 11-10\\ 
 &&&&& 100.0763 \\
\hline
\end{tabular}
}\par
ACES Spectral Configuration, including a non-exhaustive list of prominent, potentially continuum-affecting, lines.  The included lines are those that are, in at least some portion of the survey, masked out (see Section \ref{sec:continuum_selection}).  The rest frequencies of the targeted lines are given in GHz in the row below their names.  The columns $\nu_{L}$, $\nu_{U}$, and $\Delta\nu$ are the lower frequency, upper frequency, and frequency resolution, respectively.  
\end{table*}

\subsection{GBT MUSTANG-2 TENS}
\label{sec:gbtmustang2}
The ACES data, as with all \referee{interferometric} observations, lack short spacing continuum information.
While the 7m ACA array is intended to provide shorter spacings than the 12m array, its \referee{(u,v)} coverage is only marginally better, with shortest baselines $\sim10$m corresponding to largest angular scales $\sim 60\arcsec$.
In order to improve on this largest angular scale limit, we obtained single-dish data using the Green Bank Telescope's MUSTANG-2 instrument \citep{Dicker2014}, which is a 215-element bolometer array with a bandpass spanning 75-105 GHz.

We used data from the MUSTANG-2 Galactic Plane Survey \citep[MGPS90;][]{Ginsburg2020} and the new TENS (Three millimeter Expanded Nucleus Survey\referee{, which we present here}) continuum images.
The GBT's dish diameter of 100 m overlaps well with the ALMA 12 m array's short baselines (see Figure \ref{fig:uvhistograms} and Section \ref{sec:beams}).
The data reduction process, and half of the data, are described in \citet{Ginsburg2020}.
\referee{The final data product was feathered with Planck all-sky data to fill in the short spacing in the GBT MUSTANG-2 data, which is required because bolometric data processing applies a high-pass filter that behaves similarly to interferometric filtering.}

The MUSTANG-2 Galactic Plane Survey only covered $\ell\gtrsim-0.1$, which left half the ACES footprint unobserved.
We therefore acquired additional GBT observations on May 2, 2024, which covered the right half of the ACES footprint, under the TENS program.
The full area covered by TENS is $ |\ell| < 1$, $|b| < 0.5$
The MUSTANG-2 beam shape and variations in the instrument response were measured every 30 minutes using observations of the nearby radio source J1700-2610.
At the beginning and end of each observation session, the amplitude of this source was tied to multiple ALMA flux calibrators (extrapolated to match MUSTANG-2's bandpass). %
The relative elevation of the sources and the opacity calculated from archival weather data were also used to give a calibration accuracy better than 10\%. %
The calibrated timestreams were then used with the Minkasi\footnote{\url{https://github.com/sievers/minkasi}} maximum likelihood mapmaker.  As some areas of the map are signal-dominated, multiple iterations were used with the positive flux from the previous step being subtracted from the raw data before re-estimating the noise.

Further processing of these data, and their combination with the ACES data, are described in Section \ref{sec:singledish}.

\section{Data Processing}
\label{sec:dataprocessing}

We use CASA for calibration and imaging \citep{CASATeam2022}.
All of the calibration was performed using the appropriate CASA version specified in the ALMA archive data set.
We ran the \texttt{scriptForPI.py} file for each measurement observation unit set (MOUS), i.e., we did not perform any custom calibration.
For most of the ACES data, this was version \texttt{6.2.1.7}, but a few fields required reprocessing or were obtained in a later cycle and were calibrated with \texttt{6.4.1.12}.
However, at later stages, \referee{we used \texttt{6.5.5-21}}; the versions used \referee{for imaging are} reported in the FITS header under the \texttt{ORIGIN} keyword.

\subsection{Data Quality Assessment}
\label{sec:ACESQA}
We performed additional quality assessment, on top of ALMA's QA0 and QA2 processes, to check each scheduling block.
Our assessment included a thorough inspection of the weblogs \referee{(the web pages produced by the ALMA pipeline for visual inspection)} and interactive exploration of the delivered image products.
Downloading and unpacking of the weblogs was handled automatically by the ACES pipeline, and links were created to github issues, one per scheduling block.
The issues can be found on github at \url{https://github.com/ACES-CMZ/reduction_ACES/issues}, with each window labeled by its unique identifier (UID), the ACES field name (Fig. \ref{fig:fieldidlabels}), and the array name.
The full history of our team's analysis of each scheduling block can be found on those Issue pages.

\subsection{Imaging}
\label{sec:imaging}
We imaged individual \referee{mosaic fields} using parameters drawn from the ALMA pipeline.
\referee{All fields were reimaged.}
The pipeline automatically selected parameters that are often user selected, such as the number of pixels, cell size, number of iterations, and threshold\footnote{Except for two fields, \texttt{al} and \texttt{am}, which were manually imaged by the ALMA observatory.}.
We adopted the automated pipeline parameters for all cases where they were available \referee{and the manual imaging parameters otherwise}, increasing \texttt{cyclefactor} in only a few cases (\texttt{ag}, \texttt{i}, \texttt{aa}, and \texttt{z}) to improve convergence of the clean. 
The robust parameter adopted for each field was 0.0, 0.5, or 1.0, as listed in Table \ref{tab:continuum_data_summary_0-35}\referee{; these are determined by the heuristic pipeline to optimize the noise and beam size.}
\referee{We investigated changing robust parameter for the fields with the largest beams, but found no significant improvement, as these were already imaged with a low robust parameter (small beam).}
The complete set of parameters used to image each field can be found using the \texttt{merge\_tclean\_commands.py} script on the imaging code github repository\footnote{\url{https://github.com/ACES-CMZ/reduction_ACES}}.

\subsection{Mosaic Image Products}
\label{sec:imageproducts}
\begin{figure}
    \centering
    \includegraphics[width=\linewidth]{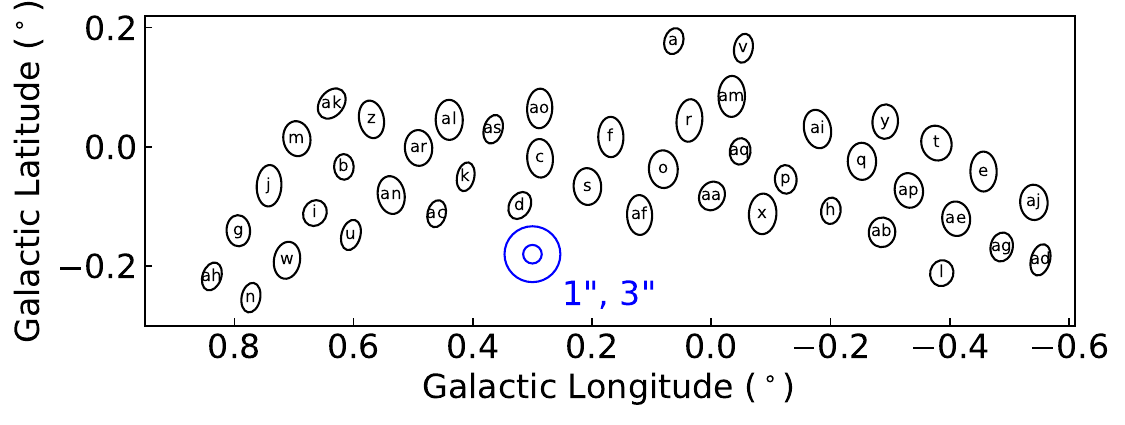}
    \caption{Map of the aggregate-resolution beam relative sizes and orientations by field.
    Each ellipse has major and minor diameter proportional to the full-width half-max (FWHM) of the corresponding beam.
    The concentric blue circles at the bottom-center show the size scale to associate with the beams; they have diameters of 1\arcsec and 3\arcsec (the beams are not shown to scale with respect to the axis labels).
    }
    \label{fig:beamshapemap}
\end{figure}

The ACES data span a frequency range of 86-102 GHz, \referee{such that the frequency range $(\nu_{max}-\nu_{min})/\nu_{max}= 16\%$}, which means the low- and high-frequency ends of the band have appreciably different central frequencies \referee{and therefore effective beam sizes} (see Section \ref{sec:continuum_selection}).
We therefore produce several different continuum mosaic images: one including all continuum-containing windows, one including only the low frequencies ($<87$ GHz, spw 25 and 27), and one including only high frequencies ($>98$ GHz, spw 33 and 35).
Additionally, we produce a mosaic of the images in their native resolution: this image should not be used for quantitative measurements, since it has inconsistent units (Jy beam$^{-1}$, but the beam size varies); instead, it is distributed so that those interested in the highest-resolution views ACES can provide are able to quickly examine the data.
The beam sizes in this final image are shown in Figure \ref{fig:beamshapemap}.
All of these images are produced with the same measurement sets, i.e., the 12m array data.

    \begin{table}[h]
    \centering
    \caption{Continuum Image Products}
    \begin{tabular}{c c c}
    \hline
    \hline
    \label{tab:imagetypes}
        Image Name                & SPWs & Beam \\
        \hline
        Low, circular & 25, 27 & 2.80\arcsec \\
        High, circular & 33, 35 & 2.14\arcsec \\
        All data, circular & 25, 27, 33, 35 & 2.30\arcsec \\
        All data, best resolution & 25, 27, 33, 35 & n/a \\
        \hline
    \end{tabular}
    \end{table}
     
Because of the differing beam size, the overall footprint of ACES is slightly frequency-dependent.
The total areal coverage of the ACES 12m data images is 1286 arcmin$^2$ at the low-frequency end, covering spectral windows 25 and 27, and 1261 arcmin$^2$  for the high-frequency window (33, 35).
The 7m array mosaic (not presented here) covers a slightly larger area.

All of the science-usable mosaic file names contain the string
\texttt{continuum\_commonbeam\_circular\_reimaged}.
The 12m data use three suffixes, \texttt{mosaic}, \texttt{spw25\_27\_mosaic}, and \texttt{spw33\_35\_mosaic}, referring to thee rows of Table \ref{tab:imagetypes}.
Additionally, we provide a file \texttt{12m\_continuum\_reimaged\_mosaic.fits} that is a mosaic of each of the fields imaged at their native resolution.
It is provided as a finder chart for the highest-resolution data.
This image should not be used for measurements because it has inconsistent resolution across the field of view.

\subsection{Angular Resolution \& Beams}
\label{sec:beams}
\begin{figure}%
    \centering
    \includegraphics[width=0.5\textwidth]{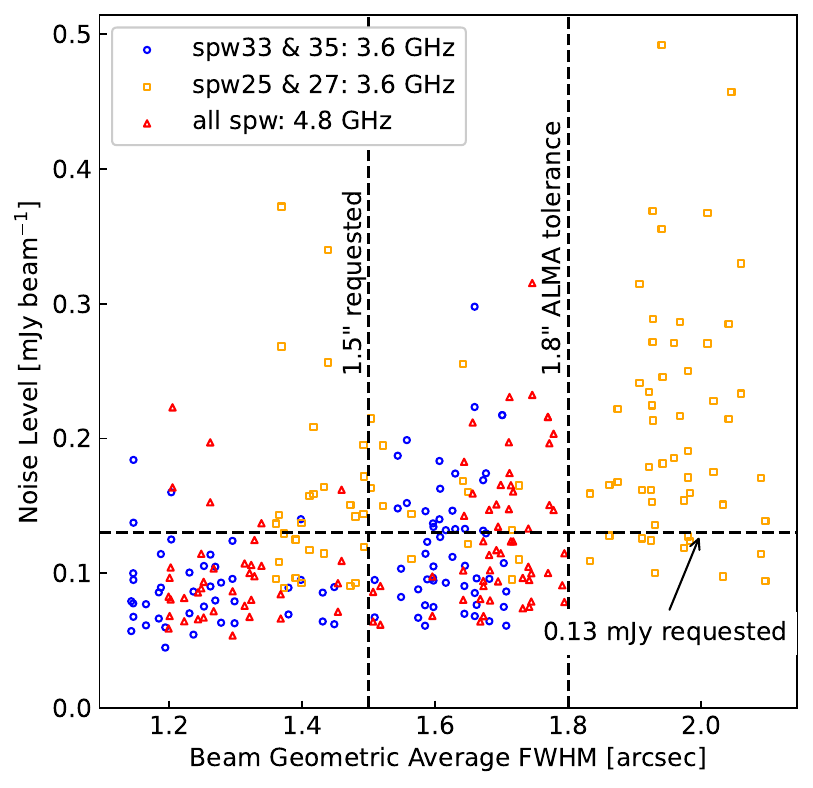}
    \includegraphics[width=0.5\textwidth]{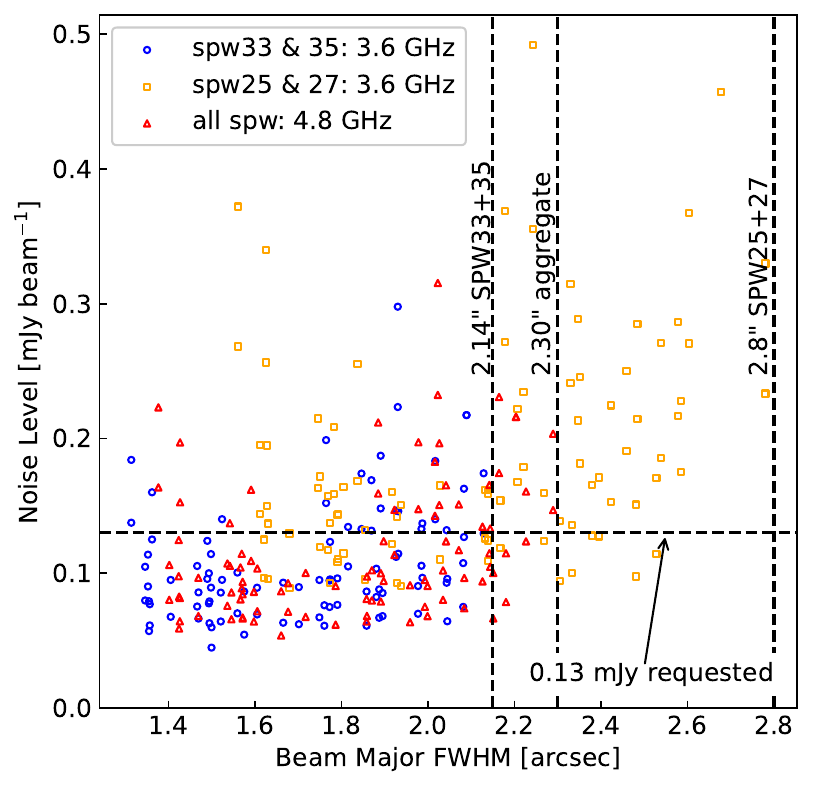}
    \caption{Noise vs beam size for each field.  The plotted beam size is the
    geometric average between the major and minor axes \referee{(top) and the major axis size (bottom)}.  The noise is the
    MAD-estimated standard deviation.}
    \label{fig:noisevsbeam}
\end{figure}

The ACES program as proposed aimed to achieve a uniform beam size of 1.5\arcsec with a noise level of 0.13 mJy beam$^{-1}$.
Because of the limitations noted in Section \ref{sec:observations}, this goal was not achieved.
ALMA's Quality Assessment Phase 2 (QA2) process allows for observations to be accepted and passed to the PI if the noise is within 10\% and the geometric average of the beam axes at the representative frequency of 100.5 GHz is within 20\% of the target\footnote{\url{https://help.almascience.org/kb/articles/what-are-the-current-qa2-requirements}}.
Figure \ref{fig:noisevsbeam} shows that all of the observations using the high-frequency data fell within the 20\% target of 1.8\arcsec using the geometric average (the noise axis will be described in Section \ref{sec:noise}).

The effective resolution of the ACES maps is set by the largest beam major axis, not the geometric average.
A uniform map of the full mosaic is needed to provide a consistently interpretable image of the CMZ.
The common beam we adopt is the smallest circular beam that encompasses all synthesized beams.
The resulting beam sizes in the combined mosaics are given in Table \ref{tab:imagetypes}.
The beam size for each field can be seen in Table \ref{tab:continuum_data_summary_0-35}; field \texttt{am} sets the limiting resolution, though field \texttt{r} has the largest geometric average beam size.

\subsection{Continuum Selection}
\label{sec:continuum_selection}

\begin{figure*}%
    \centering
    \includegraphics[width=\textwidth]{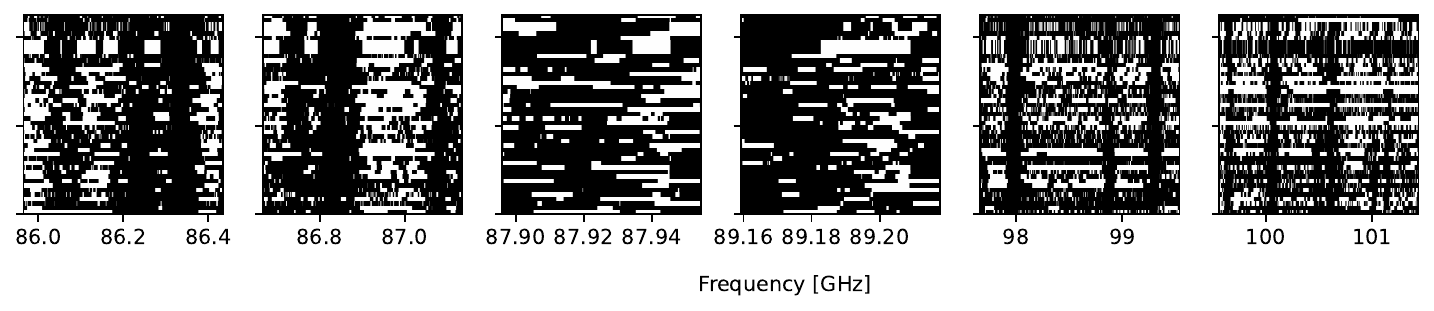}
    \includegraphics[width=\textwidth]{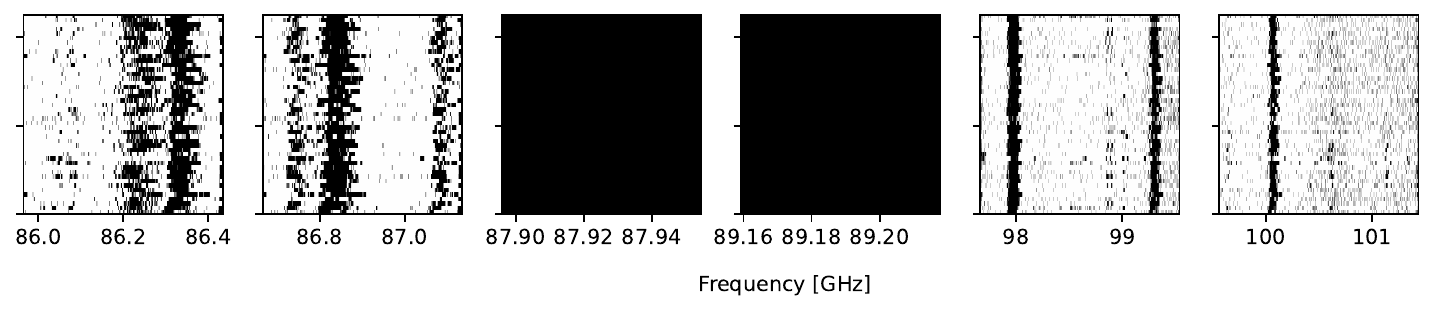}
    \caption{Continuum selection for the 12m data before (top) and after (bottom) re-estimation of the continuum selection as a function of field number (vertical) and frequency (horizontal).
    \referee{These figures give a visual indication of the amount of bandwidth recovered with the \texttt{statcont}-based continuum selection.
    It also shows that the field-to-field variation is greatly reduced using the improved approach.
    }
    Figure \ref{fig:contfrac} labels the fields in the same order as shown here; field labels are omitted in this figure because they would be too small.
    Black regions are those identified as emission line contaminated and are not used for continuum measurement; the middle two windows are entirely excluded as they have line emission at all frequencies.
    Note that the rightmost two panels comprise the majority of the total bandwidth, as the X-axes are not on the same scale.
    }
    \label{fig:continuumselection}
\end{figure*}

The continuum selection was originally performed automatically by the ALMA pipeline.
It resulted in different amounts of bandwidth being included in each field.
Figure \ref{fig:continuumselection}(top) shows the selection as a function of frequency and field number.

\begin{figure*}%
    \centering
    \includegraphics[width=\textwidth]{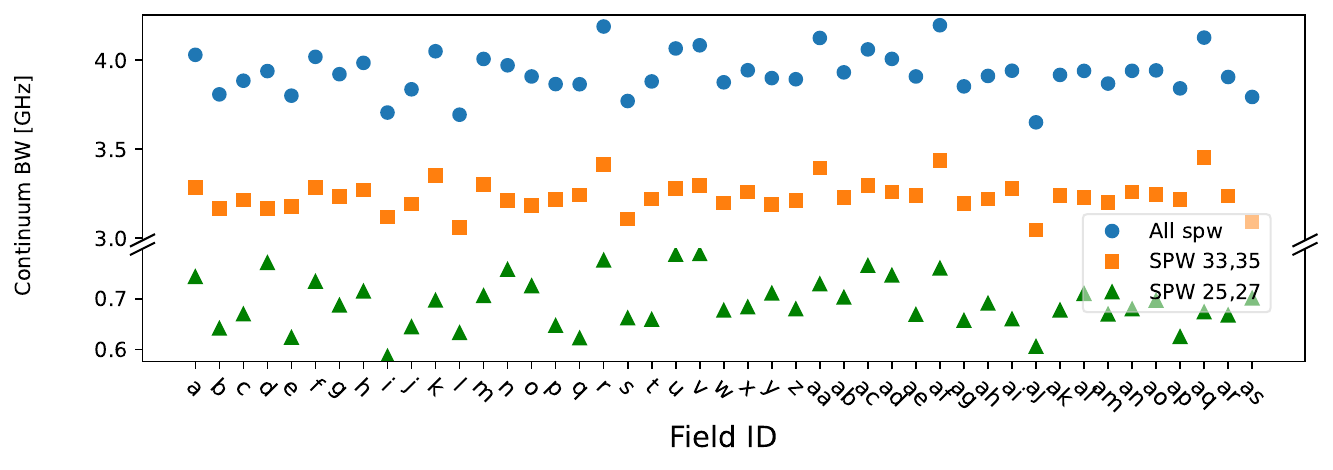}
    \caption{Continuum bandwidth used for continuum mapping vs field number.
    The three colors show the aggregate over all frequencies (blue circles), spectral windows 33+35 (orange squares), and spectral windows 25+27 (green triangles), from top to bottom, respectively.
    Note that spectral windows 25+27 are smaller than 33+35, so the y-axis label is split.
    }
    \label{fig:contfrac}
\end{figure*}

However, during our QA process, we discovered that significant bandwidth was being \referee{unnecessarily} excluded and, particularly in the two narrow windows (spw29 \& 31), some contamination was included.
We therefore recalculated the continuum selection using a process based on \citet{Sanchez-Monge2018}.
We used the fully imaged spectral cubes, which will be described in companion papers \citep{Walker2025_ACESIII,Lu2025_ACESIV,Hsieh2025_ACESV}.
For each cube, we:
\begin{enumerate}
    \item Calculate the peak intensity spectrum, i.e., the brightest pixel in each channel.
    \item Calculate the median and the standard deviation estimated from the median absolute deviation ($\sigma_{MAD}$) \footnote{$\sigma_{MAD} = 1.4286 \mathrm{~MAD}$ if the distribution is Gaussian.  We use the astropy \texttt{mad\_std} function.} of the peak intensity spectrum.
    \item Select pixels as continuum if their values in the peak intensity spectrum are within $\pm 2.5 \sigma_{MAD}$ of the median.
    \item Remove isolated continuum selections by running two iterations of binary erosion\footnote{\url{https://docs.scipy.org/doc/scipy/reference/generated/scipy.ndimage.binary_erosion.html}} followed by one iteration of binary dilation \footnote{\url{https://docs.scipy.org/doc/scipy/reference/generated/scipy.ndimage.binary_dilation.html}}.
\end{enumerate}
The resulting selections, shown in the lower panel in Figure \ref{fig:continuumselection}, are more consistent between fields than the ALMA pipeline selection.
The new selections generally include more bandwidth, and therefore the reprocessed images presented here have higher sensitivity than the archive products\footnote{\url{https://almascience.org/aq/}}.
Figure \ref{fig:contfrac} summarizes the resulting amount of bandwidth included in each field.

These continuum selections result in a variety of different effective central frequencies ($\nu_{\mathrm{obs}}$) for the image products described above (Section \ref{sec:imageproducts}).
The upper and lower frequency images each are nearly monochromatic, independent of the spectral index of the input source.
However, the aggregate bandwidth image has a $\nu_{\mathrm{obs}}$ that depends on the input spectral index.
The resulting intensity-weighted average frequency $\nu_{\mathrm{obs}}$, given by equation 
\begin{equation}
    \label{eqn:nueff}
    \nu_{\mathrm{obs}} = \frac{\int_{\mathrm{obs}} \nu^\alpha \nu d\nu}{\int_{\mathrm{obs}} \nu^\alpha d\nu}
\end{equation}
where $\alpha$ is the spectral index ($\alpha=2$ corresponds to the Rayleigh-Jeans tail of a blackbody) and the subscript $obs$ indicates that the integral is limited to the observed wavelengths included in the continuum, is given in Table \ref{tab:nueff}.

\begin{table}
\centering
\caption{Effective Central Frequencies}
\begin{tabular}{cccc}
\label{tab:nueff}
$\alpha$ & $\nu_{\mathrm{eff,agg}}$ & $\nu_{\mathrm{eff,33+35}}$ & $\nu_{\mathrm{eff,25+27}}$ \\
 & $\mathrm{GHz}$ & $\mathrm{GHz}$ & $\mathrm{GHz}$ \\
-2 & 96.674 & 99.519 & 86.536 \\
-1 & 96.984 & 99.531 & 86.537 \\
0 & 97.271 & 99.543 & 86.539 \\
1 & 97.535 & 99.555 & 86.541 \\
2 & 97.775 & 99.567 & 86.542 \\
3 & 97.994 & 99.579 & 86.544 \\
4 & 98.193 & 99.591 & 86.546 \\
\end{tabular}
\par 
\referee{This table reports the effective central frequency (Equation \ref{eqn:nueff}) for assumed power-law flux indices $\alpha$ for the aggregate (subscript agg), high-frequency (SPW 33+35), and low-frequency (SPW 25+27) windows.}
\end{table}

\subsection{Noise maps}
\label{sec:noise}
We perform noise estimation by locally measuring the RMS in the continuum images.
We operate directly on the images, not the residuals, since the CLEAN algorithm can potentially remove noise from the residual.
The noise we estimate is appropriate for the beam scale, but is not appropriate for larger angular scales.
Scales \referee{somewhat} larger than the beam generally will have lower noise, \referee{since they are sampled with the same (u,v) spacing used to produce the synthesized beam,} while those much larger than the beam will have progressively greater uncertainty \referee{because they are more sparsely sampled}.
The noise level is roughly set by the (u,v) coverage (see Figure \ref{fig:uvhistograms}), but must be empirically measured at any given scale.

The noise estimation technique we adopt minimizes the contribution of point sources to the noise, but accounts for the increased uncertainty produced by the presence of structured extended emission.
We measure the local RMS weighted with a Gaussian with width three times that of the beam major axis.
\begin{equation}
    RMS_j^2 = \Sigma_i \left[ (I_i - \bar{I}_j)^2 K_i \right]
\end{equation}
where $K_i$ is a normalized Gaussian kernel, \referee{$I_i$ is the intensity value at a given pixel, $\bar{I}_j$ is the local mean intensity value
estimated using the same kernel}, and the subscripts $i$ and $j$ are indices of
individual pixels (the evaluation is carried out in two dimensions, but we write
the one-dimensional form here for simplicity).
This calculation is implemented in the \texttt{imaging.make\_mosaic.rms\_map} function in the ACES pipeline.
To remove the contribution of point sources, we iteratively mask out pixels with \referee{surface brightness} values $S_\nu > 2.5 \sigma$ and then recompute $\sigma$ until no more significant pixels are identified.

In detail, the process is:
\begin{enumerate}
    \item Create a smoothed image \texttt{sm} by convolving the image with a Gaussian kernel with width three times that of the beam.
    \item Create a residual image from the original image minus the smoothed, \texttt{res = im - sm}.
    \item Create a variance map by squaring the residual, \texttt{var = res$^2$}.
    \item Compute the average variance over a three beam width area by convolving the variance map with the same Gaussian kernel.
    \item Compute the RMS as the square root of the variance map, $\sigma =$ \texttt{var\_sm}$^{1/2}$.
    \item Mask out pixels with values in the original image greater than 2.5$\sigma$.
    \item Repeat the above steps until no additional pixels are added to the mask.
\end{enumerate}
The convolution steps ignore masked pixels using \texttt{astropy}'s convolve functions.

\begin{figure*}%
    \centering
    \includegraphics[width=\textwidth]{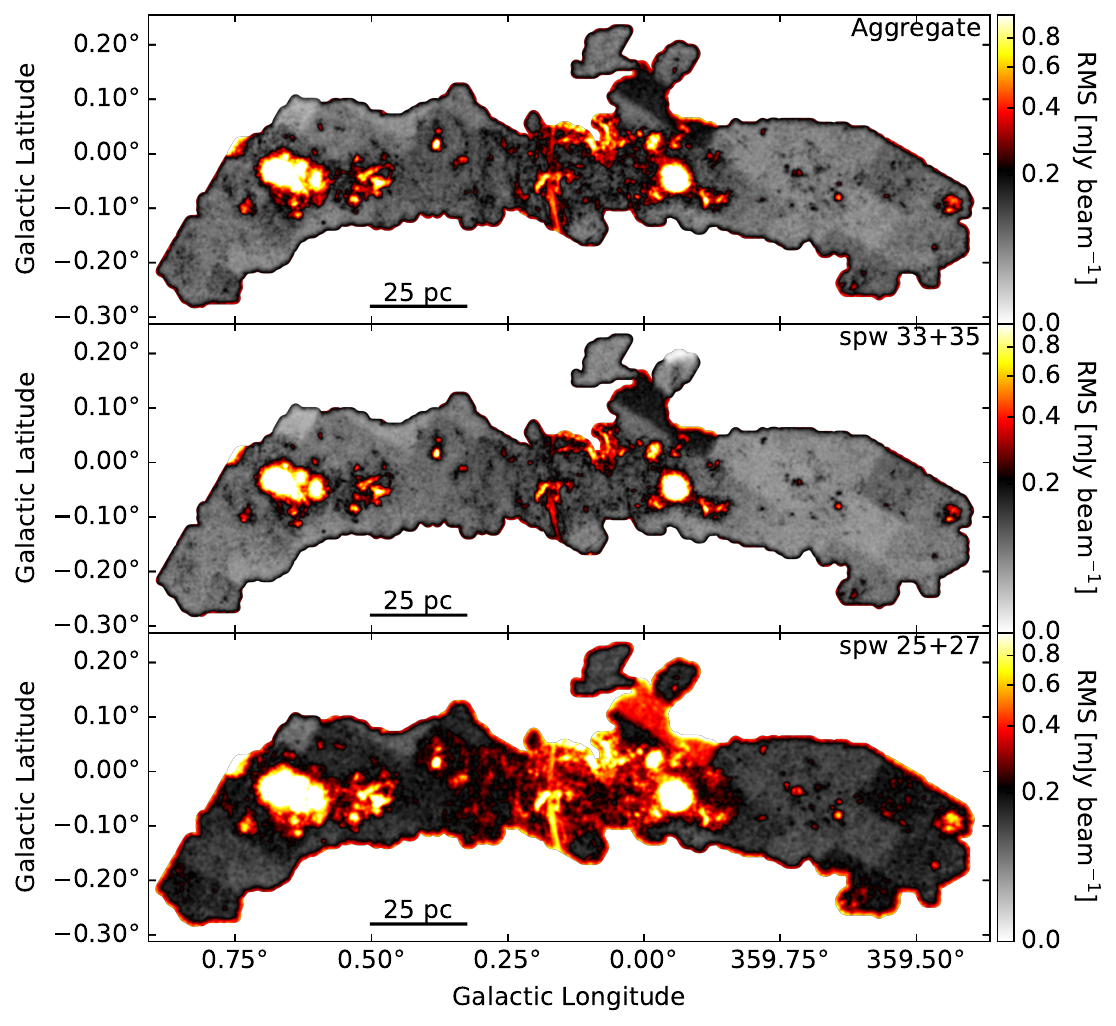}
    \caption{RMS maps.  Top is full bandwidth, middle is spw25+27, bottom is spw33+35.
    Colorbar units are mJy beam$^{-1}$.
    Contours are shown at [10,15,20,25,30] mJy beam$^{-1}$.
    }
    \label{fig:noisemaps}
\end{figure*}

\begin{figure}%
    \centering
    \includegraphics[width=0.5\textwidth]{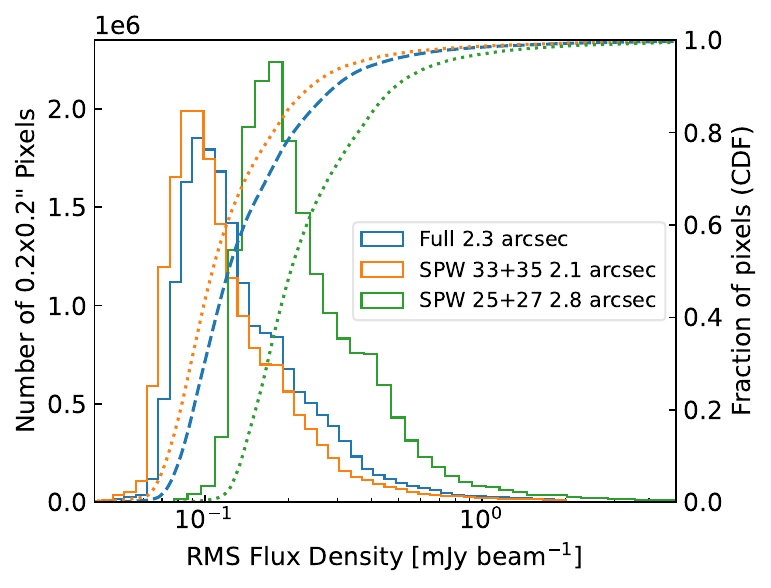}
    \caption{Histograms and cumulative distribution functions (CDFs) of the RMS maps shown in Figure \ref{fig:noisemaps}.
    For the histograms, labeled on the left axis, the X-axis is logarithmic and the Y-axis is linear.
    The CDFs are labeled on the right side and are linear.
    The inset legend labels the colors and provides the beam FWHM for each.
    }
    \label{fig:RMShistograms}
\end{figure}

The resulting noise maps are shown in Figure \ref{fig:noisemaps}, and their histograms in Figure \ref{fig:RMShistograms}.

The noise is remarkably higher in a few regions.
Most significantly, both Sgr A ($\ell\sim-0.05, b\sim-0.05$) and Sgr B2 ($\ell\sim0.68, b\sim-0.01$) have dramatically higher noise because of bright, challenging-to-clean, structure.
Both of these regions would require substantial self-calibration to reduce their noise and increase their dynamic range, but past experience suggests the noise would still remain highly elevated because of the multi-scale extended structure \citep{Ginsburg2018}.
Sgr A also contains the highly-variable point source Sgr A*, which is not well-modeled by a single flux value.
We chose not to attempt significantly deeper cleaning of these regions for this publication, in part because comparable or superior images within these high-noise areas have already been published \citep[Sgr B2;][see also Section \ref{sec:sgrb2compare}]{Ginsburg2018} or because we have not yet determined an appropriate path toward high-dynamic range imaging (Sgr A*).

The remaining high-noise areas are less extreme and simply follow extended structure: this extended structure contributes to the measured RMS, and while its presence limits our ability to identify point sources, it is not truly noise.
One feature that drives the highly structured noise is the penultimate step in the above process, in which only positive pixels are masked out.
Negative bowls surrounding bright extended structures contribute to the RMS estimate and are not masked out.
There is one apparently low-noise region at the top of field \texttt{v} in the spw33+35 maps: the signal in this region has spuriously been suppressed, and we have been unable to diagnose the issue; data in this region of the spw33+35 mosaic image should not be used.
No other regions were so affected, and it is clear that only one spectral window is affected in this particular case.

For individual fields, we present a simple summary statistic rather than a map.
We compute $\sigma_{MAD}$ for each field in their native Jy beam$^{-1}$ unit.
The field-by-field noise is summarized in Figure \ref{fig:noisevsbeam}, where it is shown as a function of beam size.
The noise in each field is a function of both integration time and cleaning depth.

\subsection{Single Dish combination}
\label{sec:singledish}
We merge our images with the MUSTANG-2 Galactic Plane Survey (MGPS90) and TENS (Three millimeter Expanded Nucleus Survey) continuum images.
The data are described in Section \ref{sec:gbtmustang2}; here we discuss the combination with different scales.

The MUSTANG-2 data are themselves insensitive to scales larger than $\sim3$ arcminutes, so to compensate for the missing large scales, we combine the MUSTANG-2 images with Planck all-sky images \citep{Planck2014}.

We combine the ACES aggregate circular beam image with the MGPS90 image.
We adopt a central frequency 97.3 GHz for the ACES data and 87.9 GHz for the MGPS90 data; these frequencies correspond to an assumed spectral index $\alpha=0$, and in adopting that assumption, no scaling of the amplitudes is needed, but steep-index sources ($\alpha>0$) will have mismatched large- and small- angular scale fluxes (see Section \ref{sec:spectralindex} for measurements of $\alpha$).
Before combination, we realigned the MGPS90 image to the ACES data using the \texttt{image\_registration} package's \texttt{chi2\_shift} tool\footnote{\url{https://image-registration.readthedocs.io/en/latest/image_registration.html}}.
We first convolved the ACES data to the MGPS90 resolution.
We found that the MGPS90 data are offset by $d\ell=-1.0, db=2.98$ arcseconds from the ACES data.
We therefore shift the MGPS90 data before combining, and we record this offset in the \texttt{SHIFTED} header keyword.
The data are combined using the feathering technique \citep{Cotton2017} as implemented in the \texttt{uvcombine} package in the \texttt{radio-astro-tools} suite\footnote{\url{https://github.com/radio-astro-tools/uvcombine}}.
\referee{The combined image has the same point spread function (PSF) as the interferometric image, i.e., the synthesized beam, since the synthesized beam is the reconstructed response of a filled aperture to a point source (see \citealt{Rau2019} for caveats about feathering).  
Unlike joint deconvolution, because the feathering is performed after cleaning, it has no effect on the shape of the fitted synthesized beam. }

\begin{figure*}%
    \centering
    \includegraphics[trim={0 50 -25 0},clip, width=\textwidth]{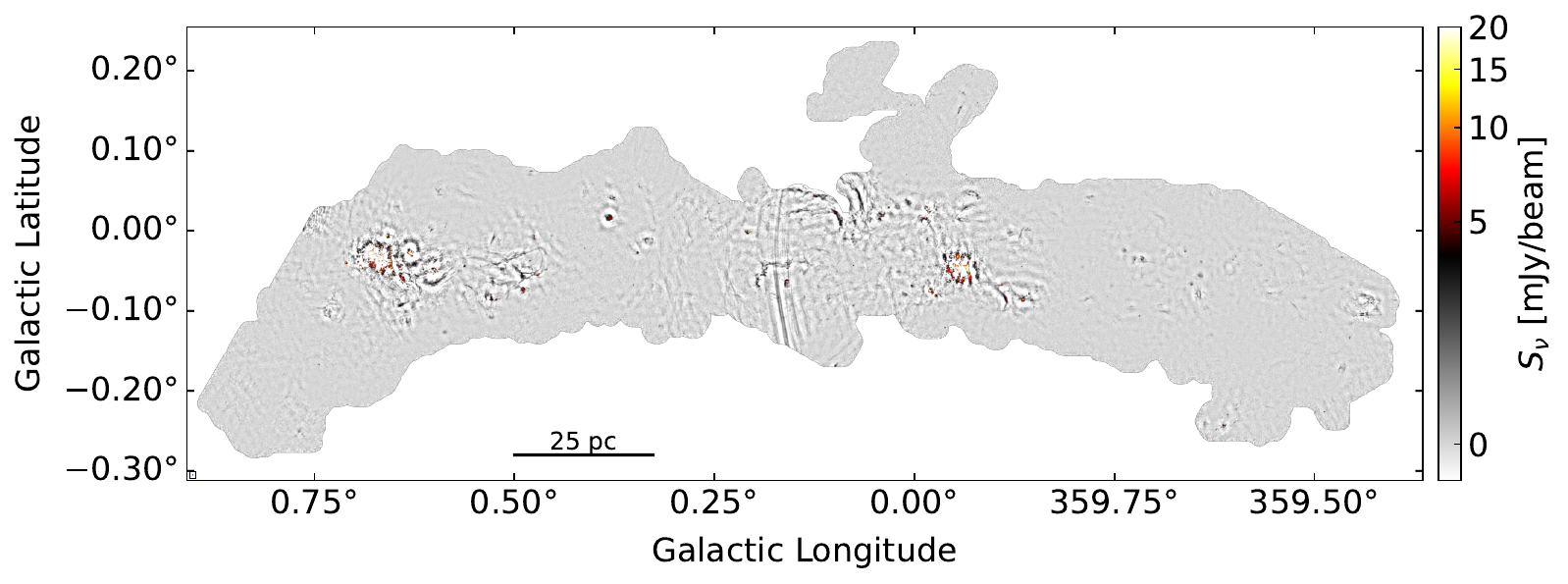}
    \includegraphics[width=\textwidth]{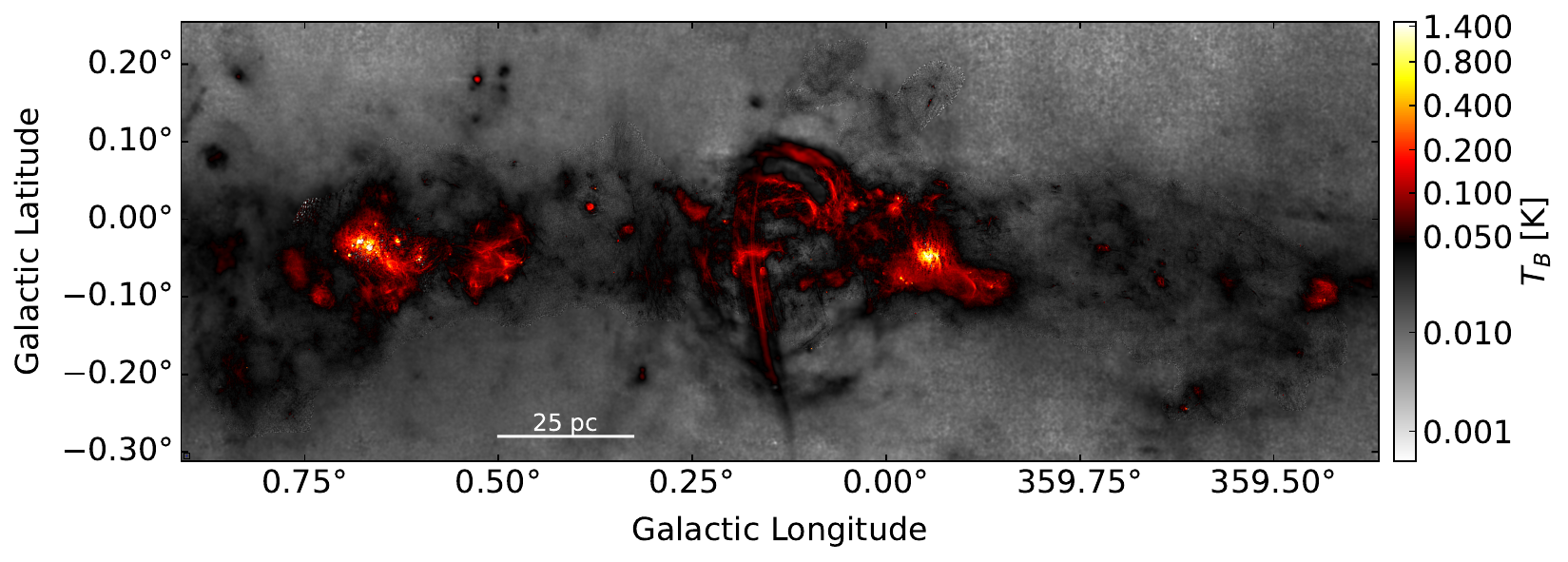}
    \caption{The full footprint mosaic ACES continuum data.
    The top figure shows the ACES 12m data, the bottom figure shows the MUSTANG-2 image feathered with the ACES data.
    The circular beam in both figures is 2.56\arcsec and is shown in the very tiny inset box, but it is a single pixel in this figure because of print and file size limitations. }
    \label{fig:fullfield}
\end{figure*}

\begin{figure*}%
    \centering
    \includegraphics[width=\textwidth]{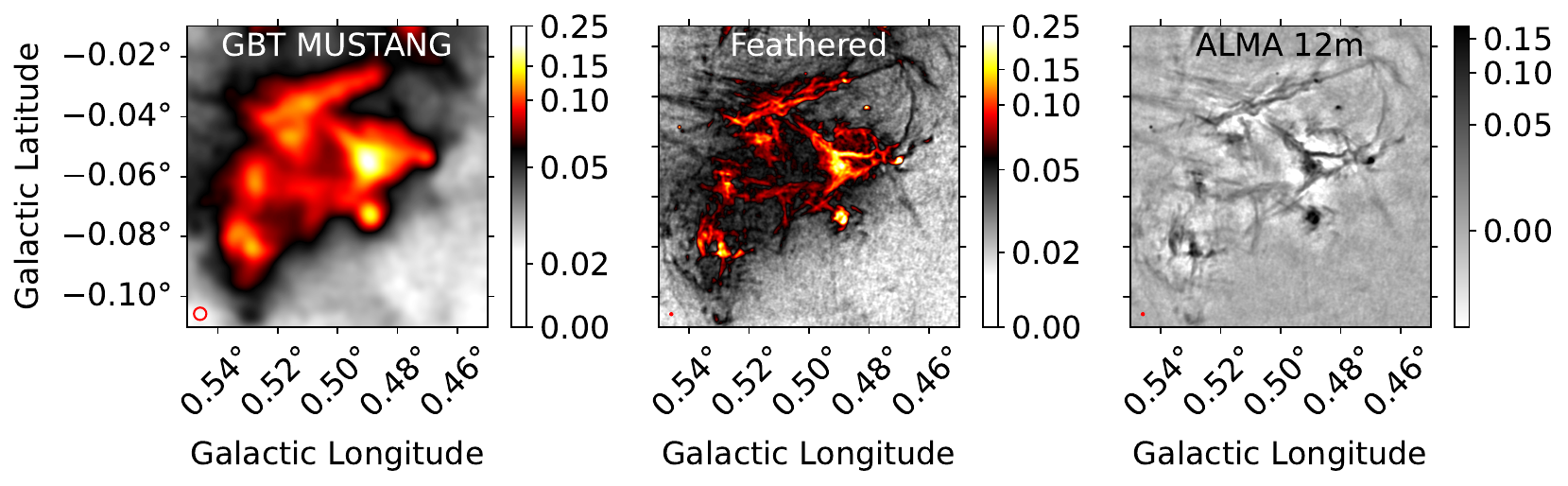}
    \caption{Comparison of the MUSTANG and ACES data and their feathered combination (see Section \ref{sec:singledish}) in a region around Sgr B1.
    The units of the colorbars are in K, and the beam size (red circles) differs in the left plot from the other two.
    The 12m data are intentionally shown in a different color scale, since the missing large angular scales result in overall fainter emission and negative backgrounds.
    }
    \label{fig:featherexample}
\end{figure*}

\referee{The ACES 12m-only image and the feathered, combined image are shown in Figure \ref{fig:fullfield}.
We compare the before and after side-by-side in Figure \ref{fig:featherexample}.}

\section{Analysis}
\label{sec:analysis}
We perform some additional minimal analysis of the data to demonstrate the possibilities, and limitations, of the ACES data.

\subsection{Spectral Index}
\label{sec:spectralindex}
As noted in Section \ref{sec:imageproducts}, the fractional bandwidth from the low end to the high end of the ACES band is $\sim16\%$, which is enough to allow an in-band spectral index measurement in some cases.
The radio spectral index $\alpha$ is defined such that 
\begin{equation}
\frac{S_1}{S_2} = \left(\frac{\nu_1}{\nu_2}\right)^{\alpha}
\end{equation}
Our images were produced with CASA's multi-term, multifrequency synthesis using two Taylor terms (\texttt{tt0} for the spectrally flat component and \texttt{tt1} for the first order term).
We convolved the \texttt{tt1} images to the common beam and stitched them together as with the \texttt{tt0} images.
We then produced a complete mosaic map of $\alpha$ directly from the modeled Taylor terms, $\alpha=\mathtt{tt}1/\mathtt{tt}0$.
Additionally, we manually computed an $\alpha$ map by combining the spw25+27 and spw33+35 images as 
\begin{equation}
\alpha_{manual} = \frac{\log (S_{33+35}/S_{25+27})}{\log\left(99.53/86.54\right)}
\end{equation}
assuming each spw image is monochromatic.

\begin{figure}
    \centering
    \includegraphics[width=\linewidth]{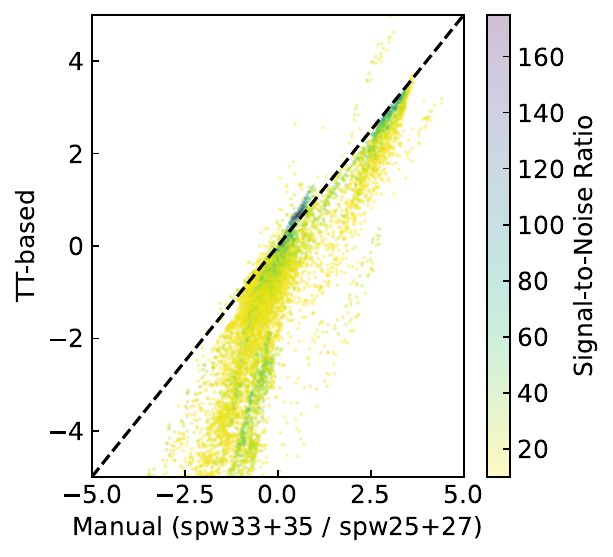}
    \caption{Spectral index $\alpha$ measurements from the Taylor Terms method
    plotted against those measured from comparison of the spw25+27 and the
    spw33+35 images.  The points are color-coded by signal-to-noise ratio.
    While there is a small tendency for higher SNR points to converge closer to the 1-1 line,
    there are many that are far from this line.
    }
    \label{fig:alpha_comparison}
\end{figure}

To assess the reliability of these measurements, we compare the pixel-by-pixel $\alpha$ measurements for all pixels in which the signal-to-noise ratio of the spw25+27 image is greater than 10.
Figure \ref{fig:alpha_comparison} shows that only a small fraction of the total $\alpha$ measurements agree between the two methods.
We have no \textit{a priori} reason to favor one estimate over another, so we recommend that, in general, in-band spectral index measurements from ACES should be used with great caution.
However, we assess the measurements in more detail before fully dismissing them.

Figures \ref{fig:northarches}-\ref{fig:G35963} show zoomed-in cutouts of individual regions.
Each four-panel figure shows the low- and high-frequency 12m images, the manually-calculated $\alpha$ image, and the feathered MUSTANG TENS plus ACES aggregate continuum image.
Much of the extended regions with significant emission exhibit negative spectral index, sometimes very negative, with notable positive-index regions in Sgr C (Fig \ref{fig:sgrc}), the 20 km/s cloud (Fig \ref{fig:20kms}), and F359 (Fig \ref{fig:f35961}), which is a foreground cloud \citep{Reid2019}.
Locations dominated by HII regions, like the rim edges in the Pistol and Sickle nebula (Fig \ref{fig:pistol}) and the compact bright source in Sgr B1 (Fig \ref{fig:sgrb1}), have relatively flat spectral indices ($\alpha\sim0$), as expected for optically thin free-free emission.
Nonthermal filaments are expected to have negative spectral indices, but the only such features with high enough signal-to-noise to measure $\alpha$ are near the Pistol \& Sickle (Fig \ref{fig:pistol}).
A common pattern is that $\alpha\sim0$ toward the central peaks, with $\alpha\rightarrow-4$ immediately around the peaks (e.g., as prominently seen in much of Sgr c; \ref{fig:sgrc}): this pattern hints that the negative indices are caused by improved flux recovery on larger angular scales at longer wavelengths.
Such a pattern is unlikely to be physical in nature.

We therefore suggest that spectral indices measured toward point sources are usable if the background is adequately removed, while spectral indices toward extended regions should generally not be trusted without corroborating data.

\begin{figure*}
    \centering
    \includegraphics[width=\linewidth]{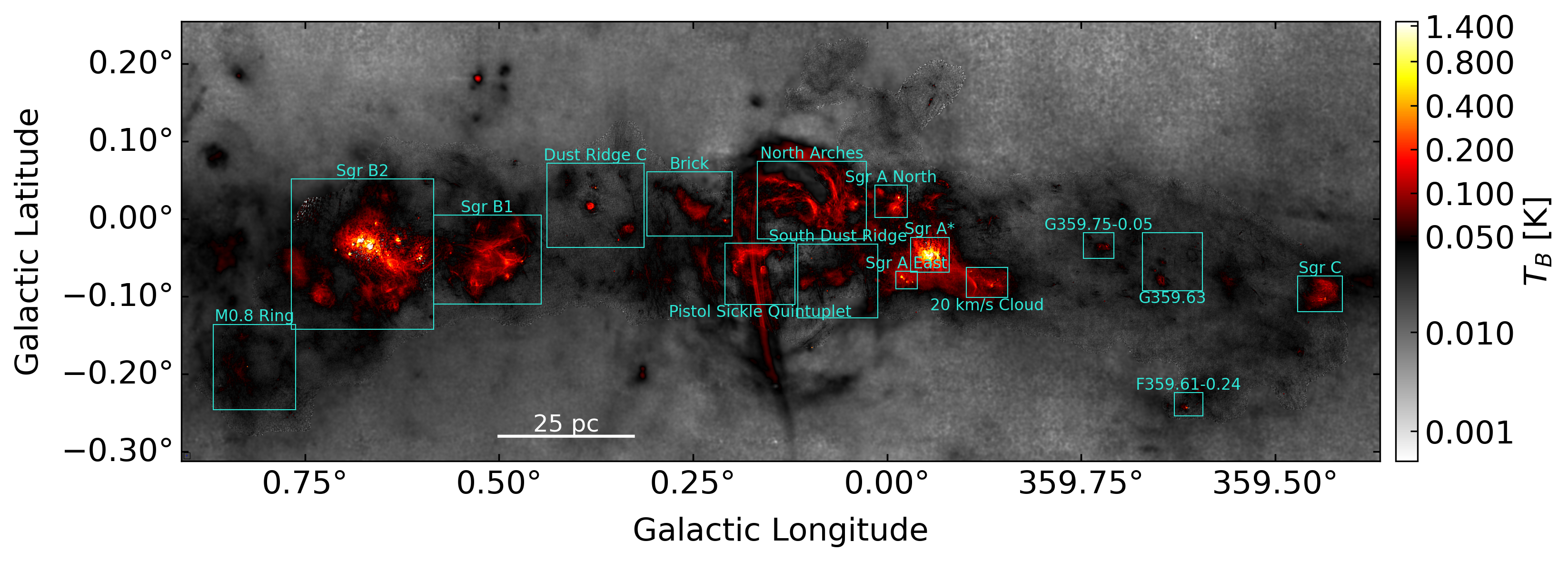}
    \caption{Repeat of Figure \ref{fig:fullfield} with added labels showing the zoom-in figure extraction locations for Figures \ref{fig:northarches}-\ref{fig:G35963}.}
    \label{fig:overviewzoomlabeled}
\end{figure*}

\subsection{Comparison to Other Data}
\label{sec:comparisons}
To validate the ACES data and explore some of the consequences of the specific configurations used in ACES, we compare the ACES images to previously published ALMA observations of Sgr B2 (Section \ref{sec:sgrb2compare}), the Brick (Section \ref{sec:brickcompare}), and other CMZ clouds (Section \ref{sec:luclouds}).
Only ALMA has observed the Galactic Center at comparable wavelength and resolution: while CMZoom with the SMA has comparable resolution, it is at 1 mm \citep{Battersby2020}, and while the CARMA CMZ survey was at a similar wavelength, it has coarser resolution \citep{Pound2018}.
In the following subsections, we compare to previously-published images or those prepared by coinvestigators on ACES.
\citet{Ginsburg2024} examined a specific continuum source in ACES and compared the ACES data to archival ALMA data; we do not repeat their analysis here, but note that that work showcases an example where it may be necessary to use data from individual ACES fields to achieve the best resolution.

\subsubsection{Sgr B2}
\label{sec:sgrb2compare}
We compare the ACES image to the 2013.1.00269.S mosaic of Sgr B2, which was tied for the largest area mapped in the CMZ prior to ACES.
The Sgr B2 image was presented in \citet{Ginsburg2018} and achieved a high dynamic range through self-calibration and the combination of data from multiple ALMA 12m array configurations and the 7m ACA.
The image has $\sim0.5\arcsec$ resolution and a central effective frequency of 96.34 GHz, a little lower than the ACES aggregate image.

\begin{figure}
    \centering
    \includegraphics[width=\linewidth]{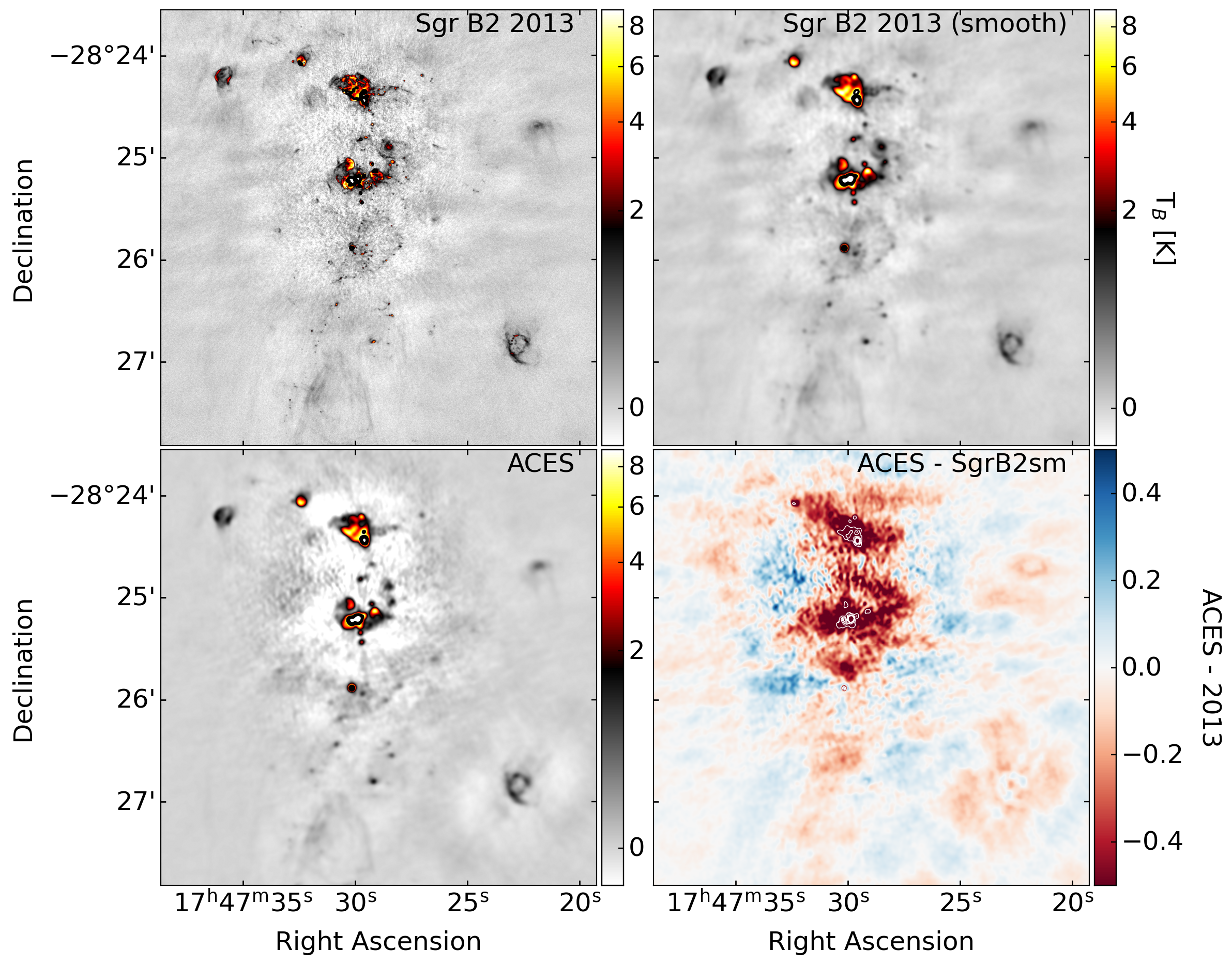}
    \caption{Comparison between the ACES 12m and Sgr B2 2013.1.00269.S \citep{Ginsburg2018} images.
    (top left) The published image from \citet{Ginsburg2018},
    (top right) the same image smoothed to ACES resolution,
    (bottom left) the ACES image,
    and (bottom right) the difference between the ACES image and the smoothed \citet{Ginsburg2018} image.
    The contours in the bottom-right panel are at [-1, -2, -3, -4, -5] K.
    }
    \label{fig:ACESvSgrB2}
\end{figure}

Figure \ref{fig:ACESvSgrB2} shows a comparison between these data sets, in which the Sgr B2 image is convolved to the ACES resolution and then subtracted.
In the difference image, it is apparent that the ACES data under-recover flux on scales of several to several tens of arcseconds, as expected given that the ACES images do not include 7m ACA data and have not been self-calibrated.
The difference at the peaks, Sgr B2 N, M, and S, may be caused by under-cleaning in ACES or a failure to reject line contaminants in the 2013 data set.
The ACES continuum selection was not optimized for the Sgr B2 hot cores, so it may include significant absorption features that reduce the overall flux.
Because of the challenge in imaging Sgr B2 in particular, we refer readers to previous, more careful measurements of these few sources \citep{Sanchez-Monge2017} for continuum analyses.

While the central cores are poorly recovered, the regions in the `extended Sgr B2' cloud far from the N/M/S ridge are well-recovered.
There are absolute differences notable in the comparison image, but they are mostly unstructured on the scales of the ACES data, indicating good recovery.
Nevertheless, for this region, it is evident that the archival data outperform the ACES data in several regards.

\subsubsection{G0.253+0.016: The Brick}
\label{sec:brickcompare}

\begin{figure*}
    \centering
    \includegraphics[width=0.8\linewidth]{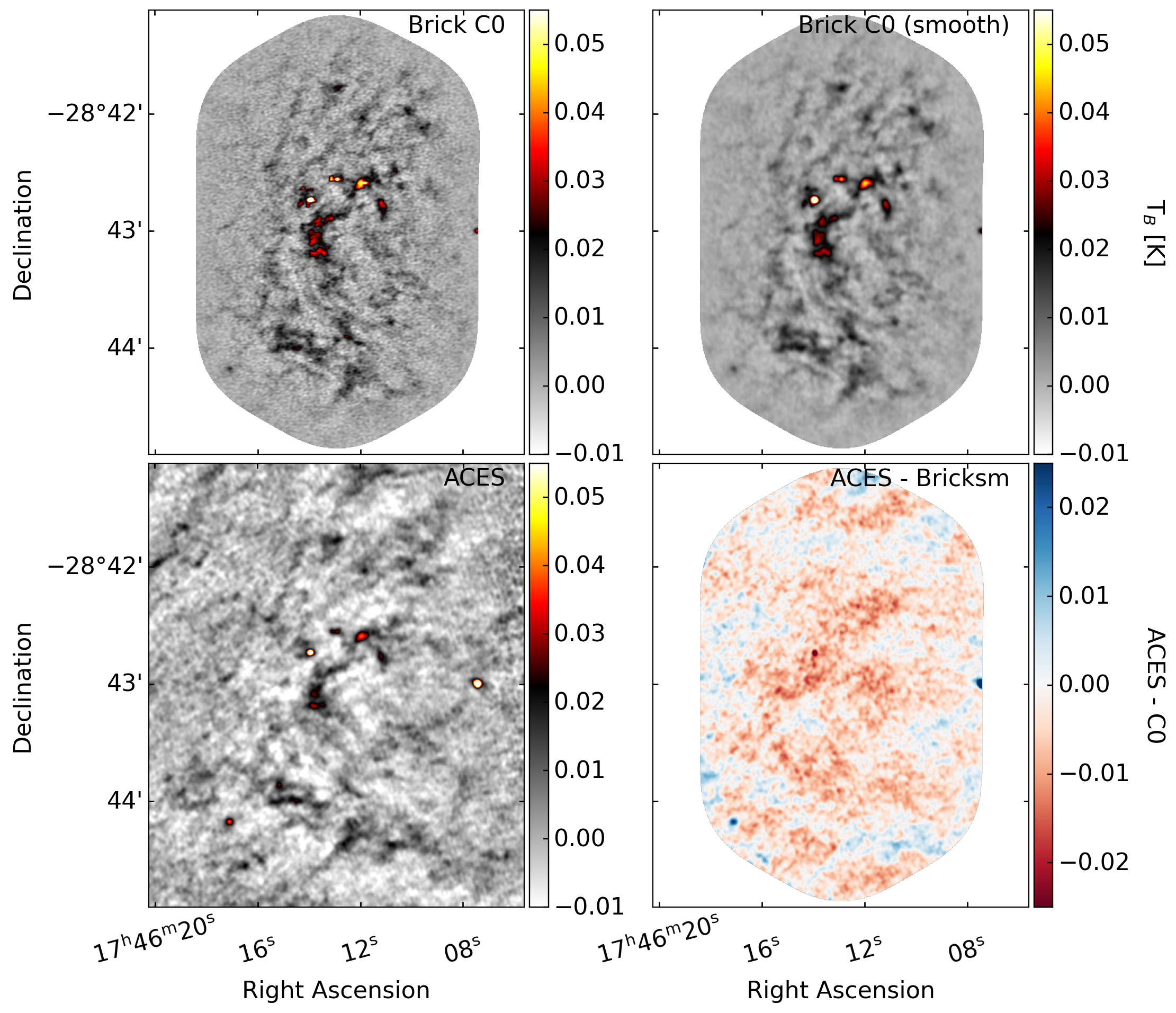}
    \caption{Comparison between ACES 12m and the Cycle 0 mosaic of the Brick \citep{Rathborne2015}.
    The Brick image is not primary-beam corrected, so the signal is artificially
    suppressed around the edges of the C0 image.
    }
    \label{fig:brickc0}
\end{figure*}

\citet{Rathborne2015} published one of the first mosaics created by ALMA, a Cycle 0 (C0) band 3 image of the Brick.
Their image had an effective central frequency $\nu_{\mathrm{obs}} = 94.125$ GHz, reported noise $\sigma_{RMS}=0.025$ mJy beam$^{-1}$ (a few times better than ACES), and $1.7$\arcsec resolution.
The better sensitivity was achieved despite having fewer antennae (25 vs 40-45 in typical ACES observations) by making a smaller mosaic (13 pointings) and revisiting the field \referee{six} times \referee{(compared to two visits spread over 95 pointings for ACES; see Table \ref{tab:observation_metadata_12m})}.
The resulting images are evidently of higher quality, with better \referee{(u,v)} coverage in addition to better sensitivity.
Figure \ref{fig:brickc0} shows the comparison between the ACES and Brick data.
Larger-scale emission is better recovered in the C0 data.

\begin{figure*}
    \centering
    \includegraphics[width=\linewidth]{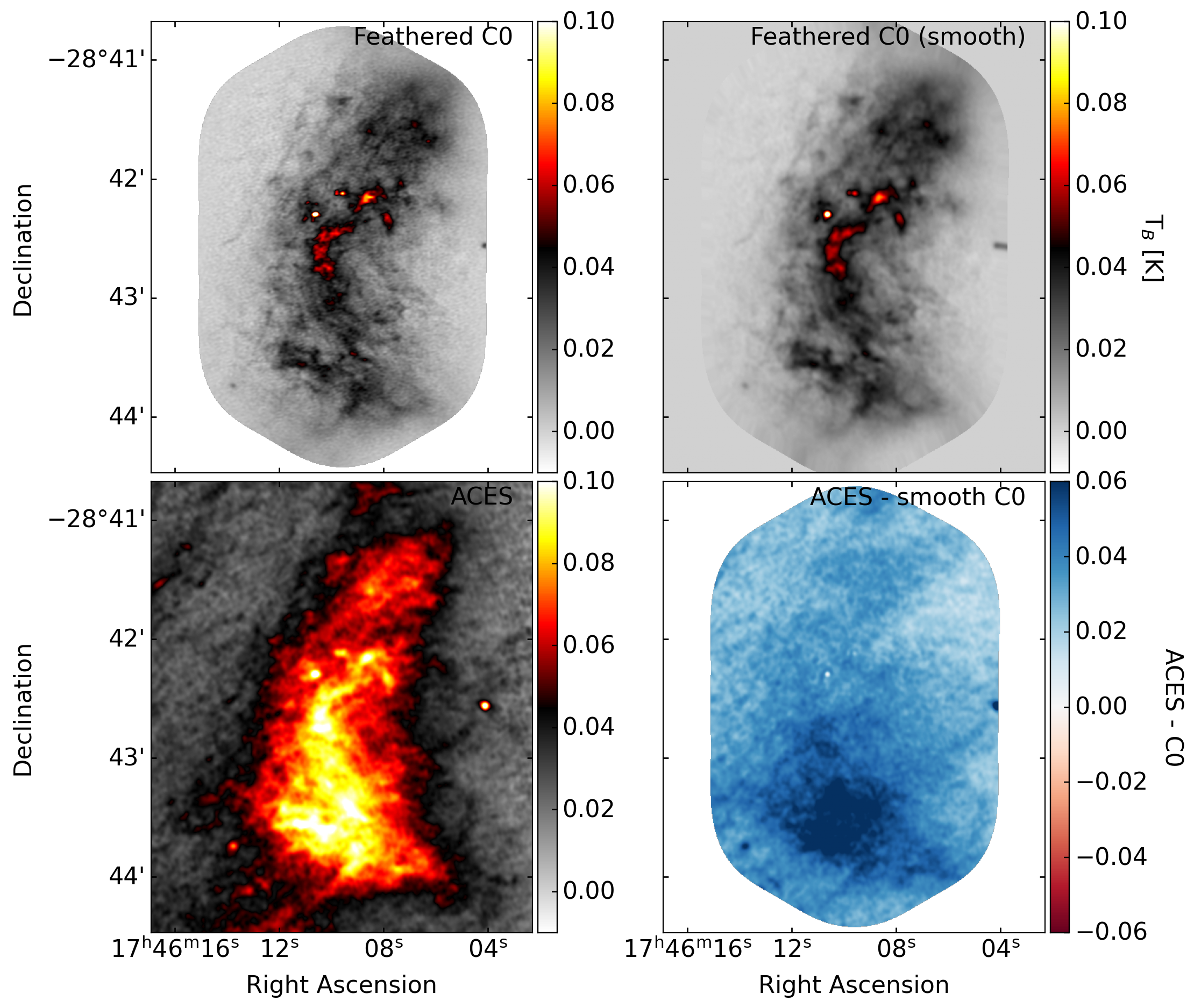}
    \caption{Comparison between ACES plus TENS image and the Cycle 0 mosaic of
    the Brick that included short-spacing \citep{Rathborne2015}.
    \referee{The bottom left panel shows the ACES minus scaled MEERKAT minus the Cycle 0 combined with rescaled Herschel image.  This is intended to show the difference between the ACES+MUSTANG-2 combination and the ALMA Cycle 0 + extrapolated Herschel data after accounting for any free-free contamination in the ACES and MUSTANG-2 data.}
    }
    \label{fig:brickc0feather}
\end{figure*}

In order to fill in the short spacing (large angular scale) emission, \citet{Rathborne2015} extrapolated from a different wavelength.
They scaled the Herschel 500 \um\, image using a fixed spectral index $\alpha=3.2$.
By contrast, we use the MUSTANG TENS 3 mm image, taken at nearly the same wavelength as the ACES data, to fill in our short spacing.
We compare these images in Figure \ref{fig:brickc0feather}, resulting in a striking difference: there is a north-south intensity gradient in the image, with the southern half of the Brick much brighter than the north.
This difference was apparent in the original publication showing the MUSTANG-2 data \citep{Ginsburg2020} but was not commented on there.
The excess emission in the south implies that there is substantial free-free contribution to the 3 mm emission in the Brick \referee{that is not accounted for in the Herschel extrapolation}.

\begin{figure*}
    \centering
    \includegraphics[width=0.75\linewidth]{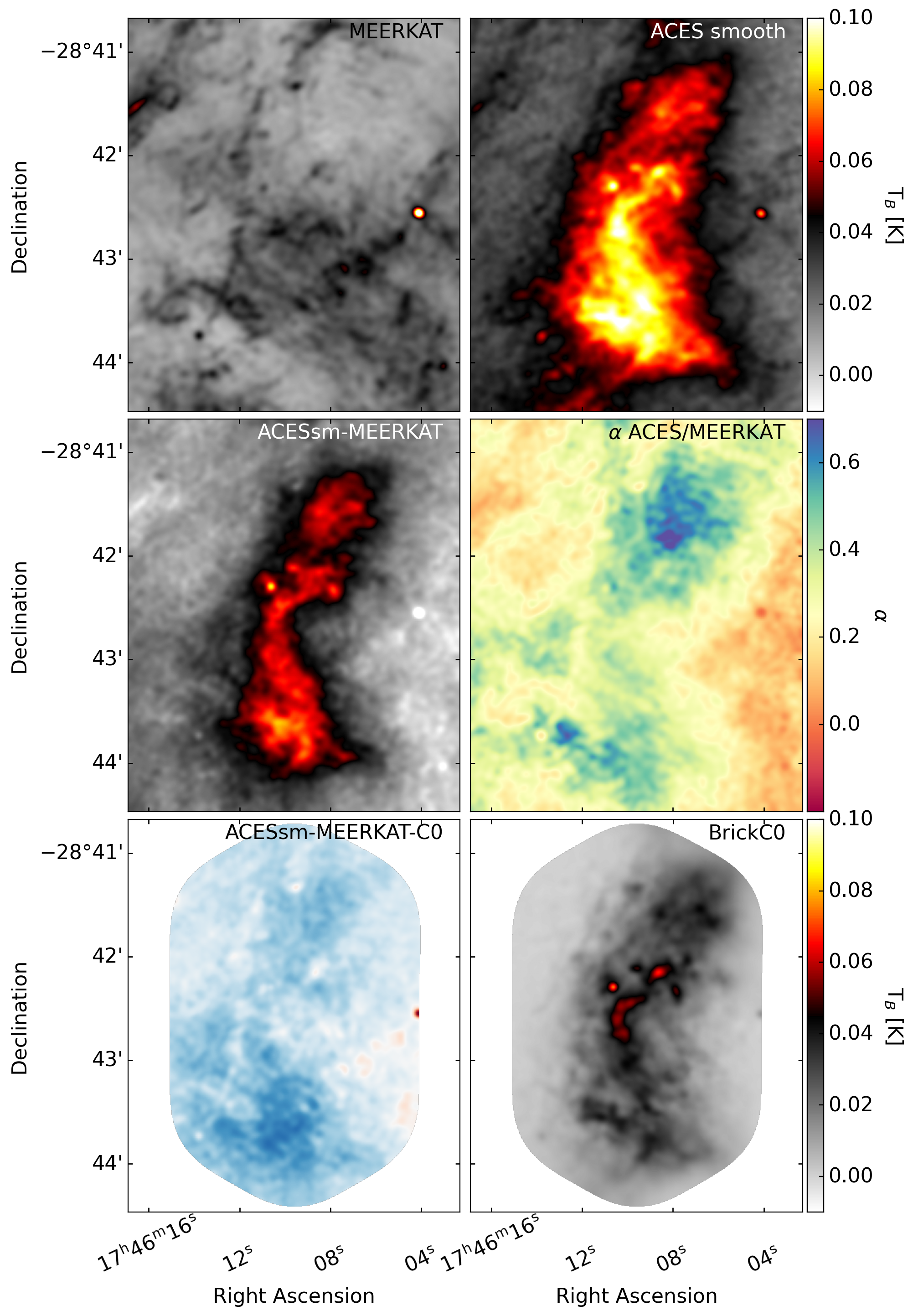}
    \caption{Comparison between ACES plus TENS image and the MEERKAT 1.4 GHz image from \citet{Heywood2022}.
    \referee{
    The MEERKAT image has been scaled by $\alpha=-0.1$ and is shown in brightness temperature units at $\nu=97.1$ GHz such that the colorbars are identical.
    The bottom-left panel shows the same image as the bottom-right panel of Figure \ref{fig:brickc0feather} after subtracting the scaled MEERKAT image.
    }
    }
    \label{fig:brickmeerkat}
\end{figure*}

To evaluate the nature of the disagreement further, we compare to the MEERKAT \referee{1.4 GHz} \citep{Heywood2022} image in Figure \ref{fig:brickmeerkat}.
\referee{
To attempt to match the MEERKAT and ACES data, we scaled the MEERKAT to 95 GHz adopting $\alpha=-0.1$, as expected for optically-thin free-free emission.
}
Figure \ref{fig:brickmeerkat} shows that subtracting the scaled MEERKAT image from the ACES+TENS image brings the latter into closer agreement with the \citet{Rathborne2015} image, but there still remains some variation, either because the spectral index spatially varies or because of a genuine excess at 3 mm in the southern portion of the Brick.

Understanding the spatial and spectral structure of the Brick specifically \citep{Rathborne2015,Federrath2016,Henshaw2019,Walker2021,Ginsburg2023,Petkova2023,Sofue2024}, and any gas structure in the CMZ more generally, clearly remains an open challenge.
We defer further analysis to future work.

\subsubsection{Other CMZ dense clouds}
\label{sec:luclouds}

\begin{figure*}
    \centering
    \includegraphics[width=0.65\linewidth]{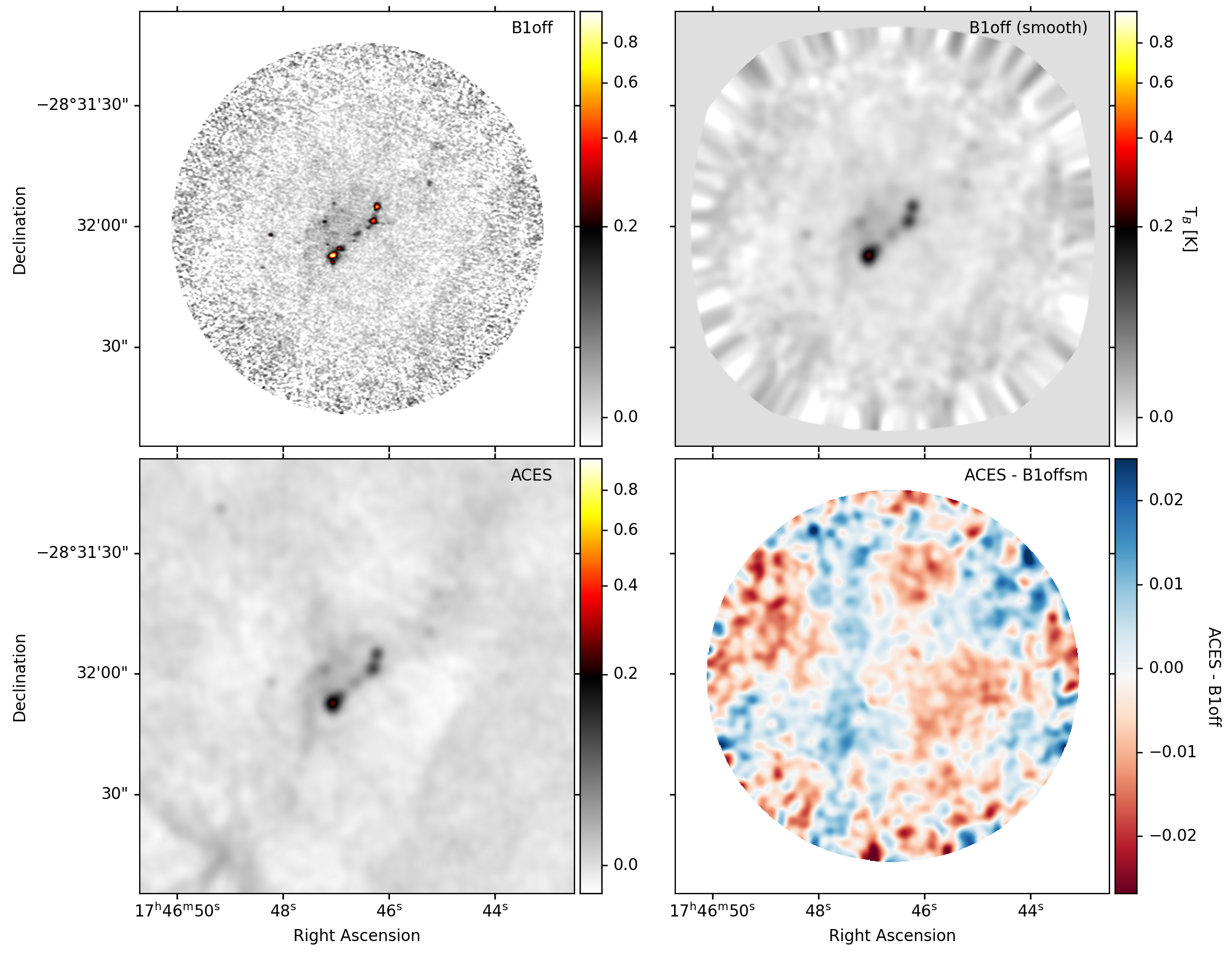}
    \includegraphics[width=0.65\linewidth]{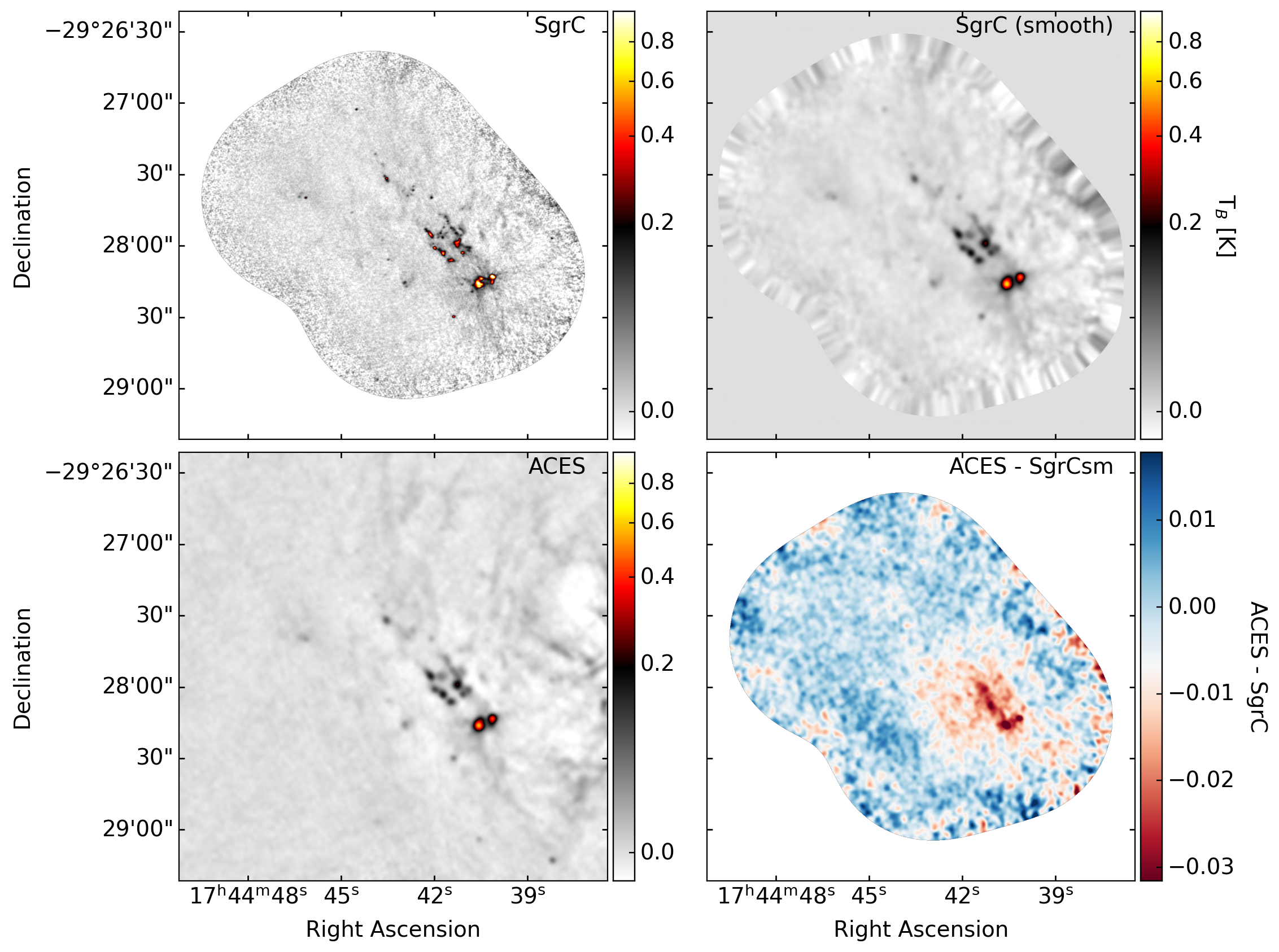}
    \caption{Comparison of ACES to \citet{Xu2025} observations of the G0.6 cloud region and Sgr C.}
    \label{fig:xucomparison}
\end{figure*}

As part of program 2018.1.01420.S (PI: Xing Lu), \citet{Xu2025} surveyed the Sgr C, Cloud E, and 20 km/s clouds using ALMA in configuration C-4 and C-7.
We compare to the shorter-baseline C-4 images in Figures \ref{fig:xucomparison}.
In the 20 km/s cloud and Cloud E, there is small disagreement on large angular scales, which we attribute to the difference in \referee{(u,v)} coverage between the ACES and Xu data sets.
In Sgr C, there is some disagreement in the small-scale structures at the few percent level; this difference is also attributable to differences in \referee{(u,v)} coverage, but may be slightly amplified by the small difference in frequency coverage of ACES vs \citeauthor{Xu2025}

While these targeted observations provide higher-resolution, more detailed views of the star-forming clumps and cores, the ACES mosaics provide a view of the large-scale context.

\begin{figure*}
    \centering
    \includegraphics[width=0.65\linewidth]{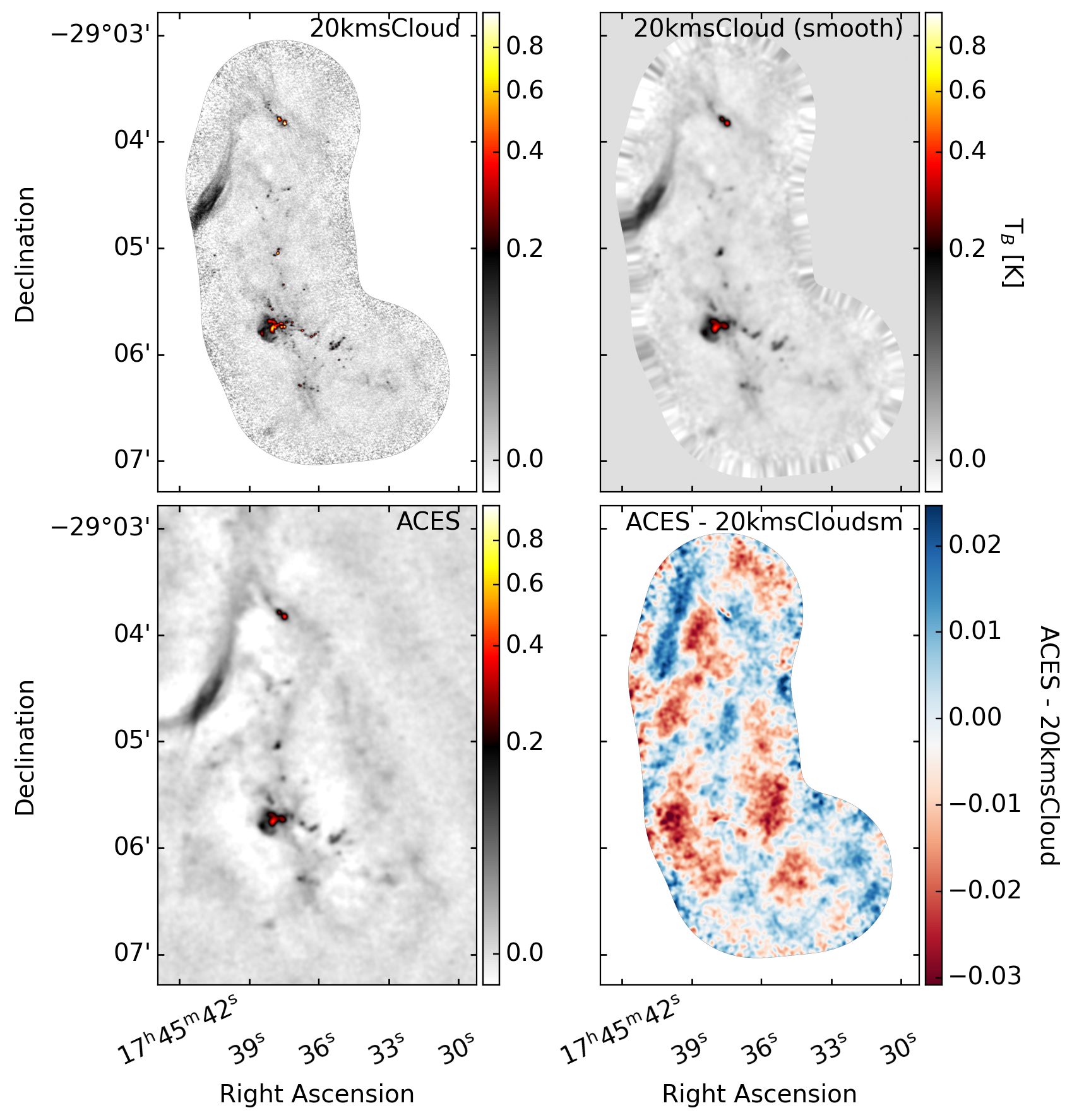}
    \caption{Comparison of ACES to \citet{Xu2025} observations of the 20 \kms cloud.}
    \label{fig:xucomparisonb}
\end{figure*}

\section{Summary}
We have presented the continuum data from the ALMA ACES Large Program supplemented by the GBT TENS MUSTANG project.

These data include:
\begin{itemize}
    \item A $\sim 3$ mm image produced using only 12m array data with $\approx0.1$ mJy beam$^{-1}$ sensitivity and 2.4\arcsec beam.  The largest angular scale recovered in these images is roughly 10-20\arcsec
    \item Two monochromatic images at 86.5 and 99.5 GHz produced from the upper- and lower- sidebands of the ALMA data, with 2.9\arcsec and 2.2\arcsec resolution and 0.2 and 0.1 mJy beam$^{-1}$ RMS, respectively.
    \item A feathered image combining the ACES aggregate continuum data with the MUSTANG-2 TENS image and Planck all-sky data.  This image is sensitive to all angular scales, but is comprised of individual images with different spatial frequency sensitivity.
    \item Images of the in-band spectral index, measured both using CASA's MTMFS imaging and by taking the ratio of the low- and high-frequency ACES mosaic images.
\end{itemize}

\referee{Appendix \ref{sec:appendix_zooms} shows zoom-in views of some of the most prominent features in the CMZ, highlighting the regions where spectral index measurements are most significantly detected.}

We performed basic quality assessment analyses by comparing the ACES data to previous ALMA observations.
We found overall good agreement \referee{with a few exceptions:
\begin{itemize}
    \item Imaging programs with better (u,v) coverage outperform ACES.
    \item Toward bright sources (Sgr B2 and Sgr A*), programs that have undergone careful self-calibration and/or time-dependent calibration outperform ACES.
\end{itemize}
Otherwise, for most of the CMZ, ACES represents the new standard for millimeter continuum observations.}

\section*{Acknowledgements}

\referee{
The paper was instigated and led by the ACES data reduction working group, which is coordinated by Adam Ginsburg, Daniel Walker, and Ashley Barnes, and includes (alphabetically) Nazar Budaiev, Laura Colzi, Claire Cooke, Savannah Gramze, Pei-Ying Hsieh, Katharina Immer, Desmond Jeff, Xing Lu, Elisabeth Mills, Jaime Pineda, Marc Pound, Álvaro Sánchez-Monge, and Qizhou Zhang.
ACES is led by Principal Investigator Steven Longmore, together with co-PIs and the management team, John Bally, Ashley Barnes, Cara Battersby, Laura Colzi, Adam Ginsburg, Jonathan Henshaw, Paul Ho, Izaskun Jiménez-Serra, Elisabeth Mills, Maya Petkova, Mattia Sormani, Robin Tress, Daniel Walker, and Jennifer Wallace. 
The ACES proposal was led by Principal Investigator Steven Longmore, together with co-PIs John Bally, Cara Battersby,  Adam Ginsburg, Jonathan Henshaw, Paul Ho, J. M. Diederik Kruijssen,  Izaskun Jiménez-Serra, and Elisabeth Mills. 
The remaining coauthors comprise the ACES team, and its members contributed to the proposal, data analysis, idea development, and/or reading and commenting on the manuscript.
}

AG acknowledges support from the NSF under grants CAREER 2142300, AAG 2008101, and particularly AAG 2206511 that supports the ACES large program.
I.J-.S., L.C., and V.M.R. acknowledge support from the grant PID2022-136814NB-I00 by the Spanish Ministry of Science, Innovation and Universities/State Agency of Research MICIU/AEI/10.13039/501100011033 and by ERDF, UE. I.J-.S. also acknowledges the ERC Consolidator grant OPENS (project number 101125858) funded by the European Union. V.M.R. also acknowledges the grant RYC2020-029387-I funded by MICIU/AEI/10.13039/501100011033 and by "ESF, Investing in your future", and from the Consejo Superior de Investigaciones Cient{\'i}ficas (CSIC) and the Centro de Astrobiolog{\'i}a (CAB) through the project 20225AT015 (Proyectos intramurales especiales del CSIC); and from the grant CNS2023-144464 funded by MICIU/AEI/10.13039/501100011033 and by “European Union NextGenerationEU/PRTR.
C.~F.~acknowledges funding provided by the Australian Research Council (Discovery Project grants~DP230102280 and~DP250101526), and the Australia-Germany Joint Research Cooperation Scheme (UA-DAAD).
The authors acknowledge UFIT Research Computing for providing computational resources and support that have contributed to the research results reported in this publication.
R.F. acknowledges support from the grants PID2023-146295NB-I00, and from the Severo Ochoa grant CEX2021-001131-S funded by MCIN/AEI/ 10.13039/501100011033 and by ``European Union NextGenerationEU/PRTR''.
D.L.W gratefully acknowledges support from the UK ALMA Regional Centre (ARC) Node, which is supported by the Science and Technology Facilities Council [grant numbers ST/Y004108/1 and ST/T001488/1]. COOL Research DAO \citep{cool_whitepaper} is a Decentralized Autonomous Organization supporting research in astrophysics aimed at uncovering our cosmic origins.
C.B.\ gratefully  acknowledges  funding  from  National  Science  Foundation  under  Award  Nos. 2108938, 2206510, and CAREER 2145689, as well as from the National Aeronautics and Space Administration through the Astrophysics Data Analysis Program under Award ``3-D MC: Mapping Circumnuclear Molecular Clouds from X-ray to Radio,” Grant No. 80NSSC22K1125.
This paper makes use of the following ALMA data: ADS/JAO.ALMA\#2021.1.00172.L, ADS/JAO.ALMA\#2013.1.00269.S, ADS/JAO.ALMA\#2011.0.00217.S. ALMA is a partnership of ESO (representing its member states), NSF (USA) and NINS (Japan), together with NRC (Canada), NSTC and ASIAA (Taiwan), and KASI (Republic of Korea), in cooperation with the Republic of Chile. The Joint ALMA Observatory is operated by ESO, AUI/NRAO and NAOJ.
The authors are grateful to the staff throughout the ALMA organisation, particularly those at the European ALMA Regional Centre, the Joint ALMA Observatory, and the UK ALMA Regional Centre Node, for their extensive support, which was essential to the success of this challenging Large Program.
The GBT data were acquired under projects GBT23A-268 \& GBT18A-014. The National Radio Astronomy Observatory and Green Bank Observatory are facilities of the U.S. National Science Foundation operated under cooperative agreement by Associated Universities, Inc.
FHL acknowledges support from the ESO Studentship Programme, the Scatcherd European Scholarship of the University of Oxford, and the European Research Council’s starting grant ERC StG-101077573 (`ISM-METALS').
J.Wallace gratefully acknowledges funding from National Science Foundation under Award Nos. 2108938 and 2206510.
E.A.C.\ Mills  gratefully  acknowledges  funding  from the National  Science  Foundation  under  Award  Nos. 1813765, 2115428, 2206509, and CAREER 2339670. F.N.-L. gratefully acknowledges financial support from grant PID2024-162148NA-I00, funded by MCIN/AEI/10.13039/501100011033 and the European Regional Development Fund (ERDF) “A way of making Europe”, from the Ramón y Cajal programme (RYC2023-044924-I) funded by MCIN/AEI/10.13039/501100011033 and FSE+, and from the Severo Ochoa grant CEX2021-001131-S, funded by MCIN/AEI/10.13039/501100011033.

\section*{Data Availability}
All data products and associated documentation can be found at \url{https://almascience.org/alma-data/lp/aces}.

All code and data processing issues are available at the public GitHub repository here: \url{https://github.com/ACES-CMZ/reduction_ACES}.

\referee{
The ACES pipeline is based on a number of open-source astronomy software packages, including \texttt{astropy} \citep{AstropyCollaboration2013,AstropyCollaboration2018,AstropyCollaboration2022}, \texttt{astroquery} \citep{Ginsburg2019astroquery}, \texttt{spectral-cube} \citep{Ginsburg2019spectralcube}, \texttt{radio-beam} \citep{Koch2025}, pvextractor \citep{Ginsburg2016pvextractor}, \texttt{reproject} \citep{Robitaille2020reproject}, \texttt{image\_registration} \citep{Ginsburg2014}, \texttt{statcont} \citep{Sanchez-Monge2018}, CASA \citep{CASATeam2022}, numpy \citep{Harris2020}, CARTA \citep{Comrie2021}, and matplotlib \citep{Hunter2007}.
}

\bibliographystyle{mnras}
\bibliography{bib}

\section*{Author Affiliations}
\printaffiliation{uflorida}{Department of Astronomy, University of Florida, P.O. Box 112055, Gainesville, FL 32611, USA}
\printaffiliation{ukarcnode}{UK ALMA Regional Centre Node, Jodrell Bank Centre for Astrophysics, The University of Manchester, Manchester M13 9PL, UK}
\printaffiliation{ice_csic}{Institut de Ci\`encies de l'Espai (ICE), CSIC, Campus UAB, Carrer de Can Magrans s/n, E-08193, Bellaterra, Barcelona, Spain}
\printaffiliation{ieec}{Institut d'Estudis Espacials de Catalunya (IEEC), E-08860, Castelldefels, Barcelona, Spain}
\printaffiliation{eso}{European Southern Observatory (ESO), Karl-Schwarzschild-Stra{\ss}e 2, 85748 Garching, Germany}
\printaffiliation{shao}{Shanghai Astronomical Observatory, Chinese Academy of Sciences, 80 Nandan Road, Shanghai 200030, P.\ R.\ China}
\printaffiliation{naoc_key}{State Key Laboratory of Radio Astronomy and Technology, A20 Datun Road, Chaoyang District, Beijing, 100101, P. R. China}
\printaffiliation{mpe}{Max-Planck-Institut f\"ur extraterrestrische Physik, Gie\ss enbachstra\ss e 1, 85748 Garching bei M\"unchen, Germany}
\printaffiliation{cfa}{Center for Astrophysics | Harvard \& Smithsonian, 60 Garden Street, Cambridge, MA 02138, USA}
\printaffiliation{colorado}{Center for Astrophysics and Space Astronomy, Department of Astrophysical and Planetary Sciences, University of Colorado, Boulder, CO 80389, USA}
\printaffiliation{cab_csic}{Centro de Astrobiolog{\'i}a (CAB), CSIC-INTA, Carretera de Ajalvir km 4, 28850 Torrej{\'o}n de Ardoz, Madrid, Spain}
\printaffiliation{ucn}{Instituto de Astronom\'ia, Universidad Cat\'olica del Norte, Av. Angamos 0610, Antofagasta, Chile}
\printaffiliation{cassaca}{Chinese Academy of Sciences South America Center for Astronomy, National Astronomical Observatories, CAS, Beijing 100101, China}
\printaffiliation{ljmu}{Astrophysics Research Institute, Liverpool John Moores University, 146 Brownlow Hill, Liverpool L3 5RF, The UK}
\printaffiliation{mpia}{{Max Planck Institute for Astronomy, K\"{o}nigstuhl 17, D-69117 Heidelberg, Germany}}
\printaffiliation{naoj}{National Astronomical Observatory of Japan, 2-21-1 Osawa, Mitaka, Tokyo 181-8588, Japan}
\printaffiliation{nrao}{National Radio Astronomy Observatory, 520 Edgemont Road, Charlottesville, VA 22903, USA}
\printaffiliation{ita_heidelberg}{Universit\"{a}t Heidelberg, Zentrum f\"{u}r Astronomie, Institut f\"{u}r Theoretische Astrophysik, Albert-Ueberle-Str 2, D-69120 Heidelberg, Germany}
\printaffiliation{izw_heidelberg}{Universit\"{a}t Heidelberg, Interdisziplin\"{a}res Zentrum f\"{u}r Wissenschaftliches Rechnen, Im Neuenheimer Feld 225, 69120 Heidelberg, Germany}
\printaffiliation{radcliffe}{Elizabeth S. and Richard M. Cashin Fellow at the Radcliffe Institute for Advanced Studies at Harvard University, 10 Garden Street, Cambridge, MA 02138, U.S.A.}
\printaffiliation{upenn}{Department of Physics and Astronomy, University of Pennsylvania, 209 S. 33rd Street, Philadelphia, PA 19104, USA}
\printaffiliation{COOL}{Cosmic Origins Of Life (COOL) Research DAO, \href{https://coolresearch.io}{https://coolresearch.io}}
\printaffiliation{iff_csic}{Instituto de Física Fundamental (CSIC), Calle Serrano 121-123, 28006, Madrid, Spain}
\printaffiliation{umass}{Department of Astronomy, University of Massachusetts, Amherst, MA 01003, USA}
\printaffiliation{kiaa_pku}{Kavli Institute for Astronomy and Astrophysics, Peking University, Beijing 100871, People's Republic of China}
\printaffiliation{pku_astro}{Department of Astronomy, School of Physics, Peking University, Beijing, 100871, People's Republic of China}
\printaffiliation{uconn}{Department of Physics, University of Connecticut, 196A Auditorium Road, Unit 3046, Storrs, CT 06269, USA}
\printaffiliation{iaa_taipei}{Academia Sinica Institute of Astronomy and Astrophysics, Astronomy-Mathematics Building, AS/NTU No.1, Sec. 4, Roosevelt Rd, Taipei 10617, Taiwan}
\printaffiliation{kansas}{Department of Physics and Astronomy, University of Kansas, 1251 Wescoe Hall Drive, Lawrence, KS 66045, USA}
\printaffiliation{chalmers}{Space, Earth and Environment Department, Chalmers University of Technology, SE-412 96 Gothenburg, Sweden}
\printaffiliation{clap}{{Como Lake centre for AstroPhysics (CLAP), DiSAT, Universit{\`a} dell’Insubria, via Valleggio 11, 22100 Como, Italy}}
\printaffiliation{iop_epfl}{Institute of Physics, Laboratory for Galaxy Evolution and Spectral Modelling, EPFL, Observatoire de Sauverny, Chemin Pegasi 51, 1290 Versoix, Switzerland}
\printaffiliation{oaq}{Observatorio Astron\'omico de Quito, Observatorio Astron\'omico Nacional, Escuela Polit\'ecnica Nacional, 170403, Quito, Ecuador}
\printaffiliation{inaf_arcetri}{INAF Arcetri Astrophysical Observatory, Largo Enrico Fermi 5, Firenze, 50125, Italy}
\printaffiliation{jbca}{Jodrell Bank Centre for Astrophysics, The University of Manchester, Manchester M13 9PL, UK}
\printaffiliation{uva}{Dept. of Astronomy, University of Virginia, Charlottesville, Virginia 22904, USA}
\printaffiliation{leiden}{Leiden Observatory, Leiden University, P.O. Box 9513, 2300 RA Leiden, The Netherlands}
\printaffiliation{iaa_csic}{Instituto de Astrof\'{i}sica de Andaluc\'{i}a, CSIC, Glorieta de la Astronomía s/n, 18008 Granada, Spain}
\printaffiliation{anu}{Research School of Astronomy and Astrophysics, Australian National University, Canberra, ACT 2611, Australia}
\printaffiliation{ias}{Institute for Advanced Study, 1 Einstein Drive, Princeton, NJ 08540, USA}
\printaffiliation{ucl}{Department of Physics and Astronomy, University College London, Gower Street, London WC1E 6BT, UK}
\printaffiliation{ari_heidelberg}{Astronomisches Rechen-Institut, Zentrum f\"{u}r Astronomie der Universit\"{a}t Heidelberg, M\"{o}nchhofstra\ss e 12-14, D-69120 Heidelberg, Germany}
\printaffiliation{eso_chile}{European Southern Observatory, Alonso de C\'ordova, 3107, Vitacura, Santiago 763-0355, Chile}
\printaffiliation{jao}{Joint ALMA Observatory, Alonso de C\'ordova, 3107, Vitacura, Santiago 763-0355, Chile}
\printaffiliation{ipm}{Institute for Research in Fundamental Sciences (IPM), School of Astronomy, Tehran, Iran}
\printaffiliation{nanjing}{School of Astronomy and Space Science, Nanjing University, 163 Xianlin Avenue, Nanjing 210023, P.R.China}
\printaffiliation{nanjing_key}{Key Laboratory of Modern Astronomy and Astrophysics (Nanjing University), Ministry of Education, Nanjing 210023, P.R.China}
\printaffiliation{villanova}{Department of Physics, Villanova University, 800 E. Lancaster Ave., Villanova, PA 19085, USA}
\printaffiliation{mit}{Haystack Observatory, Massachusetts Institute of Technology, 99 Millstone Road, Westford, MA 01886, USA}
\printaffiliation{umd}{University of Maryland, Department of Astronomy, College Park, MD 20742-2421, USA}
\printaffiliation{ulaserena}{Departamento de Astronom\'ia, Universidad de La Serena, Ra\'ul Bitr\'an 1305, La Serena, Chile}
\printaffiliation{gbo}{Green Bank Observatory, P.O. Box 2, Green Bank, WV 24944, USA}
\printaffiliation{utokyo}{Institute of Astronomy, The University of Tokyo, Mitaka, Tokyo 181-0015, Japan}
\printaffiliation{aberystwyth}{Department of Physics, Aberystwyth University, Ceredigion, Cymru, SY23 3BZ, UK}

\onecolumn
\appendix
\section{Continuum zoom-ins}
\label{sec:appendix_zooms}

\begin{figure*}
    \includegraphics[width=\textwidth]{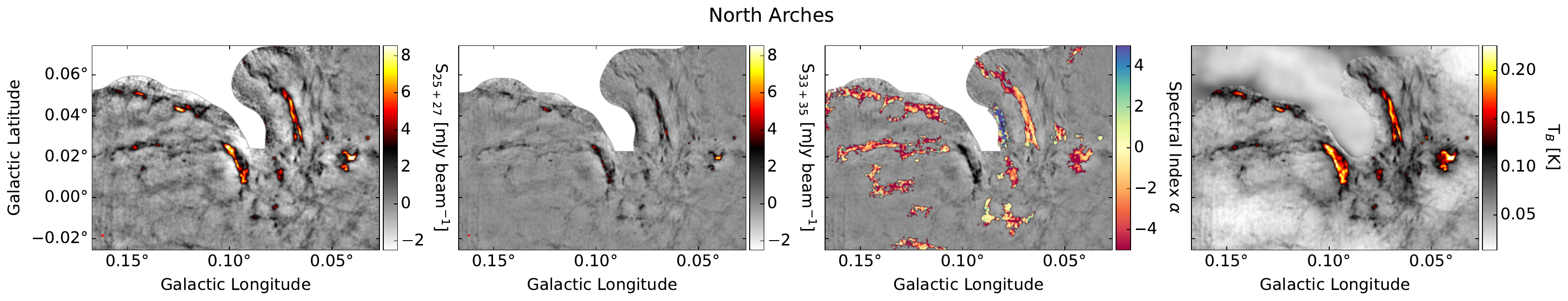}
    \caption{Cutout image from the full mosaic showing the spw25+27 image (left), spw33+35 (center-left), spectral index $\alpha$ (center-right), and the feathered aggregate continuum plus MUSTANG (right).
    This cutout shows the northern Arched Filaments.
    These are HII regions dominated by free-free emission, but their spectral indices appear to be predominantly highly negative.
    }
    \label{fig:northarches}
\end{figure*}

\begin{figure*}
    \centering
    \includegraphics[width=\textwidth]{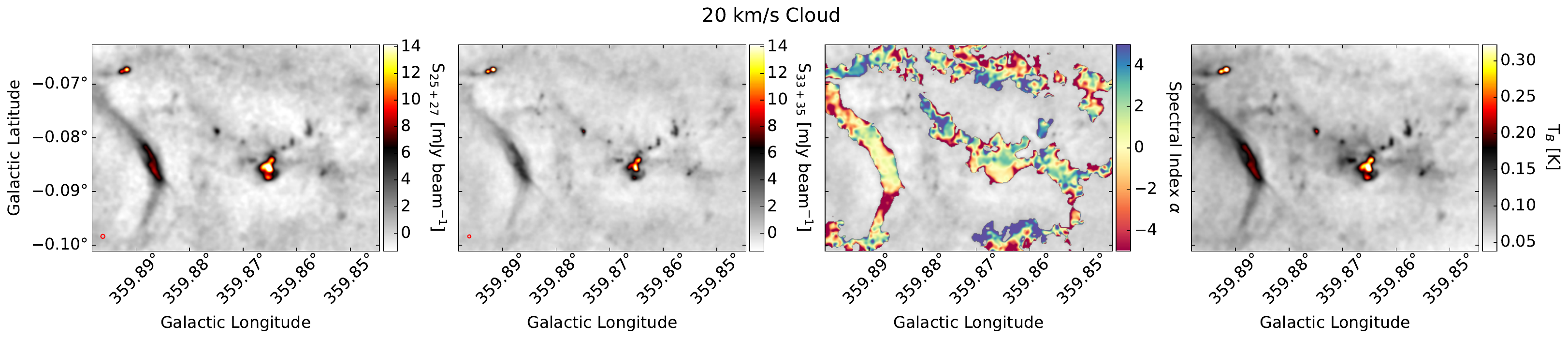}
    \caption{Like Figure \ref{fig:northarches}, but for the 20 km/s cloud.  
    High spectral index ($\alpha\sim3-4$), dust-dominated regions are associated with compact, clumpy emission, while flatter indexes are associated with the filament on the left, suggesting that it is dominated by emission from hot plasma.}
    \label{fig:20kms}
\end{figure*}

\begin{figure*}
    \centering
    \includegraphics[width=\textwidth]{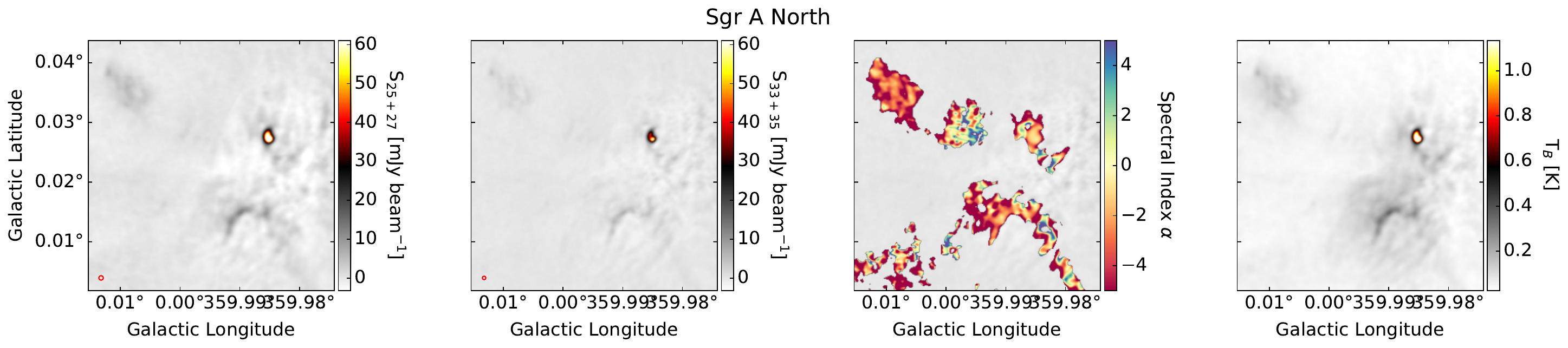}
    \caption{Like Figure \ref{fig:northarches}, but for the region north of Sgr A* (see Figure \ref{fig:overviewzoomlabeled}).
    Most of the emission is flat or negative spectral index, indicating that there is little dust in this region.
    }
    \label{fig:sgrAnorth}
\end{figure*}

\begin{figure*}
    \centering
    \includegraphics[width=\textwidth]{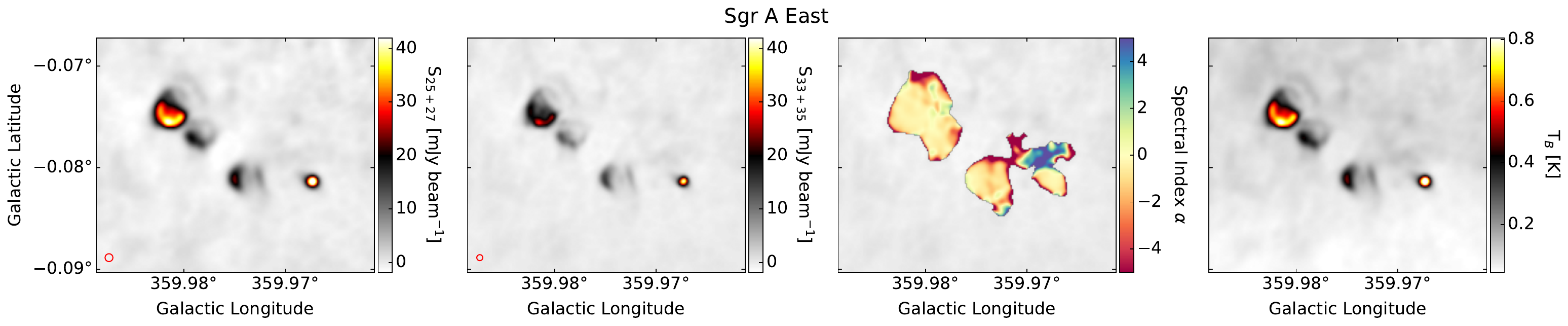}
    \caption{Like Figure \ref{fig:northarches}, but for Sgr A East (see Figure \ref{fig:overviewzoomlabeled}).
    The circular objects are known HII regions, and they correspondingly exhibit flat spectral indices, though there is also a faint ridge of high-index (dust-dominated) emission connecting the rightmost two.
    }
    \label{fig:sgrAsouth}
\end{figure*}

\begin{figure*}
    \centering
    \includegraphics[width=\textwidth]{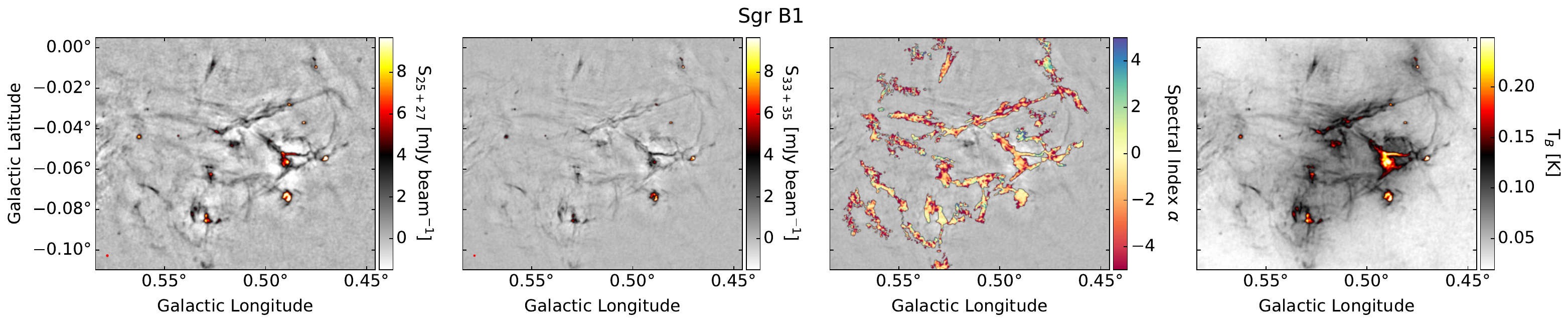}
    \caption{Like Figure \ref{fig:northarches}, but for Sgr B1 (see Figure \ref{fig:overviewzoomlabeled}).
    The Sgr B1 region is comprised primarily of extended HII regions, and the spectral index is correspondingly flat or negative.
    }
    \label{fig:sgrb1}
\end{figure*}

\begin{figure*}
    \centering
    \includegraphics[width=\textwidth]{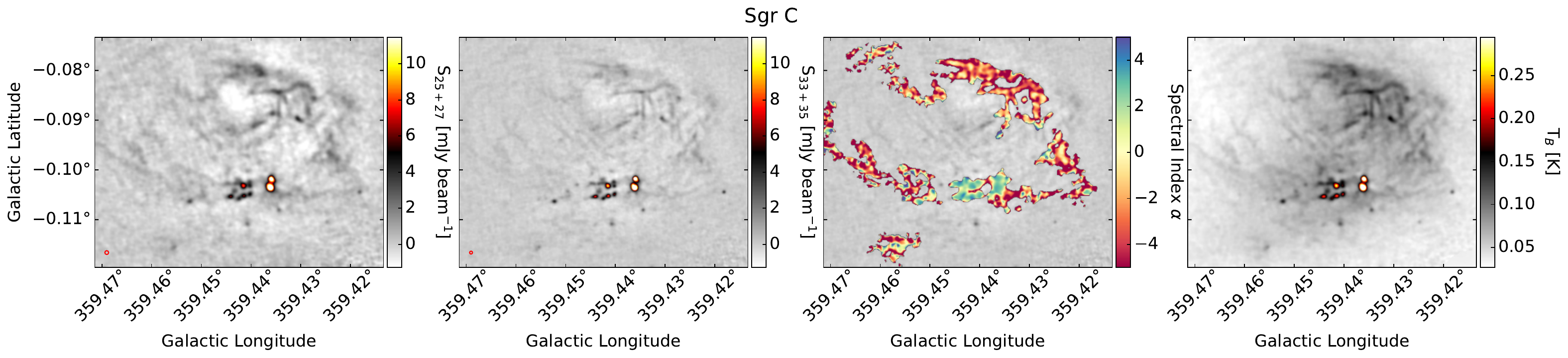}
    \caption{Like Figure \ref{fig:northarches}, but for Sgr C (see Figure \ref{fig:overviewzoomlabeled}).
    The dust cores in the bottom center are clearly steep index ($\alpha\sim3-4$) and dust-dominated, while the remainder of the extended emission is produced by plasma emission \citep{Bally2024}.
    }
    \label{fig:sgrc}
\end{figure*}

\begin{figure*}
    \centering
    \includegraphics[width=\textwidth]{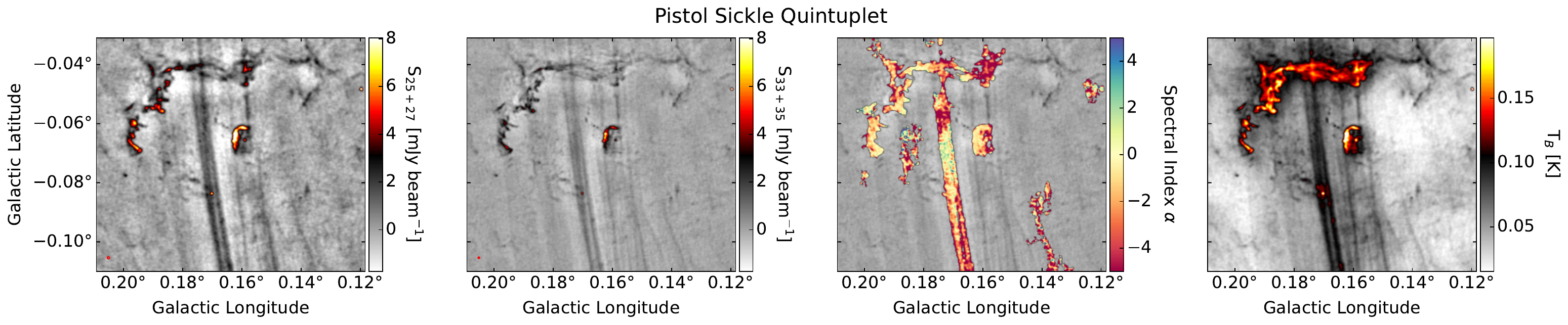}
    \caption{Like Figure \ref{fig:northarches}, but for the Pistol, Sickle, and Quintuplet region (see Figure \ref{fig:overviewzoomlabeled}).
    The large-scale arced feature is the Sickle HII region, while the vertical streaked features are part of the nonthermal arch filaments.
    The emission in this region is mostly flat or negative ($\alpha < 2$), consistent with plasma emission.
    }
    \label{fig:pistol}
\end{figure*}

\begin{figure*}
    \centering
    \includegraphics[width=\textwidth]{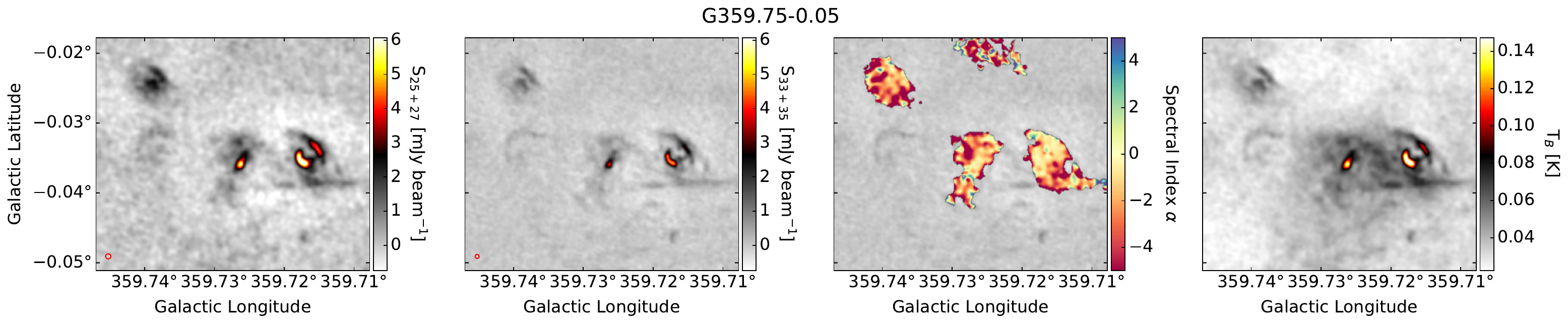}
    \caption{Like Figure \ref{fig:northarches}, but for the G359.75-0.05 region (see Figure \ref{fig:overviewzoomlabeled}).
    Because of the flat spectral indices ($\alpha\lesssim0$), this region appears to be dominated by plasma emission.
    }
    \label{fig:g35975}
\end{figure*}

\begin{figure*}
    \centering
    \includegraphics[width=\textwidth]{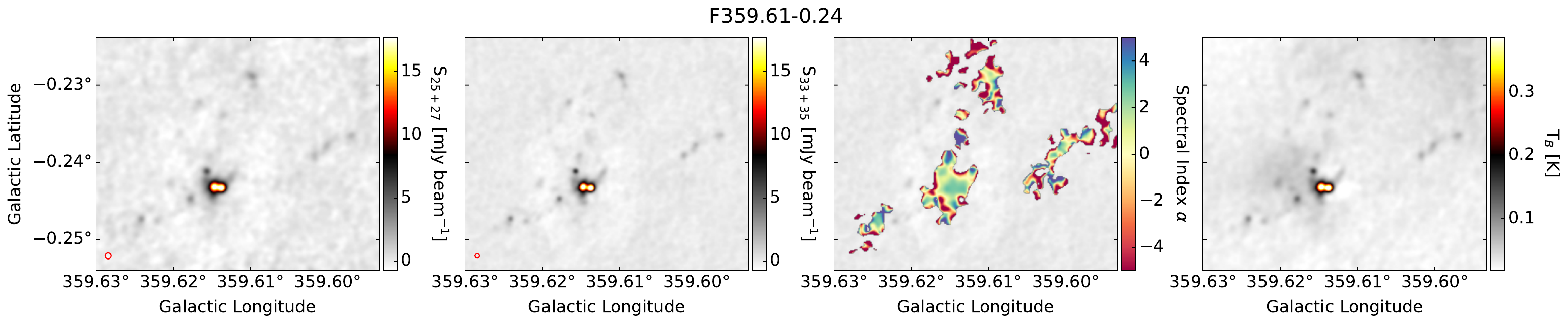}
    \caption{Like Figure \ref{fig:northarches}, but for the G359.61-0.24 region (see Figure \ref{fig:overviewzoomlabeled}).
    This region is a known foreground cloud \citep{Reid2019} and is dust-dominated based on its spectral indexes.
    }
    \label{fig:f35961}
\end{figure*}

\begin{figure*}
    \centering
    \includegraphics[width=\textwidth]{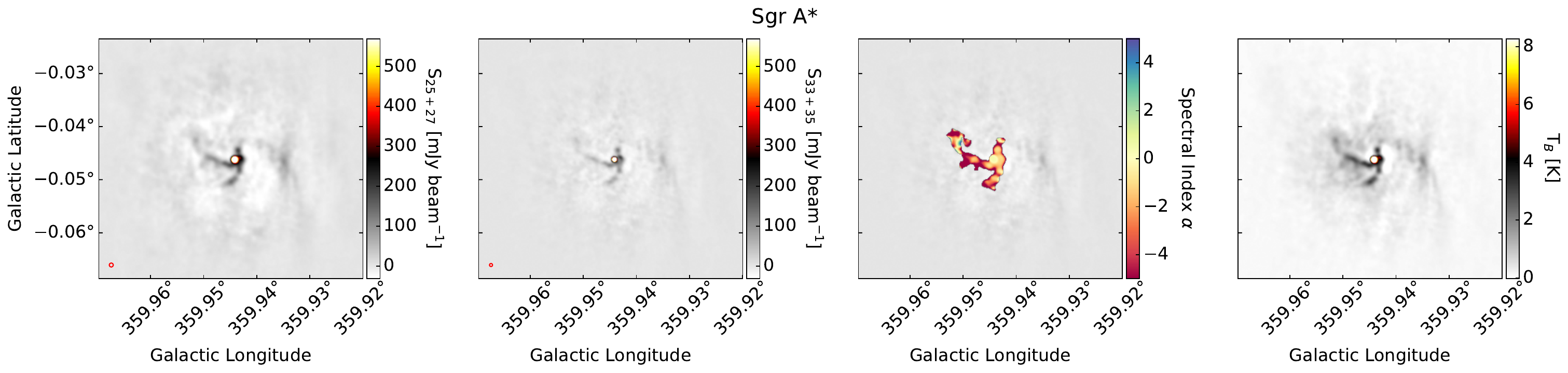}
    \caption{Like Figure \ref{fig:northarches}, but for Sgr A* (see Figure \ref{fig:overviewzoomlabeled}).
    The negative spectral index observed in the minispiral is consistent with previous measurements \citep{Kunneriath2012}.
    }
    \label{fig:sgrastar}
\end{figure*}

\begin{figure*}
    \centering
    \includegraphics[width=\textwidth]{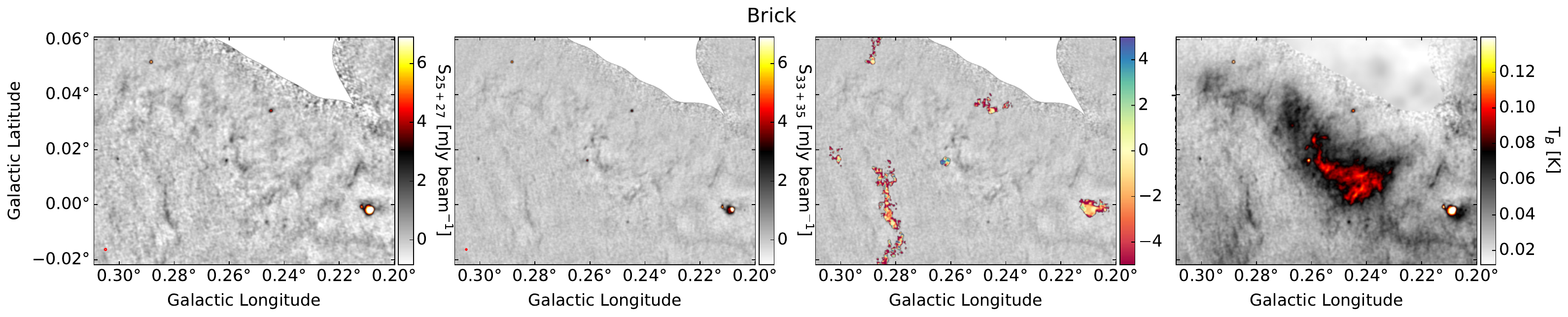}
    \caption{Like Figure \ref{fig:northarches}, but for The Brick (see Figure \ref{fig:overviewzoomlabeled}).
    As noted in Section \ref{sec:brickcompare}, the Brick is mostly resolved out in ACES images, so there is no measurement of the spectral index within the main body of the cloud.
    Only the known small star-cluster-forming region \citep{Walker2021,Bulatek2023} has a dustlike index.
    }
    \label{fig:brick}
\end{figure*}

\begin{figure*}
    \centering
    \includegraphics[width=\textwidth]{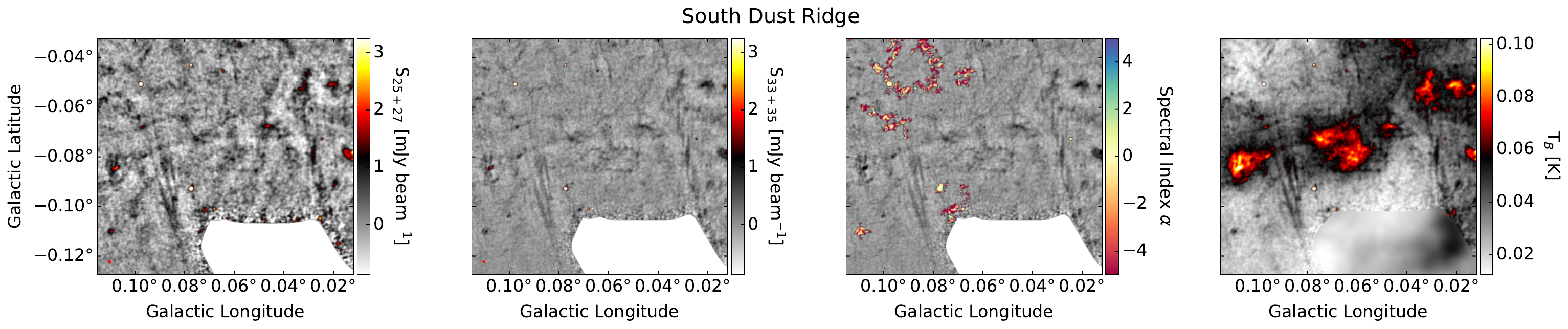}
    \caption{Like Figure \ref{fig:northarches}, but for the southern dust ridge  region, also know as the `Three Little Pigs' \citep[see Figure \ref{fig:overviewzoomlabeled}][]{Battersby2020}.
    Like The Brick (Fig. \ref{fig:brick}), most of the cloud emission is resolved out.
    The extended emission with detected spectral index is most likely plasma.
    }
    \label{fig:southdustridge}
\end{figure*}

\begin{figure*}
    \centering
    \includegraphics[width=\textwidth]{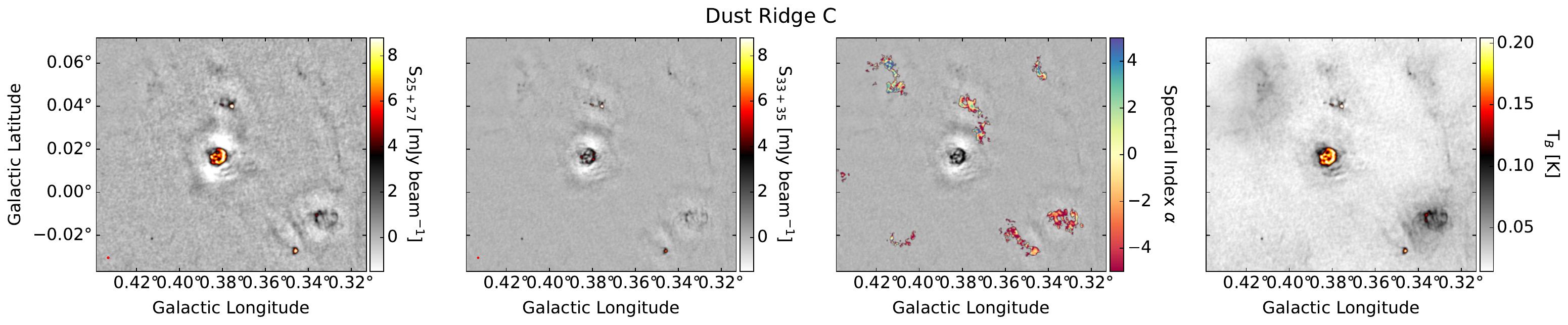}
    \caption{Like Figure \ref{fig:northarches}, but for Dust Ridge Cloud C (see Figure \ref{fig:overviewzoomlabeled}).
    Because of the deep negative bowls around the central HII region in this image, it did not pass the signal-to-noise threshold used to make the $\alpha$ maps.
    }
    \label{fig:dustridgeC}
\end{figure*}

\begin{figure*}
    \centering
    \includegraphics[width=\textwidth]{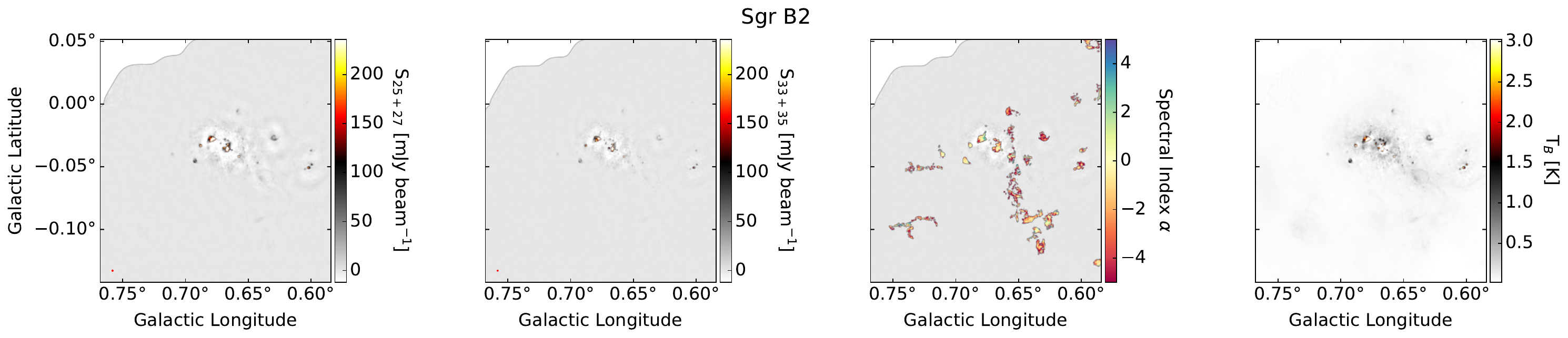}
    \caption{Like Figure \ref{fig:northarches}, but for the Sgr B2 region (see Figure \ref{fig:overviewzoomlabeled}).
    }
    \label{fig:sgrb2}
\end{figure*}

\begin{figure*}
    \centering
    \includegraphics[width=\textwidth]{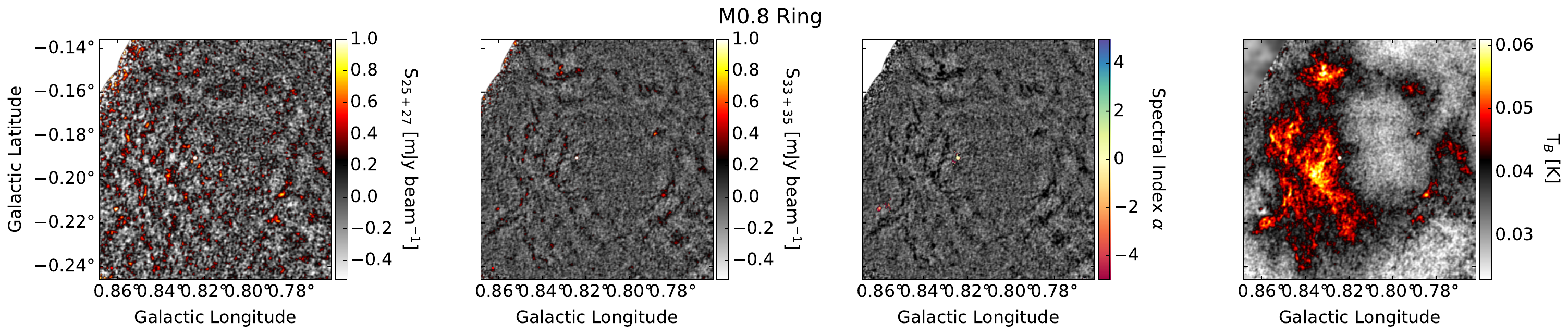}
    \caption{Like Figure \ref{fig:northarches}, but for the \citet{Nonhebel2024} ring region (see Figure \ref{fig:overviewzoomlabeled}).
    Like the Brick (Fig \ref{fig:brick}) and the southern dust ridge (Fig \ref{fig:southdustridge}), most of the cloud emission exists only on large angular scales and is resolved out.
    The rightmost panel convincingly demonstrates that the ring does indeed exist in dust, but is difficult to pick out in 12m-only data.
    }
    \label{fig:g0pt8ring}
\end{figure*}

\begin{figure*}
    \centering
    \includegraphics[width=\textwidth]{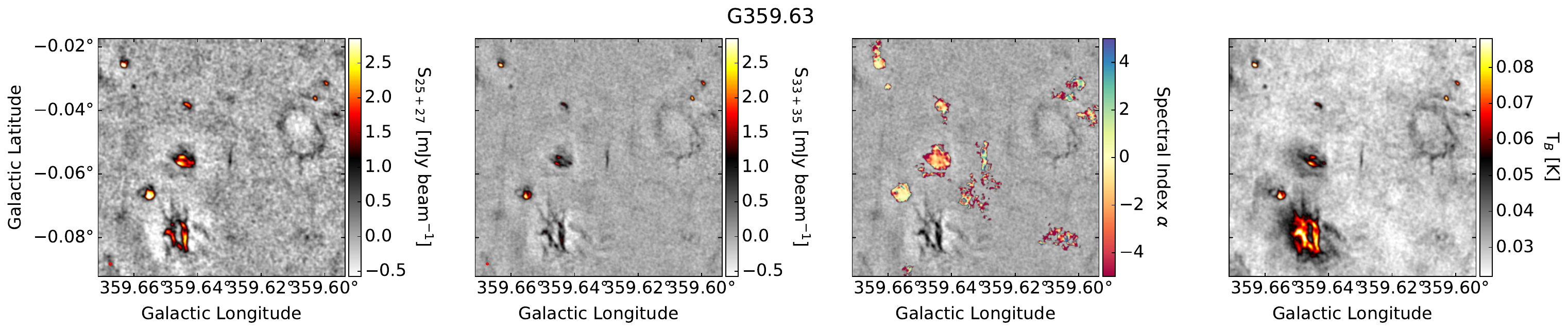}
    \caption{Like Figure \ref{fig:northarches}, but for the region G359.63 (see Figure \ref{fig:overviewzoomlabeled})}
    \label{fig:G35963}
\end{figure*}

We show a series of zoom figures cut out from the regions shown in Figure \ref{fig:overviewzoomlabeled} in this Appendix \referee{in Figures \ref{fig:northarches}-\ref{fig:G35963}}.
These regions highlight most of the high signal-to-noise extended emission regions with significant measurements of the spectral index (see Section \ref{sec:spectralindex}).
\referee{The spectral index measurement used in these figures is the `manual' index from dividing the high-frequency (spw33+35) by the low-frequency (spw25+27) ACES images.}
A brief analysis of each figure is given in its caption.

\section{Observation Metadata Table}
\referee{
The full metadata table, Table \ref{tab:observation_metadata_12m}, is presented here.
The table includes the field name, time of each execution, the configuration,
the precipitable water vapor (PWV) during the execution, the number of pointings in the scheduling block, the exposure time per mosaic pointing,
and the expected resolution and largest angular scale.
}

\begin{table*}[htp]

\caption{12m Observation Metadata}
\resizebox{\textwidth}{!}{
\begin{tabular}{rrrrrrrrrrrrrrrrrrr}
\label{tab:observation_metadata_12m}
Center & Field & Execution Dates & EB ID & Configuration & PWV & Time & N(P) & Res. & LAS \\
 &  &  &  &  & $\mathrm{mm}$ & $\mathrm{min}$ &  & $\mathrm{{}^{\prime\prime}}$ & $\mathrm{{}^{\prime\prime}}$ \\
\hline
G359.448-00.183 & ad & \makecell[l]{2022-01-05 19:14:51.991. \vspace{1.5mm}} & \makecell[l]{\texttt{uid://A002/Xf49cca/Xde0f}. \vspace{1.5mm}} & C-4 & \makecell[l]{1.09} & \makecell[l]{70.2} & 61 & 1.1 & 17 \\
G359.463-00.090 & aj & \makecell[l]{2022-05-09 09:34:09.882,\\ 2022-05-09 07:19:14.47,\\ 2022-05-08 09:39:43.05. \vspace{1.5mm}} & \makecell[l]{\texttt{uid://A002/Xf8b429/X2c01},,\\ \texttt{uid://A002/Xf8b429/X1ea4},,\\ \texttt{uid://A002/Xf89be2/Xe4fd}. \vspace{1.5mm}} & C-3 & \makecell[l]{0.35,\\ 0.40,\\ 1.68} & \makecell[l]{65.5,\\65.5,\\65.4} & 147 & 1.6 & 22 \\
G359.511-00.166 & ag & \makecell[l]{2022-05-16 05:15:44.486,\\ 2022-05-15 10:21:36.497. \vspace{1.5mm}} & \makecell[l]{\texttt{uid://A002/Xf8f6a9/X1912b},,\\ \texttt{uid://A002/Xf8f6a9/X12db3}. \vspace{1.5mm}} & C-3 & \makecell[l]{1.12,\\ 0.38} & \makecell[l]{63.0,\\62.8} & 140 & 1.3 & 19 \\
G359.543-00.041 & e & \makecell[l]{2022-09-22 01:23:55.951,\\ 2022-09-21 00:52:07.005. \vspace{1.5mm}} & \makecell[l]{\texttt{uid://A002/Xfeb5e6/Xb30},,\\ \texttt{uid://A002/Xfe90b7/Xbd5d}. \vspace{1.5mm}} & C-4 & \makecell[l]{0.67,\\ 0.96} & \makecell[l]{65.1,\\65.1} & 147 & 1.1 & 21 \\
G359.590-00.121 & ae & \makecell[l]{2022-05-13 07:23:53.702,\\ 2022-05-13 06:18:54.818. \vspace{1.5mm}} & \makecell[l]{\texttt{uid://A002/Xf8f6a9/X1947},,\\ \texttt{uid://A002/Xf8f6a9/Xe40}. \vspace{1.5mm}} & C-3 & \makecell[l]{0.30,\\ 0.32} & \makecell[l]{64.2,\\64.4} & 144 & 1.6 & 18 \\
G359.602-00.207 & l & \makecell[l]{2022-05-16 06:22:59.613,\\ 2022-05-15 09:14:38.483. \vspace{1.5mm}} & \makecell[l]{\texttt{uid://A002/Xf8f6a9/X199a6},,\\ \texttt{uid://A002/Xf8f6a9/X1251d}. \vspace{1.5mm}} & C-3 & \makecell[l]{1.24,\\ 0.38} & \makecell[l]{66.1,\\66.3} & 149 & 1.2 & 18 \\
G359.622+00.003 & t & \makecell[l]{2022-04-30 08:27:09.933,\\ 2022-04-29 07:31:20.612. \vspace{1.5mm}} & \makecell[l]{\texttt{uid://A002/Xf84513/Xd44b},,\\ \texttt{uid://A002/Xf84513/X1a3e}. \vspace{1.5mm}} & C-3 & \makecell[l]{1.33,\\ 1.56} & \makecell[l]{60.7,\\60.8} & 134 & 1.6 & 20 \\
G359.668-00.073 & ap & \makecell[l]{2022-05-08 08:34:14.943,\\ 2022-05-08 07:29:00.701,\\ 2022-05-08 06:24:05.21. \vspace{1.5mm}} & \makecell[l]{\texttt{uid://A002/Xf89be2/Xdba9},,\\ \texttt{uid://A002/Xf89be2/Xd5f3},,\\ \texttt{uid://A002/Xf89be2/Xcf2b}. \vspace{1.5mm}} & C-3 & \makecell[l]{1.82,\\ 1.68,\\ 1.87} & \makecell[l]{65.2,\\64.9,\\65.1} & 146 & 1.6 & 19 \\
G359.704+00.031 & y & \makecell[l]{2022-01-25 15:02:47.947,\\ 2022-01-22 15:59:25.572,\\ 2022-01-20 17:22:19.999. \vspace{1.5mm}} & \makecell[l]{\texttt{uid://A002/Xf53eeb/X3071},,\\ \texttt{uid://A002/Xf531c1/X14f4},,\\ \texttt{uid://A002/Xf512ae/X64a7}. \vspace{1.5mm}} & C-3 & \makecell[l]{7.35,\\ 5.84,\\ 6.46} & \makecell[l]{46.4,\\46.7,\\36.3} & 74 & 1.5 & 18 \\
G359.717-00.135 & ab & \makecell[l]{2022-05-16 08:22:51.756,\\ 2022-05-12 08:36:36.281. \vspace{1.5mm}} & \makecell[l]{\texttt{uid://A002/Xf8f6a9/X1ab3f},,\\ \texttt{uid://A002/Xf8d822/X6bdc}. \vspace{1.5mm}} & C-3 & \makecell[l]{1.23,\\ 0.86} & \makecell[l]{55.0,\\54.8} & 92 & 1.2 & 20 \\
G359.748-00.025 & q & \makecell[l]{2022-05-02 09:47:27.976,\\ 2022-05-02 07:36:30.219,\\ 2022-04-29 08:36:06.955. \vspace{1.5mm}} & \makecell[l]{\texttt{uid://A002/Xf859f0/X43e5},,\\ \texttt{uid://A002/Xf859f0/X3e35},,\\ \texttt{uid://A002/Xf84513/X23b6}. \vspace{1.5mm}} & C-3 & \makecell[l]{2.13,\\ 1.99,\\ 1.45} & \makecell[l]{64.9,\\64.5,\\64.7} & 145 & 1.7 & 20 \\
G359.799-00.102 & h & \makecell[l]{2022-05-22 08:58:23.814,\\ 2022-05-18 10:12:05.715. \vspace{1.5mm}} & \makecell[l]{\texttt{uid://A002/Xf96bbc/X3e74},,\\ \texttt{uid://A002/Xf934b1/X30dd}. \vspace{1.5mm}} & C-3 & \makecell[l]{1.17,\\ 0.48} & \makecell[l]{65.0,\\64.3} & 145 & 1.1 & 18 \\
G359.822+00.016 & ai & \makecell[l]{2022-05-14 06:45:09.401,\\ 2022-05-12 09:54:33.106. \vspace{1.5mm}} & \makecell[l]{\texttt{uid://A002/Xf8f6a9/X85e2},,\\ \texttt{uid://A002/Xf8d822/X728a}. \vspace{1.5mm}} & C-3 & \makecell[l]{0.53,\\ 0.78} & \makecell[l]{69.9,\\69.7} & 124 & 1.5 & 22 \\
G359.874-00.055 & p & \makecell[l]{2022-05-25 09:08:16.264,\\ 2022-05-25 08:03:12.566. \vspace{1.5mm}} & \makecell[l]{\texttt{uid://A002/Xf99390/X1b76},,\\ \texttt{uid://A002/Xf99390/X12c7}. \vspace{1.5mm}} & C-4 & \makecell[l]{0.58,\\ 0.56} & \makecell[l]{64.1,\\64.3} & 144 & 1.1 & 17 \\
G359.915-00.110 & x & \makecell[l]{2022-06-12 06:27:16.623,\\ 2022-05-15 10:18:23.434,\\ 2022-05-15 09:04:25.963. \vspace{1.5mm}} & \makecell[l]{\texttt{uid://A002/Xfa2f45/X258d},,\\ \texttt{uid://A002/Xf8f6a9/X12c6a},,\\ \texttt{uid://A002/Xf8f6a9/X122d3}. \vspace{1.5mm}} & C-3 & \makecell[l]{-,\\ -,\\ -} & \makecell[l]{76.6,\\73.9,\\76.7} & 31 & 13 & 91 \\
G359.945+00.165 & v & \makecell[l]{2022-09-01 03:02:48.482. \vspace{1.5mm}} & \makecell[l]{\texttt{uid://A002/Xfdd416/Xfeb}. \vspace{1.5mm}} & C-4 & \makecell[l]{2.16} & \makecell[l]{68.1} & 60 & 1.2 & 17 \\
G359.952-00.007 & aq & \makecell[l]{2022-04-06 10:23:42.108,\\ 2022-04-06 08:27:11.543,\\ 2022-04-05 10:26:55.772,\\ 2022-01-27 15:24:49.778,\\ 2022-01-27 14:08:34.173,\\ 2022-01-22 13:39:45.577,\\ 2022-01-08 14:19:32.254. \vspace{1.5mm}} & \makecell[l]{\texttt{uid://A002/Xf73568/X1553},,\\ \texttt{uid://A002/Xf73568/Xbfc},,\\ \texttt{uid://A002/Xf7294e/X2b09},,\\ \texttt{uid://A002/Xf53eeb/X10748},,\\ \texttt{uid://A002/Xf53eeb/Xfcbb},,\\ \texttt{uid://A002/Xf531c1/X545},,\\ \texttt{uid://A002/Xf49cca/X23d82}. \vspace{1.5mm}} & C-4 & \makecell[l]{-,\\ -,\\ -,\\ -,\\ -,\\ -,\\ -} & \makecell[l]{71.4,\\75.1,\\74.9,\\75.3,\\75.7,\\75.7,\\76.1} & 52 & 12 & 76 \\
G359.972+00.089 & am & \makecell[l]{2022-09-20 02:09:48.971,\\ 2022-09-20 01:02:43.885. \vspace{1.5mm}} & \makecell[l]{\texttt{uid://A002/Xfe90b7/X823e},,\\ \texttt{uid://A002/Xfe90b7/X773a}. \vspace{1.5mm}} & C-3 & \makecell[l]{1.23,\\ 1.44} & \makecell[l]{65.4,\\65.7} & 148 & 1.7 & 23 \\
G359.999-00.077 & aa & \makecell[l]{2022-09-16 01:30:14.236,\\ 2022-09-15 01:25:32.273. \vspace{1.5mm}} & \makecell[l]{\texttt{uid://A002/Xfe62c1/Xa4d3},,\\ \texttt{uid://A002/Xfe62c1/X2274}. \vspace{1.5mm}} & C-3 & \makecell[l]{0.50,\\ 0.37} & \makecell[l]{67.9,\\68.1} & 120 & 1.3 & 19 \\
G000.029+00.041 & r & \makecell[l]{2022-09-24 01:08:56.699,\\ 2022-09-23 01:09:43.032. \vspace{1.5mm}} & \makecell[l]{\texttt{uid://A002/Xfec857/X9f6},,\\ \texttt{uid://A002/Xfeb5e6/X4fa5}. \vspace{1.5mm}} & C-3 & \makecell[l]{0.52,\\ 0.71} & \makecell[l]{63.2,\\63.5} & 141 & 1.7 & 21 \\
G000.073+00.183 & a & \makecell[l]{2022-08-27 03:03:13.83,\\ 2022-08-26 23:54:12.505,\\ 2022-08-24 02:21:34.477. \vspace{1.5mm}} & \makecell[l]{\texttt{uid://A002/Xfd8c42/X15d3},,\\ \texttt{uid://A002/Xfd8c42/X525},,\\ \texttt{uid://A002/Xfd6eb8/Xafb}. \vspace{1.5mm}} & C-4 & \makecell[l]{-,\\ -,\\ -} & \makecell[l]{75.5,\\74.8,\\74.4} & 45 & 12 & 88 \\
G000.082-00.037 & o & \makecell[l]{2022-09-30 00:44:54.677,\\ 2022-09-28 23:24:31.463. \vspace{1.5mm}} & \makecell[l]{\texttt{uid://A002/Xfef6d0/X269f},,\\ \texttt{uid://A002/Xfef6d0/Xd8}. \vspace{1.5mm}} & C-3 & \makecell[l]{0.45,\\ 1.05} & \makecell[l]{63.1,\\63.2} & 141 & 1.7 & 23 \\
G000.123-00.115 & af & \makecell[l]{2022-09-17 02:00:32.793,\\ 2022-09-17 00:48:37.892,\\ 2022-09-16 22:21:51.301. \vspace{1.5mm}} & \makecell[l]{\texttt{uid://A002/Xfe83cd/X125c},,\\ \texttt{uid://A002/Xfe83cd/X286},,\\ \texttt{uid://A002/Xfe62c1/X10de7}. \vspace{1.5mm}} & C-3 & \makecell[l]{1.07,\\ 0.57,\\ 0.51} & \makecell[l]{58.0,\\58.1,\\45.8} & 130 & 1.7 & 20 \\
G000.162+00.010 & f & \makecell[l]{2022-09-28 00:20:09.967,\\ 2022-09-27 00:34:32.054. \vspace{1.5mm}} & \makecell[l]{\texttt{uid://A002/Xfee03e/X822c},,\\ \texttt{uid://A002/Xfee03e/X2d13}. \vspace{1.5mm}} & C-3 & \makecell[l]{0.56,\\ 0.64} & \makecell[l]{62.7,\\62.9} & 140 & 1.7 & 22 \\
G000.208-00.063 & s & \makecell[l]{2022-09-26 23:26:43.52,\\ 2022-09-25 00:57:01.376. \vspace{1.5mm}} & \makecell[l]{\texttt{uid://A002/Xfee03e/X2787},,\\ \texttt{uid://A002/Xfed4ee/X1e3}. \vspace{1.5mm}} & C-3 & \makecell[l]{0.70,\\ 0.66} & \makecell[l]{61.6,\\61.2} & 136 & 1.7 & 21 \\
\hline
\end{tabular}
}\par

        The \emph{Time} column gives the execution time of each execution block.
        The \emph{PWV} is the median of the water column in the \texttt{ASDM\_CALWVR} table.
        \emph{Res.} is the resolution in arcseconds.
        \emph{N(P)} is the number of pointings.
        
\end{table*}
\begin{table*}[htp]
\addtocounter{table}{-1}
\caption{12m Observation Metadata continued}
\resizebox{\textwidth}{!}{
\begin{tabular}{rrrrrrrrrrrrrrrrrrr}
\label{tab:observation_metadata_12m}
Center & Field & Execution Dates & EB ID & Configuration & PWV & Time & N(P) & Res. & LAS \\
 &  &  &  &  & $\mathrm{mm}$ & $\mathrm{min}$ &  & $\mathrm{{}^{\prime\prime}}$ & $\mathrm{{}^{\prime\prime}}$ \\
\hline
G000.286-00.019 & c & \makecell[l]{2022-09-19 23:57:01.095,\\ 2022-09-18 00:40:55.253. \vspace{1.5mm}} & \makecell[l]{\texttt{uid://A002/Xfe90b7/X7016},,\\ \texttt{uid://A002/Xfe90b7/X2ab}. \vspace{1.5mm}} & C-3 & \makecell[l]{1.64,\\ 2.34} & \makecell[l]{64.7,\\64.5} & 145 & 1.7 & 18 \\
G000.300+00.063 & ao & \makecell[l]{2022-01-25 16:00:32.766,\\ 2022-01-22 16:56:41.945. \vspace{1.5mm}} & \makecell[l]{\texttt{uid://A002/Xf53eeb/X323e},,\\ \texttt{uid://A002/Xf531c1/X16b2}. \vspace{1.5mm}} & C-3 & \makecell[l]{7.06,\\ 5.89} & \makecell[l]{57.0,\\57.1} & 95 & 1.6 & 19 \\
G000.323-00.089 & d & \makecell[l]{2022-09-15 00:16:13.897,\\ 2022-09-12 02:37:20.469. \vspace{1.5mm}} & \makecell[l]{\texttt{uid://A002/Xfe62c1/X1871},,\\ \texttt{uid://A002/Xfe3986/X9083}. \vspace{1.5mm}} & C-3 & \makecell[l]{0.37,\\ 2.06} & \makecell[l]{68.1,\\67.4} & 119 & 1.2 & 19 \\
G000.367+00.029 & as & \makecell[l]{2022-05-23 09:12:08.001,\\ 2022-05-22 10:18:49.464. \vspace{1.5mm}} & \makecell[l]{\texttt{uid://A002/Xf96bbc/Xbcc0},,\\ \texttt{uid://A002/Xf96bbc/X41f2}. \vspace{1.5mm}} & C-3 & \makecell[l]{0.84,\\ 0.70} & \makecell[l]{65.1,\\64.8} & 146 & 1.1 & 19 \\
G000.412-00.050 & k & \makecell[l]{2022-08-29 03:46:11.755,\\ 2022-05-27 09:23:02.46. \vspace{1.5mm}} & \makecell[l]{\texttt{uid://A002/Xfda0c7/Xacb5},,\\ \texttt{uid://A002/Xf99bb0/Xf742}. \vspace{1.5mm}} & C-4 & \makecell[l]{3.55,\\ 0.41} & \makecell[l]{63.9,\\63.8} & 143 & 1.1 & 18 \\
G000.453+00.060 & al & \makecell[l]{2022-05-13 08:17:48.547,\\ 2022-05-10 09:05:23.071. \vspace{1.5mm}} & \makecell[l]{\texttt{uid://A002/Xf8f6a9/X216b},,\\ \texttt{uid://A002/Xf8b429/Xa4c5}. \vspace{1.5mm}} & C-3 & \makecell[l]{0.29,\\ 0.63} & \makecell[l]{48.6,\\48.6} & 78 & 1.6 & 20 \\
G000.464-00.104 & ac & \makecell[l]{2022-08-30 02:48:07.786. \vspace{1.5mm}} & \makecell[l]{\texttt{uid://A002/Xfdb69b/X1e74}. \vspace{1.5mm}} & C-4 & \makecell[l]{1.80} & \makecell[l]{66.1} & 58 & 1.2 & 16 \\
G000.491-00.002 & ar & \makecell[l]{2022-05-04 08:57:44.492,\\ 2022-05-02 08:42:30.42. \vspace{1.5mm}} & \makecell[l]{\texttt{uid://A002/Xf87688/X931},,\\ \texttt{uid://A002/Xf859f0/X4107}. \vspace{1.5mm}} & C-3 & \makecell[l]{0.46,\\ 2.05} & \makecell[l]{66.0,\\66.0} & 148 & 1.6 & 18 \\
G000.538-00.079 & an & \makecell[l]{2022-05-14 07:53:05.542,\\ 2022-05-13 10:21:58.756. \vspace{1.5mm}} & \makecell[l]{\texttt{uid://A002/Xf8f6a9/X8d0c},,\\ \texttt{uid://A002/Xf8f6a9/X2f1d}. \vspace{1.5mm}} & C-3 & \makecell[l]{0.50,\\ 0.30} & \makecell[l]{62.0,\\61.6} & 137 & 1.5 & 21 \\
G000.569+00.044 & z & \makecell[l]{2022-05-14 09:28:16.65,\\ 2022-05-13 09:20:16.997. \vspace{1.5mm}} & \makecell[l]{\texttt{uid://A002/Xf8f6a9/X9732},,\\ \texttt{uid://A002/Xf8f6a9/X269f}. \vspace{1.5mm}} & C-3 & \makecell[l]{0.50,\\ 0.30} & \makecell[l]{61.4,\\61.6} & 136 & 1.6 & 19 \\
G000.601-00.131 & u & \makecell[l]{2022-09-03 03:19:48.814. \vspace{1.5mm}} & \makecell[l]{\texttt{uid://A002/Xfddca6/X8451}. \vspace{1.5mm}} & C-4 & \makecell[l]{0.60} & \makecell[l]{44.2} & 36 & 1.2 & 18 \\
G000.616-00.033 & b & \makecell[l]{2022-05-26 09:44:28.024,\\ 2022-05-26 08:38:20.722,\\ 2022-05-26 04:46:55.553. \vspace{1.5mm}} & \makecell[l]{\texttt{uid://A002/Xf99bb0/X4ae1},,\\ \texttt{uid://A002/Xf99bb0/X3f39},,\\ \texttt{uid://A002/Xf99bb0/X2491}. \vspace{1.5mm}} & C-4 & \makecell[l]{0.77,\\ 0.70,\\ 0.53} & \makecell[l]{65.2,\\65.3,\\65.7} & 147 & 1.1 & 18 \\
G000.637+00.070 & ak & \makecell[l]{2022-04-07 08:35:00.62,\\ 2022-04-06 12:46:15.491,\\ 2022-04-06 11:43:25.533. \vspace{1.5mm}} & \makecell[l]{\texttt{uid://A002/Xf73ead/X1d0f},,\\ \texttt{uid://A002/Xf73568/X1f60},,\\ \texttt{uid://A002/Xf73568/X1c47}. \vspace{1.5mm}} & C-2 & \makecell[l]{-,\\ -,\\ -} & \makecell[l]{64.9,\\62.8,\\65.0} & 25 & 12 & 92 \\
G000.665-00.111 & i & \makecell[l]{2022-05-16 09:28:22.185,\\ 2022-05-16 07:27:40.23. \vspace{1.5mm}} & \makecell[l]{\texttt{uid://A002/Xf8f6a9/X1b140},,\\ \texttt{uid://A002/Xf8f6a9/X1a382}. \vspace{1.5mm}} & C-3 & \makecell[l]{1.18,\\ 1.30} & \makecell[l]{64.4,\\64.6} & 144 & 1.2 & 18 \\
G000.692+00.005 & m & \makecell[l]{2022-01-27 15:01:24.852,\\ 2022-01-26 15:28:51.686. \vspace{1.5mm}} & \makecell[l]{\texttt{uid://A002/Xf53eeb/X1055f},,\\ \texttt{uid://A002/Xf53eeb/Xa0e8}. \vspace{1.5mm}} & C-3 & \makecell[l]{8.34,\\ 5.72} & \makecell[l]{65.2,\\65.8} & 112 & 1.6 & 20 \\
G000.715-00.180 & w & \makecell[l]{2022-05-16 10:36:44.557,\\ 2022-05-10 08:03:41.495. \vspace{1.5mm}} & \makecell[l]{\texttt{uid://A002/Xf8f6a9/X1b8a0},,\\ \texttt{uid://A002/Xf8b429/X991b}. \vspace{1.5mm}} & C-3 & \makecell[l]{1.13,\\ 0.60} & \makecell[l]{67.4,\\67.7} & 118 & 1.4 & 20 \\
G000.741-00.064 & j & \makecell[l]{2022-05-04 11:22:09.805,\\ 2022-05-04 10:01:48.267. \vspace{1.5mm}} & \makecell[l]{\texttt{uid://A002/Xf87688/Xf8e},,\\ \texttt{uid://A002/Xf87688/Xc13}. \vspace{1.5mm}} & C-3 & \makecell[l]{0.35,\\ 0.41} & \makecell[l]{63.0,\\63.2} & 141 & 1.7 & 22 \\
G000.774-00.244 & n & \makecell[l]{2022-08-29 02:41:37.35. \vspace{1.5mm}} & \makecell[l]{\texttt{uid://A002/Xfda0c7/Xa597}. \vspace{1.5mm}} & C-4 & \makecell[l]{3.44} & \makecell[l]{60.1} & 51 & 1.1 & 16 \\
G000.792-00.142 & g & \makecell[l]{2022-05-15 07:15:22.553,\\ 2022-05-14 10:34:49.257. \vspace{1.5mm}} & \makecell[l]{\texttt{uid://A002/Xf8f6a9/X113a4},,\\ \texttt{uid://A002/Xf8f6a9/X9e67}. \vspace{1.5mm}} & C-3 & \makecell[l]{0.47,\\ 0.50} & \makecell[l]{63.5,\\63.1} & 141 & 1.3 & 21 \\
G000.835-00.215 & ah & \makecell[l]{2022-08-30 22:57:42.302,\\ 2022-08-30 03:10:02.834,\\ 2022-08-29 03:30:50.762,\\ 2022-08-29 02:12:44.32. \vspace{1.5mm}} & \makecell[l]{\texttt{uid://A002/Xfdb69b/X7a3c},,\\ \texttt{uid://A002/Xfdb69b/X20cc},,\\ \texttt{uid://A002/Xfda0c7/Xa983},,\\ \texttt{uid://A002/Xfda0c7/X9fc0}. \vspace{1.5mm}} & C-4 & \makecell[l]{-,\\ -,\\ -,\\ -} & \makecell[l]{79.0,\\76.4,\\76.6,\\78.4} & 49 & 13 & 92 \\
\hline
\end{tabular}
}\par

        The \emph{Time} column gives the execution time of each execution block.
        The \emph{PWV} is the median of the water column in the \texttt{ASDM\_CALWVR} table.
        \emph{Res.} is the resolution in arcseconds.
        \emph{N(P)} is the number of pointings.
        
\end{table*}
 
\section{Continuum data summary table}
\referee{A summary of the continuum imaging is reported in Table \ref{tab:continuum_data_summary_0-35}.
This table reports the properties of the individual continuum images that comprise the full ACES mosaic.
The naming convention is given in \S \ref{subsec:data_appendix}.
}
\begin{table*}[htp]

\caption{Continuum Data Summary}
\resizebox{\textwidth}{!}{
\begin{tabular}{llllllllll}
\label{tab:continuum_data_summary_0-35}
Region & $\theta_{\mathrm{maj}}$ & $\theta_{\mathrm{min}}$ & BPA & Robust & SPWs & $S_{\mathrm{peak}}$ & $\sigma_{\mathrm{MAD}}$ & $T_{\mathrm{B,peak}}$ & $\sigma_{\mathrm{MAD,mK}}$ \\
 & $\mathrm{{}^{\prime\prime}}$ & $\mathrm{{}^{\prime\prime}}$ & $\mathrm{{}^{\circ}}$ &  &  & $\mathrm{mJy\,beam^{-1}}$ & $\mathrm{mJy\,beam^{-1}}$ & $\mathrm{K}$ & $\mathrm{mK}$ \\
\hline
G359.448-00.183 & 1.9 & 1.1 & -75 & r1.0 & 25,27 & 3.2 & 0.093 & 0.24 & 6.9 \\
G359.448-00.183 & 1.7 & 0.98 & -75 & r1.0 & 33,35 & 2.5 & 0.063 & 0.19 & 4.8 \\
G359.448-00.183 & 1.7 & 1.0 & -75 & r1.0 & all & 2.8 & 0.068 & 0.20 & 5.0 \\
G359.463-00.090 & 2.2 & 1.6 & 86 & r0.0 & 25,27 & 14 & 0.17 & 0.67 & 7.8 \\
G359.463-00.090 & 1.8 & 1.4 & 86 & r0.0 & 33,35 & 19 & 0.095 & 0.91 & 4.7 \\
G359.463-00.090 & 1.9 & 1.5 & 86 & r0.0 & all & 17 & 0.094 & 0.76 & 4.4 \\
G359.511-00.166 & 1.8 & 1.4 & -79 & r0.5 & 25,27 & 1.3 & 0.11 & 0.086 & 7.4 \\
G359.511-00.166 & 1.5 & 1.1 & -78 & r0.5 & 33,35 & 0.71 & 0.063 & 0.052 & 4.6 \\
G359.511-00.166 & 1.6 & 1.2 & -79 & r0.5 & all & 0.79 & 0.066 & 0.054 & 4.6 \\
G359.543-00.041 & 2.6 & 1.6 & 90 & r0.0 & 25,27 & 2.5 & 0.18 & 0.099 & 7.0 \\
G359.543-00.041 & 2.0 & 1.4 & 89 & r0.0 & 33,35 & 3.5 & 0.11 & 0.16 & 4.8 \\
G359.543-00.041 & 2.1 & 1.4 & 90 & r0.0 & all & 3.1 & 0.10 & 0.13 & 4.5 \\
G359.590-00.121 & 2.1 & 1.7 & 82 & r0.5 & 25,27 & 0.69 & 0.13 & 0.031 & 5.6 \\
G359.590-00.121 & 1.8 & 1.4 & 83 & r0.5 & 33,35 & 0.56 & 0.061 & 0.027 & 3.0 \\
G359.590-00.121 & 1.9 & 1.5 & 83 & r0.5 & all & 0.49 & 0.064 & 0.023 & 3.0 \\
G359.602-00.207 & 1.6 & 1.4 & -77 & r0.5 & 25,27 & 12 & 0.14 & 0.86 & 11 \\
G359.602-00.207 & 1.3 & 1.2 & -74 & r0.5 & 33,35 & 15 & 0.080 & 1.2 & 6.1 \\
G359.602-00.207 & 1.4 & 1.2 & -76 & r0.5 & all & 13 & 0.080 & 0.99 & 5.9 \\
G359.622+00.003 & 2.2 & 1.8 & 74 & r0.5 & 25,27 & 1.4 & 0.12 & 0.060 & 5.0 \\
G359.622+00.003 & 1.8 & 1.5 & 73 & r0.5 & 33,35 & 2.1 & 0.077 & 0.093 & 3.4 \\
G359.622+00.003 & 1.9 & 1.6 & 73 & r0.5 & all & 1.9 & 0.079 & 0.081 & 3.4 \\
G359.668-00.073 & 2.1 & 1.7 & 79 & r0.5 & 25,27 & 2.5 & 0.12 & 0.11 & 5.5 \\
G359.668-00.073 & 1.8 & 1.4 & 80 & r0.5 & 33,35 & 1.9 & 0.075 & 0.094 & 3.6 \\
G359.668-00.073 & 1.9 & 1.5 & 80 & r0.5 & all & 2.1 & 0.080 & 0.097 & 3.6 \\
G359.704+00.031 & 2.1 & 1.6 & -86 & r0.5 & 25,27 & 0.66 & 0.11 & 0.032 & 5.3 \\
G359.704+00.031 & 1.7 & 1.3 & -86 & r0.5 & 33,35 & 0.69 & 0.067 & 0.037 & 3.6 \\
G359.704+00.031 & 1.9 & 1.4 & -86 & r0.5 & all & 0.63 & 0.068 & 0.032 & 3.5 \\
G359.717-00.135 & 1.9 & 1.6 & -87 & r0.5 & 25,27 & 3.6 & 0.095 & 0.20 & 5.3 \\
G359.717-00.135 & 1.5 & 1.3 & -86 & r0.5 & 33,35 & 2.9 & 0.064 & 0.17 & 3.9 \\
G359.717-00.135 & 1.6 & 1.4 & -84 & r0.5 & all & 3.1 & 0.064 & 0.17 & 3.7 \\
G359.748-00.025 & 2.3 & 1.7 & 88 & r0.5 & 25,27 & 4.4 & 0.12 & 0.18 & 5.1 \\
G359.748-00.025 & 1.9 & 1.5 & 89 & r0.5 & 33,35 & 3.5 & 0.068 & 0.16 & 3.1 \\
G359.748-00.025 & 2.0 & 1.5 & 89 & r0.5 & all & 3.7 & 0.075 & 0.16 & 3.2 \\
G359.799-00.102 & 1.6 & 1.2 & -83 & r0.5 & 25,27 & 1.5 & 0.096 & 0.13 & 8.1 \\
G359.799-00.102 & 1.4 & 1.0 & -82 & r0.5 & 33,35 & 1.0 & 0.061 & 0.093 & 5.6 \\
\hline
\end{tabular}
}\par

\end{table*}
\begin{table*}[htp]
\addtocounter{table}{-1}
\caption{Continuum Data Summary continued}
\resizebox{\textwidth}{!}{
\begin{tabular}{llllllllll}
\label{tab:continuum_data_summary_35-70}
Region & $\theta_{\mathrm{maj}}$ & $\theta_{\mathrm{min}}$ & BPA & Robust & SPWs & $S_{\mathrm{peak}}$ & $\sigma_{\mathrm{MAD}}$ & $T_{\mathrm{B,peak}}$ & $\sigma_{\mathrm{MAD,mK}}$ \\
 & $\mathrm{{}^{\prime\prime}}$ & $\mathrm{{}^{\prime\prime}}$ & $\mathrm{{}^{\circ}}$ &  &  & $\mathrm{mJy\,beam^{-1}}$ & $\mathrm{mJy\,beam^{-1}}$ & $\mathrm{K}$ & $\mathrm{mK}$ \\
\hline
G359.799-00.102 & 1.4 & 1.0 & -82 & r0.5 & all & 1.2 & 0.064 & 0.11 & 5.6 \\
G359.822+00.016 & 2.4 & 1.6 & 78 & r0.5 & 25,27 & 8.3 & 0.13 & 0.34 & 5.3 \\
G359.822+00.016 & 2.0 & 1.4 & 77 & r0.5 & 33,35 & 9.3 & 0.070 & 0.42 & 3.2 \\
G359.822+00.016 & 2.1 & 1.4 & 78 & r0.5 & all & 8.9 & 0.074 & 0.38 & 3.2 \\
G359.874-00.055 & 1.7 & 1.3 & 85 & r1.0 & 25,27 & 9.6 & 0.16 & 0.70 & 12 \\
G359.874-00.055 & 1.5 & 1.1 & 87 & r1.0 & 33,35 & 14 & 0.096 & 1.0 & 7.0 \\
G359.874-00.055 & 1.5 & 1.2 & 86 & r1.0 & all & 12 & 0.11 & 0.86 & 7.6 \\
G359.915-00.108 & 2.5 & 1.7 & -88 & r0.5 & 25,27 & 1.3 & 0.11 & 0.048 & 4.3 \\
G359.915-00.108 & 2.1 & 1.4 & -86 & r0.5 & 33,35 & 0.63 & 0.075 & 0.027 & 3.2 \\
G359.915-00.108 & 2.2 & 1.5 & -87 & r0.5 & all & 0.59 & 0.079 & 0.024 & 3.2 \\
G359.945+00.165 & 1.8 & 1.1 & -80 & r0.5 & 25,27 & 6.0 & 0.11 & 0.48 & 9.1 \\
G359.945+00.165 & 1.5 & 0.95 & -79 & r0.5 & 33,35 & 4.1 & 0.045 & 0.36 & 3.9 \\
G359.945+00.165 & 1.6 & 1.0 & -79 & r0.5 & all & 5.0 & 0.067 & 0.41 & 5.5 \\
G359.952-00.007 & 1.6 & 1.3 & -84 & r0.5 & 25,27 & 2400 & 0.26 & 190 & 20 \\
G359.952-00.007 & 1.4 & 1.1 & -82 & r0.5 & 33,35 & 2600 & 0.13 & 220 & 11 \\
G359.952-00.007 & 1.4 & 1.1 & -82 & r0.5 & all & 2500 & 0.15 & 200 & 12 \\
G359.999-00.077 & 1.8 & 1.5 & -71 & r0.5 & 25,27 & 33 & 0.17 & 2.0 & 10 \\
G359.999-00.077 & 1.5 & 1.3 & -61 & r0.5 & 33,35 & 27 & 0.095 & 1.7 & 6.0 \\
G359.999-00.077 & 1.6 & 1.3 & -63 & r0.5 & all & 28 & 0.11 & 1.7 & 6.6 \\
G000.029+00.041 & 2.8 & 1.5 & -85 & r0.0 & 25,27 & 44 & 0.23 & 1.7 & 9.0 \\
G000.029+00.041 & 2.1 & 1.3 & -85 & r0.0 & 33,35 & 33 & 0.13 & 1.5 & 5.7 \\
G000.029+00.041 & 2.3 & 1.4 & -85 & r0.0 & all & 37 & 0.15 & 1.5 & 6.0 \\
G000.073+00.184 & 1.6 & 1.1 & -76 & r0.5 & 25,27 & 0.56 & 0.095 & 0.050 & 8.4 \\
G000.073+00.184 & 1.4 & 0.96 & -73 & r0.5 & 33,35 & 0.67 & 0.057 & 0.064 & 5.4 \\
G000.073+00.184 & 1.4 & 1.0 & -74 & r0.5 & all & 0.74 & 0.059 & 0.067 & 5.3 \\
G000.082-00.037 & 2.5 & 1.7 & 88 & r0.0 & 25,27 & 21 & 0.21 & 0.81 & 8.4 \\
G000.082-00.037 & 1.9 & 1.5 & 87 & r0.0 & 33,35 & 17 & 0.13 & 0.75 & 5.8 \\
G000.082-00.037 & 2.0 & 1.5 & 87 & r0.0 & all & 18 & 0.15 & 0.75 & 6.2 \\
G000.123-00.115 & 2.5 & 1.5 & -90 & r0.0 & 25,27 & 35 & 0.19 & 1.5 & 7.9 \\
G000.123-00.115 & 2.0 & 1.3 & 88 & r0.0 & 33,35 & 30 & 0.097 & 1.5 & 4.7 \\
G000.123-00.115 & 2.1 & 1.3 & 88 & r0.0 & all & 32 & 0.11 & 1.4 & 5.1 \\
G000.162+00.010 & 2.6 & 1.6 & -90 & r0.0 & 25,27 & 22 & 0.27 & 0.90 & 11 \\
G000.162+00.010 & 2.0 & 1.3 & 89 & r0.0 & 33,35 & 17 & 0.14 & 0.83 & 6.7 \\
G000.162+00.010 & 2.2 & 1.4 & 89 & r0.0 & all & 19 & 0.17 & 0.83 & 7.7 \\
G000.208-00.063 & 2.3 & 1.6 & 88 & r0.0 & 25,27 & 8.1 & 0.21 & 0.36 & 9.4 \\
\hline
\end{tabular}
}\par

\end{table*}
\begin{table*}[htp]
\addtocounter{table}{-1}
\caption{Continuum Data Summary continued}
\resizebox{\textwidth}{!}{
\begin{tabular}{llllllllll}
\label{tab:continuum_data_summary_70-105}
Region & $\theta_{\mathrm{maj}}$ & $\theta_{\mathrm{min}}$ & BPA & Robust & SPWs & $S_{\mathrm{peak}}$ & $\sigma_{\mathrm{MAD}}$ & $T_{\mathrm{B,peak}}$ & $\sigma_{\mathrm{MAD,mK}}$ \\
 & $\mathrm{{}^{\prime\prime}}$ & $\mathrm{{}^{\prime\prime}}$ & $\mathrm{{}^{\circ}}$ &  &  & $\mathrm{mJy\,beam^{-1}}$ & $\mathrm{mJy\,beam^{-1}}$ & $\mathrm{K}$ & $\mathrm{mK}$ \\
\hline
G000.208-00.063 & 1.8 & 1.4 & 85 & r0.0 & 33,35 & 5.8 & 0.13 & 0.27 & 6.2 \\
G000.208-00.063 & 2.0 & 1.5 & 86 & r0.0 & all & 6.4 & 0.15 & 0.28 & 6.5 \\
G000.286-00.019 & 2.5 & 1.6 & 87 & r0.0 & 25,27 & 4.5 & 0.19 & 0.19 & 7.9 \\
G000.286-00.019 & 1.9 & 1.3 & 86 & r0.0 & 33,35 & 3.4 & 0.11 & 0.17 & 5.6 \\
G000.286-00.019 & 2.1 & 1.4 & 86 & r0.0 & all & 3.6 & 0.12 & 0.16 & 5.3 \\
G000.300+00.063 & 2.4 & 1.5 & -87 & r0.5 & 25,27 & 6.5 & 0.15 & 0.29 & 6.7 \\
G000.300+00.063 & 2.0 & 1.3 & -88 & r0.5 & 33,35 & 5.8 & 0.093 & 0.28 & 4.4 \\
G000.300+00.063 & 2.1 & 1.4 & -87 & r0.5 & all & 6.2 & 0.094 & 0.28 & 4.2 \\
G000.323-00.089 & 1.8 & 1.3 & -63 & r0.5 & 25,27 & 5.8 & 0.12 & 0.43 & 8.7 \\
G000.323-00.089 & 1.5 & 1.1 & -61 & r0.5 & 33,35 & 4.1 & 0.075 & 0.32 & 5.9 \\
G000.323-00.089 & 1.5 & 1.1 & -61 & r0.5 & all & 4.7 & 0.076 & 0.35 & 5.7 \\
G000.367+00.029 & 1.8 & 1.1 & -74 & r0.5 & 25,27 & 30 & 0.16 & 2.4 & 13 \\
G000.367+00.029 & 1.5 & 0.94 & -74 & r0.5 & 33,35 & 43 & 0.089 & 3.7 & 7.8 \\
G000.367+00.029 & 1.6 & 0.99 & -74 & r0.5 & all & 37 & 0.089 & 3.1 & 7.4 \\
G000.412-00.050 & 1.8 & 1.0 & -79 & r0.5 & 25,27 & 4.7 & 0.11 & 0.41 & 9.5 \\
G000.412-00.050 & 1.5 & 0.88 & -78 & r0.5 & 33,35 & 3.6 & 0.078 & 0.34 & 7.3 \\
G000.412-00.050 & 1.6 & 0.92 & -78 & r0.5 & all & 4.0 & 0.081 & 0.35 & 7.2 \\
G000.453+00.060 & 2.5 & 1.7 & 88 & r0.5 & 25,27 & 1.4 & 0.098 & 0.057 & 3.9 \\
G000.453+00.060 & 2.0 & 1.4 & 89 & r0.5 & 33,35 & 1.3 & 0.064 & 0.057 & 2.8 \\
G000.453+00.060 & 2.2 & 1.5 & 89 & r0.5 & all & 1.3 & 0.067 & 0.055 & 2.7 \\
G000.464-00.104 & 1.7 & 1.1 & -77 & r0.5 & 25,27 & 1.7 & 0.089 & 0.15 & 7.7 \\
G000.464-00.104 & 1.4 & 0.93 & -76 & r0.5 & 33,35 & 0.82 & 0.068 & 0.077 & 6.4 \\
G000.464-00.104 & 1.5 & 0.98 & -77 & r0.5 & all & 1.1 & 0.068 & 0.10 & 6.1 \\
G000.491-00.002 & 2.2 & 1.7 & 84 & r0.5 & 25,27 & 9.8 & 0.18 & 0.43 & 7.9 \\
G000.491-00.002 & 1.8 & 1.4 & 87 & r0.5 & 33,35 & 11 & 0.11 & 0.51 & 5.1 \\
G000.491-00.002 & 1.9 & 1.5 & 87 & r0.5 & all & 9.2 & 0.11 & 0.42 & 5.2 \\
G000.538-00.079 & 2.4 & 1.6 & 82 & r0.5 & 25,27 & 11 & 0.18 & 0.49 & 7.8 \\
G000.538-00.079 & 1.9 & 1.4 & 80 & r0.5 & 33,35 & 8.7 & 0.11 & 0.40 & 5.2 \\
G000.538-00.079 & 2.0 & 1.4 & 81 & r0.5 & all & 9.6 & 0.12 & 0.42 & 5.4 \\
G000.569+00.044 & 2.4 & 1.5 & 81 & r0.0 & 25,27 & 1.9 & 0.13 & 0.087 & 6.0 \\
G000.569+00.044 & 1.9 & 1.3 & 78 & r0.0 & 33,35 & 1.8 & 0.082 & 0.092 & 4.2 \\
G000.569+00.044 & 2.0 & 1.3 & 79 & r0.0 & all & 1.6 & 0.080 & 0.074 & 3.8 \\
G000.601-00.131 & 1.9 & 1.1 & -75 & r0.5 & 25,27 & 0.87 & 0.091 & 0.065 & 6.8 \\
G000.601-00.131 & 1.6 & 0.97 & -75 & r0.5 & 33,35 & 0.58 & 0.054 & 0.047 & 4.4 \\
G000.601-00.131 & 1.7 & 1.0 & -75 & r0.5 & all & 0.62 & 0.054 & 0.047 & 4.1 \\
\hline
\end{tabular}
}\par

\end{table*}
\begin{table*}[htp]
\addtocounter{table}{-1}
\caption{Continuum Data Summary continued}
\resizebox{\textwidth}{!}{
\begin{tabular}{llllllllll}
\label{tab:continuum_data_summary_105-140}
Region & $\theta_{\mathrm{maj}}$ & $\theta_{\mathrm{min}}$ & BPA & Robust & SPWs & $S_{\mathrm{peak}}$ & $\sigma_{\mathrm{MAD}}$ & $T_{\mathrm{B,peak}}$ & $\sigma_{\mathrm{MAD,mK}}$ \\
 & $\mathrm{{}^{\prime\prime}}$ & $\mathrm{{}^{\prime\prime}}$ & $\mathrm{{}^{\circ}}$ &  &  & $\mathrm{mJy\,beam^{-1}}$ & $\mathrm{mJy\,beam^{-1}}$ & $\mathrm{K}$ & $\mathrm{mK}$ \\
\hline
G000.616-00.033 & 1.6 & 1.2 & 87 & r0.5 & 25,27 & 2600 & 0.27 & 230 & 23 \\
G000.616-00.033 & 1.3 & 1.0 & 89 & r0.5 & 33,35 & 2500 & 0.14 & 230 & 13 \\
G000.616-00.033 & 1.4 & 1.1 & 88 & r0.5 & all & 2500 & 0.16 & 220 & 15 \\
G000.637+00.068 & 2.3 & 1.9 & 78 & r0.5 & 25,27 & 0.52 & 0.094 & 0.019 & 3.5 \\
G000.637+00.068 & 1.9 & 1.6 & 80 & r0.5 & 33,35 & 0.33 & 0.061 & 0.014 & 2.6 \\
G000.637+00.068 & 2.0 & 1.6 & 79 & r0.5 & all & 0.39 & 0.064 & 0.016 & 2.6 \\
G000.665-00.111 & 1.6 & 1.4 & -65 & r0.5 & 25,27 & 5.2 & 0.15 & 0.37 & 11 \\
G000.665-00.111 & 1.4 & 1.2 & -63 & r0.5 & 33,35 & 2.7 & 0.090 & 0.21 & 7.0 \\
G000.665-00.111 & 1.4 & 1.2 & -64 & r0.5 & all & 3.3 & 0.098 & 0.24 & 7.2 \\
G000.692+00.005 & 2.2 & 1.7 & 81 & r0.0 & 25,27 & 1900 & 0.27 & 86 & 12 \\
G000.692+00.005 & 1.8 & 1.4 & 82 & r0.0 & 33,35 & 1700 & 0.15 & 85 & 7.7 \\
G000.692+00.005 & 1.9 & 1.5 & 82 & r0.0 & all & 1700 & 0.16 & 82 & 7.5 \\
G000.715-00.180 & 2.3 & 1.6 & -82 & r0.5 & 25,27 & 2.1 & 0.10 & 0.092 & 4.4 \\
G000.715-00.180 & 1.9 & 1.3 & -79 & r0.5 & 33,35 & 1.3 & 0.067 & 0.066 & 3.3 \\
G000.715-00.180 & 2.0 & 1.4 & -80 & r0.5 & all & 1.6 & 0.068 & 0.074 & 3.1 \\
G000.741-00.064 & 2.6 & 1.5 & -87 & r0.0 & 25,27 & 71 & 0.22 & 3.0 & 9.1 \\
G000.741-00.064 & 2.1 & 1.2 & -87 & r0.0 & 33,35 & 35 & 0.13 & 1.7 & 6.1 \\
G000.741-00.064 & 2.2 & 1.3 & -87 & r0.0 & all & 49 & 0.12 & 2.1 & 5.4 \\
G000.774-00.244 & 1.8 & 1.1 & -80 & r1.0 & 25,27 & 2.4 & 0.093 & 0.20 & 7.7 \\
G000.774-00.244 & 1.6 & 0.97 & -79 & r1.0 & 33,35 & 3.8 & 0.070 & 0.31 & 5.7 \\
G000.774-00.244 & 1.6 & 1.0 & -79 & r1.0 & all & 3.3 & 0.072 & 0.26 & 5.8 \\
G000.792-00.142 & 1.9 & 1.4 & 86 & r0.5 & 25,27 & 0.67 & 0.12 & 0.040 & 7.3 \\
G000.792-00.142 & 1.6 & 1.2 & 86 & r0.5 & 33,35 & 0.47 & 0.069 & 0.031 & 4.5 \\
G000.792-00.142 & 1.7 & 1.3 & 87 & r0.5 & all & 0.48 & 0.071 & 0.029 & 4.4 \\
G000.835-00.216 & 1.8 & 1.1 & -72 & r0.5 & 25,27 & 30 & 0.12 & 2.5 & 9.6 \\
G000.835-00.216 & 1.5 & 0.95 & -71 & r0.5 & 33,35 & 30 & 0.066 & 2.6 & 5.8 \\
G000.835-00.216 & 1.5 & 1.0 & -71 & r0.5 & all & 30 & 0.066 & 2.5 & 5.5 \\
\hline
\end{tabular}
}\par

\end{table*}

\section{Data release summary}
\label{subsec:data_appendix}
\noindent
The data products from the ACES survey follow a uniform naming convention. The filenames for the data products presented in this paper (continuum) can be constructed using the templates described below, in conjunction with the information provided in Table \ref{tab:observation_metadata_12m}. 

The following subsections are divided according to different product types which occur at the Member ObsUnitSet (MOUS) and Group ObsUnitSet (GOUS) levels\footnote{For a description of ALMA project structures and naming conventions, please refer to Section 2 of the Cycle 9 QA2 guide: \url{https://almascience.eso.org/documents-and-tools/cycle9/alma-qa2-data-products-for-cycle-9}}.

\begin{table}
\centering
\caption{Field information with central coordinates and MOUS IDs}
\label{tab:field_info}
\begin{tabular}{lllll}
\hline
Central Coordinates & Field ID & 12m MOUS ID & 7m MOUS ID & TP MOUS ID \\
\hline
G359.448-0.183 & ad & \texttt{X15a0\_X14e} & \texttt{X15a0\_X150} & \texttt{X15a0\_X152} \\
G359.463-0.090 & aj & \texttt{X15a0\_X172} & \texttt{X15b4\_X47} & \texttt{X15b4\_X49} \\
G359.511-0.166 & ag & \texttt{X15a0\_X160} & \texttt{X15a0\_X162} & \texttt{X15a0\_X164} \\
G359.543-0.041 & e & \texttt{X15a0\_Xb8} & \texttt{X15b4\_X39} & \texttt{X15a0\_Xbc} \\
G359.590-0.121 & ae & \texttt{X15a0\_X154} & \texttt{X15a0\_X156} & \texttt{X15a0\_X158} \\
G359.602-0.207 & l & \texttt{X15a0\_Xe2} & \texttt{X15a0\_Xe4} & \texttt{X15a0\_Xe6} \\
G359.622+0.003 & t & \texttt{X15a0\_X112} & \texttt{X15a0\_X114} & \texttt{X15a0\_X116} \\
G359.668-0.073 & ap & \texttt{X15a0\_X196} & \texttt{X15a0\_X198} & \texttt{X3577\_X6c5} \\
G359.704+0.031 & y & \texttt{X15a0\_X130} & \texttt{X15a0\_X132} & \texttt{X15a0\_X134} \\
G359.717-0.135 & ab & \texttt{X15a0\_X142} & \texttt{X15a0\_X144} & \texttt{X15a0\_X146} \\
G359.748-0.025 & q & \texttt{X15a0\_X100} & \texttt{X15a0\_X102} & \texttt{X15a0\_X104} \\
G359.799-0.102 & h & \texttt{X15a0\_Xca} & \texttt{X15a0\_Xcc} & \texttt{X15a0\_Xce} \\
G359.822+0.016 & ai & \texttt{X15a0\_X16c} & \texttt{X15a0\_X16e} & \texttt{X15a0\_X170} \\
G359.874-0.055 & p & \texttt{X15a0\_Xfa} & \texttt{X15a0\_Xfc} & \texttt{X15a0\_Xfe} \\
G359.915-0.108 & x & \texttt{X15a0\_X12a} & \texttt{X15a0\_X12c} & \texttt{X15a0\_X12e} \\
G359.945+0.165 & v & \texttt{X15a0\_X11e} & \texttt{X15a0\_X120} & \texttt{X15a0\_X122} \\
G359.952-0.007 & aq & \texttt{X15a0\_X19c} & \texttt{X15a0\_X19e} & \texttt{X15a0\_X1a0} \\
G359.972+0.089 & am & \texttt{X15a0\_X184} & \texttt{X15a0\_X186} & \texttt{X15a0\_X188} \\
G359.999-0.077 & aa & \texttt{X15a0\_X13c} & \texttt{X15a0\_X13e} & \texttt{X15a0\_X140} \\
G000.029+0.041 & r & \texttt{X15a0\_X106} & \texttt{X15a0\_X108} & \texttt{X15a0\_X10a} \\
G000.073+0.184 & a & \texttt{X1590\_X30aa} & \texttt{X1590\_X30ac} & \texttt{X1590\_X30ae} \\
G000.082-0.037 & o & \texttt{X15a0\_Xf4} & \texttt{X15a0\_Xf6} & \texttt{X15a0\_Xf8} \\
G000.123-0.115 & af & \texttt{X15a0\_X15a} & \texttt{X15a0\_X15c} & \texttt{X15a0\_X15e} \\
G000.162+0.010 & f & \texttt{X15a0\_Xbe} & \texttt{X15a0\_Xc0} & \texttt{X15a0\_Xc2} \\
G000.208-0.063 & s & \texttt{X15a0\_X10c} & \texttt{X15a0\_X10e} & \texttt{X15a0\_X110} \\
G000.286-0.019 & c & \texttt{X15a0\_Xac} & \texttt{X15a0\_Xae} & \texttt{X15a0\_Xb0} \\
G000.300+0.063 & ao & \texttt{X15a0\_X190} & \texttt{X15a0\_X192} & \texttt{X15a0\_X194} \\
G000.323-0.089 & d & \texttt{X15a0\_Xb2} & \texttt{X15a0\_Xb4} & \texttt{X15a0\_Xb6} \\
G000.367+0.029 & as & \texttt{X15a0\_X1a8} & \texttt{X15a0\_X1aa} & \texttt{X15a0\_X1ac} \\
G000.412-0.050 & k & \texttt{X15a0\_Xdc} & \texttt{X15a0\_Xde} & \texttt{X15a0\_Xe0} \\
G000.453+0.060 & al & \texttt{X15a0\_X17e} & \texttt{X15a0\_X180} & \texttt{X15a0\_X182} \\
G000.464-0.104 & ac & \texttt{X15a0\_X148} & \texttt{X15a0\_X14a} & \texttt{X15a0\_X14c} \\
G000.491-0.002 & ar & \texttt{X15a0\_X1a2} & \texttt{X15b4\_X4f} & \texttt{X15a0\_X1a6} \\
G000.538-0.079 & an & \texttt{X15a0\_X18a} & \texttt{X15a0\_X18c} & \texttt{X15a0\_X18e} \\
G000.569+0.044 & z & \texttt{X15a0\_X136} & \texttt{X15a9\_X12d3} & \texttt{X15a0\_X13a} \\
G000.601-0.131 & u & \texttt{X15a0\_X118} & \texttt{X15a0\_X11a} & \texttt{X15a0\_X11c} \\
G000.616-0.033 & b & \texttt{X15a0\_Xa6} & \texttt{X15b4\_X37} & \texttt{X15a0\_Xaa} \\
G000.637+0.068 & ak & \texttt{X15b4\_X4b} & \texttt{X15a9\_X12d5} & \texttt{X15b4\_X4d} \\
G000.665-0.111 & i & \texttt{X15a0\_Xd0} & \texttt{X15b4\_Xc5} & \texttt{X15a0\_Xd4} \\
G000.692+0.005 & m & \texttt{X15b4\_X41} & \texttt{X15b4\_X43} & \texttt{X15b4\_X45} \\
G000.715-0.180 & w & \texttt{X15a0\_X124} & \texttt{X15a0\_X126} & \texttt{X15a0\_X128} \\
G000.741-0.064 & j & \texttt{X15b4\_X3d} & \texttt{X15a0\_Xd8} & \texttt{X15b4\_X3f} \\
G000.774-0.244 & n & \texttt{X15a0\_Xee} & \texttt{X15a0\_Xf0} & \texttt{X15a0\_Xf2} \\
G000.792-0.142 & g & \texttt{X15a0\_Xc4} & \texttt{X15a0\_Xc6} & \texttt{X15a0\_Xc8} \\
G000.835-0.216 & ah & \texttt{X15a0\_X166} & \texttt{X15a0\_X168} & \texttt{X15a0\_X16a} \\
\hline
\end{tabular}
\par Only the 12m MOUS UIDs are used for file naming in this paper; the 7m and TP data set IDs are included for reference.
\end{table}
 
\subsection{Member-level products (12m)}
\label{sec:memberlevel}
We release all continuum images for the 45 ACES regions. These are member-level products, and correspond to the images for the 12m array for each field prior to mosaicing.  
As described in \S \ref{sec:imageproducts}, for each field, we produce three images as outlined in Table \ref{tab:filenaming} and \ref{tab:field_info}.

\begin{samepage}
\begin{table}%
    \begin{tabular}{l|l|l}
        \texttt{frqcen}  & \texttt{bw}        & SPWs     \\
        \hline
        \texttt{99.6GHz} & \texttt{bw3.8GHz}  & 33,35       \\
        \texttt{93.7GHz} & \texttt{bw4.7GHz}  & 25,27       \\
        \texttt{86.6GHz} & \texttt{bw1.0GHz}  & 25,27,33,35 \\
    \end{tabular}
    \caption{Image file name convention information.}
    
    \label{tab:filenaming}
\end{table}
\end{samepage}

Following the requirements of the ASA, the filename template is:

\begin{verbatim}
member.uid___A001_{mous-id}.lp_slongmore.{field_coords}.{array}.{frqcen}_{bw}.cont.{contsuffix}.fits
\end{verbatim}

The placeholder template components should be replaced as follows:
\begin{itemize}
    \item \texttt{\{\textbf{mous-id}\}}: The MOUS ID for a given field, as listed in {Table \ref{tab:field_info}}.
    \item \texttt{\{\textbf{field\_coords}\}}: The central coordinates of the field from {Table \ref{tab:field_info} (e.g., \texttt{G000.073+0.184})}.
    \item \texttt{\{\textbf{array}\}}: The ALMA array.  For the continuum data, only \texttt{12m} is distributed.
    \item \texttt{\{\textbf{frqcen}\}}: The frequency center, which corresponds to the spectral windows used (see Table \ref{tab:filenaming}):
    \item \texttt{\{\textbf{bw}\}}: The bandwidth, which also corresponds to the spectral windows used (see Table \ref{tab:filenaming}):
    \item \texttt{\{\textbf{contsuffix}\}}: The continuum file name suffix.  This is in two parts, the Taylor term (either \texttt{tt0} or \texttt{tt1}) and the image type.  Distributed products include \texttt{image}, \texttt{image.pbcor}, \texttt{model}, and \texttt{residual}. 
    
\end{itemize}

For example,
\begin{verbatim}
member.uid___A001_X15b4_X4b.lp_slongmore.G000.64+00.07.12m.cont.99.6GHz_bw3.8GHz.tt1.residual.fits
\end{verbatim}
is the Taylor-term 1 (slope) residual image for the high-frequency (spw33+35) image of G0.64+0.07, field \texttt{ak}.

\clearpage
\subsection{Group-level products (Full CMZ mosaics)}
\label{sec:grouplevel}
We release the full, contiguous mosaics covering the entire ACES footprint. Since these result from the combination of many MOUSs, the full mosaics are at the group level.

The filename template for the full mosaic cubes and associated products is:

\begin{verbatim}
group.uid___A001_X1590_X30a9.lp_slongmore.cmz_mosaic.{arrays}.cont.{frqcen}_{bw}.{suffix}
\end{verbatim}

\begin{samepage}
Where the template components are:
\begin{itemize}
    \item \texttt{\{\textbf{arrays}\}} can be either \texttt{12m} or \texttt{12mGBT-MUSTANG}
    \item \texttt{\{\textbf{frqcen}\}} and \texttt{\{\textbf{bw}\}} are as described in \S \ref{sec:memberlevel}
    \item \texttt{\{\textbf{suffix}\}} defines the specific data product. Products released with this paper are, for each spectral window set:
    \begin{itemize}
        \item \texttt{pbcor.fits}: primary beam corrected image mosaic.
        \item \texttt{maskedrms.pbcor.fits}: the corresponding uncertainty map.
    \end{itemize}

    Additionally, there are \texttt{alpha} maps that only correspond to the aggregate bandwidth \texttt{93.7GHz\_bw4.7GHz}
    \begin{itemize}
        \item \texttt{alpha.fits}: 
        \item \texttt{alpha.maskedrms.fits}: 
        \item \texttt{alpha\_manual.fits}: 
        \item \texttt{alpha\_manual.maskedrms.fits}: 
    \end{itemize}
\end{itemize}

An example for the low-frequency composite 12m-only image is:
\begin{verbatim}
group.uid___A001_X1590_X30a9.lp_slongmore.cmz_mosaic.12m.cont.86.6GHz_bw1.0GHz.pbcor.fits
\end{verbatim}
\end{samepage}

\bsp
\label{lastpage}
\end{document}